%% file: Thesis.tex
\documentclass[a4paper,12pt]{report} 
\usepackage[dvips]{graphicx}
\usepackage{psfrag}
\usepackage{cite}
\usepackage{enumerate}
\usepackage{amssymb}
\usepackage{amsmath}
\usepackage{textcomp}
\usepackage{epsfig}
\usepackage{fancyhdr}

\newlength{\mylength}	
\makeatletter			
\renewcommand\@biblabel[1]{\parbox[c]{\mylength}{\hfill[#1]}}  
\makeatother

\def\nc{\newcommand}

\def\phm{\phantom{-}}
\def\prop{{\cal G}}
\def\self{\Sigma}
\def\sfrac#1#2{{\textstyle\frac#1#2}}
\def\apriori{{\em a priori }}
\def\lsim{\mathrel{\raise.3ex\hbox{$<$\kern-.75em\lower1ex\hbox{$\sim$}}}}
\def\gsim{\mathrel{\raise.3ex\hbox{$>$\kern-.75em\lower1ex\hbox{$\sim$}}}}

\nc{\pp}{{\scriptscriptstyle ||}}
\nc{\deldag}{\mathbin{\partial\mkern-10.5mu\big/}}
\nc{\kdag}{\mathbin{k\mkern-10mu\big/}}
\nc{\shalf}{\ensuremath{\textstyle \frac{1}{2}}}

\newcommand{\blankpage}{	
	\newpage \phantom{blank} \thispagestyle{empty}
}
\newcommand{\emptypage}{	
	\newpage \phantom{empty}
}

\newcommand{\eg}{{\em e.g.}~}
\newcommand{\ie}{{\em i.e.}~}

\newcommand{\beq} {\begin{equation}}
\newcommand{\eeq} {\end{equation}}
\newcommand{\beqa}{\begin{eqnarray}}
\newcommand{\eeqa}{\end{eqnarray}}

\author{Matti Herranen}
\title{Quantum transport theory with nonlocal coherence}
\date{\today}

\begin{document}

\input{titlepage.tex}
\thispagestyle{empty}
\blankpage
\blankpage

\input{Preface.tex}
\thispagestyle{empty}
\blankpage


\input{Listofpub.tex}
\pagenumbering{roman}
\setcounter{page}{1}	
\emptypage              

\tableofcontents        
                        
\newpage

\pagenumbering{arabic} \setcounter{page}{1} 
\input{Introduction.tex}


\input{Chapter1.tex}

\emptypage		

\input{Chapter2.tex}

\input{Chapter3.tex}

\input{Chapter4.tex}
\emptypage		

\input{Conclusions.tex}



{\small

\input{References.tex}
}


\end{document}

%% file: titlepage.tex
\begin{center}

DEPARTMENT OF PHYSICS \\
UNIVERSITY OF JYV\"ASKYL\"A \\
RESEARCH REPORT No. 3/2009

\bigskip
\bigskip
\bigskip

{\Large

{\bf QUANTUM KINETIC THEORY WITH NONLOCAL COHERENCE}

\bigskip
\bigskip

{\bf BY \\ MATTI HERRANEN}

\bigskip
\bigskip
\bigskip

}

Academic Dissertation \\
for the Degree of \\
Doctor of Philosophy 

\bigskip
\bigskip

\emph{
To be presented, by permission of the\\
Faculty of Mathematics and Natural Sciences\\
of the University of Jyv\"askyl\"a,\\
for public examination in Auditorium FYS-1 of the\\
University of Jyv\"askyl\"a on June 12, 2009\\
at 12 o'clock noon
}

\end{center}

\bigskip

\begin{figure}[b]
\begin{center}
\scalebox{.15}{\includegraphics{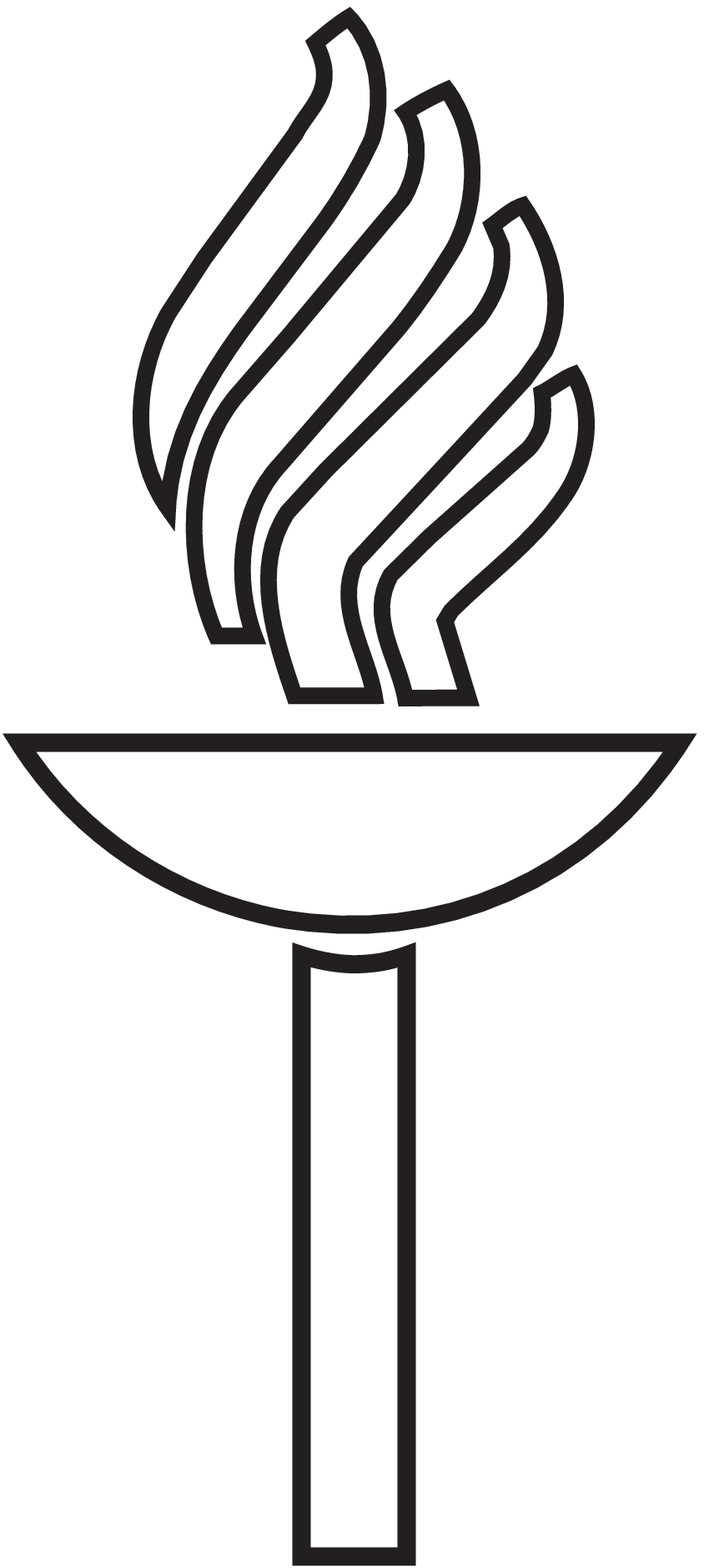}}

\medskip

\begin{minipage}[b]{3.5cm}
\begin{center}
Jyv\"askyl\"a, Finland \\
May 2009
\end{center}
\end{minipage}

\end{center}
\end{figure}

%% file: Preface.tex
\begin{center}
\vspace{1.9cm}
\section*{Preface}
\end{center}

\noindent The work reviewed in this thesis has been carried out during
the years 2005-2009 at the Department of Physics in the University of
Jyv\"askyl\"a.

I am most grateful to my supervisor Doc. Kimmo Kainulainen for
excellent guidance and support during these years. I have been
privileged to share his profound ideas and wide knowledge of particle
physics and cosmology. A major part of this work has been done
in collaboration with Pyry Rahkila, to whom I want to extend my
warmest thanks. I also want to express my gratitude to
Prof. emer. Vesa Ruuskanen who introduced me to Kimmo many years ago,
and to Prof. Mikko Laine and Prof. Kari Rummukainen for careful
reading of the manuscript and valuable comments. I also wish to thank
the staff and friends at the Department of Physics for creating an
inspiring and pleasant working atmosphere.

The financial support from the Jenny and Antti Wihuri foundation, the
Finnish Cultural Foundation, \,the Graduate School of Particle and
Nuclear Physics (GRASPANP), and the Helsinki Institute of Physics
(HIP) is gratefully acknowledged.

Finally, I wish to thank Jenni and my family for their love, support
and patience.


%% file: Listofpub.tex
\chapter*{List of publications}
\addcontentsline{toc}{chapter}{List of publications}

\noindent This thesis is based on the work contained within the following publications:

\begin{enumerate}[I]

\item {\bf Towards a kinetic theory for fermions with quantum coherence} \\
  M.~Herranen, K.~Kainulainen and P.~M.~Rahkila, \\
  Nucl.\ Phys.\  B {\bf 810} (2009) 389
  [arXiv:0807.1415 [hep-ph]].

\item {\bf Quantum kinetic theory for fermions in temporally varying backgrounds} \\
  M.~Herranen, K.~Kainulainen and P.~M.~Rahkila, \\
  JHEP {\bf 0809} (2008) 032
  [arXiv:0807.1435 [hep-ph]].

\item {\bf Kinetic theory for scalar fields with nonlocal quantum coherence} \\
  M.~Herranen, K.~Kainulainen and P.~M.~Rahkila, \\
  JHEP {\bf 0905} (2009) 119
  [arXiv:0812.4029v2 [hep-ph]].

\end{enumerate}

The author has participated equally with K~.Kainulainen and
P.~M.~Rahkila in the development of papers I-III.
A large part of the analytical calculations, in particular in papers
II and III, was carried out by the author. The draft versions of papers II
and III were largely written by the author.
 

%% file: Introduction.tex
%
%

\chapter{Introduction}

A wide range of problems in modern high energy physics and
cosmology involves the dynamics of quantum fields in highly out-of-equilibrium
conditions, including relativistic heavy ion collisions
\cite{Shuryak88,Csernai94,Wong95},
quantum fluctuations in inflationary cosmology
\cite{KolTur90,Linde90,BasTsuWan05}, preheating
after inflation \cite{BasTsuWan05}, and
baryogenesis \cite{DinKus04}. It turns out that the traditional methods of
(vacuum or thermal) quantum field theory (QFT) are not suited to
describe these complicated non-thermal processes, and a new
framework of nonequilibrium QFT \cite{BirDav82,CalHu08} is
needed. Even though a knowledge of the complete dynamics of
interacting quantum fields is clearly beyond reach, there is a
variety of approximative methods that are able catch the essentials of the
problems under study. In a so called kinetic regime with weak
interactions and slowly varying classical backgrounds the standard
methods of quantum kinetic theory reduce the problem considerably to
solving the famous (quantum) Boltzmann transport equations. For many
problems of interest these transport equations provide a remarkably
good approximation for the essentials of nonequilibrium quantum
dynamics. However, certain problems are inherently very sensitive to
quantum coherence (or interference), quantum reflection from a
potential being a typical example. The Boltzmann equation approach
inevitably loses the effects of nonlocal quantum coherence, and thus
is not very well suited to study for example the quantum reflection
problem in electroweak baryogenesis or the particle production in preheating.

In this thesis we present a novel approximation scheme related to the
quantum kinetic theory, that enables us to treat nonlocal quantum coherence in
the presence of decohering collisions with simple enough
Boltzmannian-type transport equations. The key element in our scheme
is the finding of new singular shell solutions in the phase space of
2-point correlation function, that are located at $k_0 = 0$ for
spatially homogeneous problems and at $k_z = 0$ for a static planar
symmetric case. When the complete phase space structure, including
these new coherence solutions in addition to the standard mass-shell
contribution, is inserted in the Kadanoff-Baym (KB) equations for the
correlator, we obtain a closed set of transport equations for the
corresponding on-shell distribution functions, thus giving an extension to the
standard quantum Boltzmann equation to include nonlocal coherence. 

The thesis consists of three original research papers
\cite{paper1}-\cite{paper3} and an introductory and summary part
presented below. In chapter \ref{chap:early_universe} we introduce the
basic mechanisms of (electroweak) baryogenesis and preheating, that
are good examples of highly nonequilibrium processes in the early
universe. In chapter \ref{chap:basic_formalism} we present the basic
formalism of nonequilibrium QFT needed to derive the KB-equations
for two point correlation functions, using the two-particle
irreducible (2PI) effective action method. Chapter \ref{chap:scheme} then
presents a detailed survey of the main contents of our work, the novel
approximation scheme. In chapter \ref{chap:applications} we review a
few applications that we have so far considered using our formalism,
including the Klein problem, (collisionless) quantum reflection from a
$CP$-violating mass wall, and examples of coherent production of
decaying fermionic and scalar particles relevant for preheating.   
Finally, chapter \ref{chap:discussion} contains conclusions and
outlook.

%% file: Chapter1.tex
%
%

\chapter{Nonequilibrium processes in the early universe}
\label{chap:early_universe}

According to modern theories of cosmology and particle physics the
expanding universe has once been in an extremely dense and hot
state (Hot Big Bang scenario), consisting of quantum plasma, which
during the major part of the early evolution is very close to thermal
equilibrium. Besides this overall picture of thermal plasma, however, 
 many crucial processes in the early universe are
 inherently highly non-thermal, including inflation, preheating, and
 baryogenesis. The careful understanding of these processes is of
 primary importance in modern cosmology, providing an important field of
 applications to the methods of nonequilibrium quantum
 field theory. In this chapter we introduce the basic mechanisms of
 baryogenesis, focusing on a model called electroweak baryogenesis (EWBG),
 and the process of preheating after inflation.

%
%

\section{Baryogenesis}

The visible matter content of the universe, such as planets, stars and
interstellar gas, consists of protons, neutrons and electrons. In
astrophysics it is classified as baryonic matter, since the bulk
of the mass is in protons and neutrons that are baryons. There is
strong evidence that no large domains of antimatter exists in the
universe \cite{Steigman76,CohRujGla97}, implying that the universe
has an excess of baryons compared to antibaryons which is called 
baryon asymmetry. The combination of data including several experiments of
the fluctuations of cosmic microwave background (CMB) gives the following
experimental measure of this asymmetry, the average baryon to photon number
ratio in the universe \cite{Bennett_etal03}:
\begin{equation}
\frac{n_B}{n_{\gamma}} = \left(6.1_{-0.2}^{+0.3}\right) \times 10^{-10}\,. 
\label{asymmetry}
\end{equation}
Of course it might be possible that the baryon asymmetry is an initial
condition in the evolution of the universe. Despite being very
unnatural, this explanation for the baryon asymmetry is not
consistent with the cosmological inflation \cite{KolTur90,Linde90},
which is one of the
backbones of modern cosmology explaining the homogeneity and flatness
of the universe as well as the primordial density fluctuations that will
give rise to structure formation and the observed fluctuations in the
CMB spectrum. The problem is that the exponential increase in the size
of the universe by at least a factor $e^{60}$ during the inflationary
period dilutes any prior baryon number to totally negligible level. 
For this reason, we are very tempted to seek out different ways of
creating the baryon asymmetry in the universe after the period
of inflation. 

A process that gives rise to a permanent baryon asymmetry at
cosmological scales is called {\em baryogenesis}. The idea of such a
process originates from Sakharov \cite{Sakharov67}, who presented
three conditions that any model for baryogenesis should necessarily
fulfill: 
\begin{itemize}
\item[1.] Baryon number violation.

\item[2.] $C$ and $CP$ symmetry violations.  

\item[3.] Departure from thermal equilibrium.
\end{itemize} 
The first condition is obvious. If the second is not fulfilled, then
for every reaction producing particles there is a counter-reaction that
produces antiparticles at the same rate. The third condition is the
most interesting for the scope of this work. It follows from the
$CPT$-theorem that the masses of particles and antiparticles are equal,
and consequently the thermal average of the baryon number will vanish
in equilibrium. We conclude that every scenario for baryogenesis must
be a nonequilibrium process.

Several models for baryogenesis have been proposed that fulfill the
Sakharov conditions (for a recent review see \eg. \cite{DinKus04}),
including GUT baryogenesis \cite{KolTur90},
electroweak baryogenesis \cite{KuzRubSha85}, leptogenesis
\cite{FukYan86} and Afflect-Dine baryogenesis \cite{AffDin85} as the
most prominent candidates. The first of these, GUT baryogenesis, is
based on decays of heavy gauge bosons with masses of order
$M_{\textrm{GUT}} \approx 10^{16} \textrm{\ GeV}$. While it provides a
scenario fulfilling all the Sakharov conditions, it has serious
problems with inflationary models related to the high reheating temperature
required, and the consequent overproduction of gravitinos
\cite{DinKus04}. The latter models, electroweak baryogenesis,
leptogenesis and some variants of Affleck-Dine baryogenesis are based
(directly or indirectly) on electroweak baryon number violation
\cite{'tHooft76}, which is a quantum anomaly in the electroweak sector
of the standard model allowing the baryon number to be badly violated at high
temperatures. These models differ however substantially in the
mechanisms of how the required out-of-equilibrium conditions are reached.      
In what follows we will focus on EWBG in more detail, trying to
elaborate the basic mechanism and the necessity to use the methods of
nonequilibrium quantum field theory in its study.

\subsection{Electroweak baryogenesis}
\label{sec:EWBG}

\subsubsection{Electroweak baryon number violation}

Baryon and lepton numbers are classically conserved in the standard
model. However, at the quantum level this is not the case. It can be
shown that the axial current in the electroweak sector of the standard model and
consequently the total baryon and lepton number currents $j_B^{\mu}$ and
$j_L^{\mu}$ are anomalous \ie not exactly conserved \cite{Adler69,BelJac69}: 
\begin{equation}
\partial_{\mu}j_B^{\mu} = \partial_{\mu}j_L^{\mu} = N_f
\left(\frac{g^2}{32\pi^2} \tilde{W}_{\mu\nu}^a W^{a,\mu\nu} -
  \frac{g'^2}{32\pi^2} F_{\mu\nu} \tilde{F}^{\mu\nu}\right)\,,
\label{anomaly}
\end{equation}
where $N_f$ is the number of fermionic families, $W_{\mu\nu}^a$ and
$F_{\mu\nu}$ are the field strength tensors of the $SU(2)_L$ and $U(1)_Y$
gauge symmetries with the duals $\tilde{W}^{a,\mu\nu} =
\frac{1}{2}\epsilon^{\mu\nu\rho\sigma}W_{\rho\sigma}^a$ and an
analogous expression for $\tilde{F}$, and $g$ and $g'$ are the
associated coupling constants, respectively. Equation (\ref{anomaly})
implies that the total change in baryon (lepton) number from time $t=0$ to some
arbitrary final time $t_f$ is given by:
\begin{equation}
\Delta B = \Delta L = N_f\big[N_{CS}(t_f)-N_{CS}(0)\big] -
N_f\big[n_{CS}(t_f)-n_{CS}(0)\big]\,,
\label{baryon_change}
\end{equation}
where
\begin{eqnarray}
N_{CS} &=& \frac{g^2}{32\pi^2}\int d^3 x \epsilon^{ijk}\left(W_{ij}^a
  A_{k}^a - \frac{1}{3}g \epsilon_{abc} A_{i}^a A_{j}^b A_{k}^c
\right) 
\nonumber\\ 
n_{CS} &=& \frac{g'^2}{32\pi^2}\int d^3 x \epsilon^{ijk} F_{ij} B_{k}
\label{Chern}
\end{eqnarray}
are called the Chern-Simons numbers of the $SU(2)_L$ and $U(1)_Y$
gauge symmetries. We proceed by considering the vacuum structure of the
gauge fields $A_{\mu}^a$ and $B_{\mu}$. It turns out that the abelian
$U(1)_Y$ sector has a trivial nondegenerate vacuum with $\vec{B} = 0$,
but the $SU(2)_L$ sector instead has a discrete set of degenerate vacua with   
\begin{equation}
\vec{A} = \frac{1}{i} g_n^{-1}\vec{\nabla}g_n\,,
\label{pure_gauge}
\end{equation}
where $g_n(\vec{x}) = e^{i n f(\vec{x})\hat{x} \cdot \tau /2}$, $n$
is an integer and $\tau_i$ are the generators of the $SU(2)$ gauge
group. Using the vacuum structure Eq.~(\ref{pure_gauge}) in
Eqs.~(\ref{baryon_change})-(\ref{Chern}) we see that the change in 
baryon number in transitions between the different vacua is given by 
\begin{eqnarray}
\Delta B &=& N_f \Delta N_{CS} = N_f\frac{g^2}{32\pi^2}\int
d^3 x\,\epsilon^{ijk}\textrm{Tr}\left[ g_n^{-1} \partial_i g_n g_n^{-1}
  \partial_j g_n g_n^{-1} \partial_k g_n\right] 
\nonumber\\ 
&=& N_f n\,.
\end{eqnarray}
\begin{figure}
\centering
\includegraphics[width=0.85\textwidth]{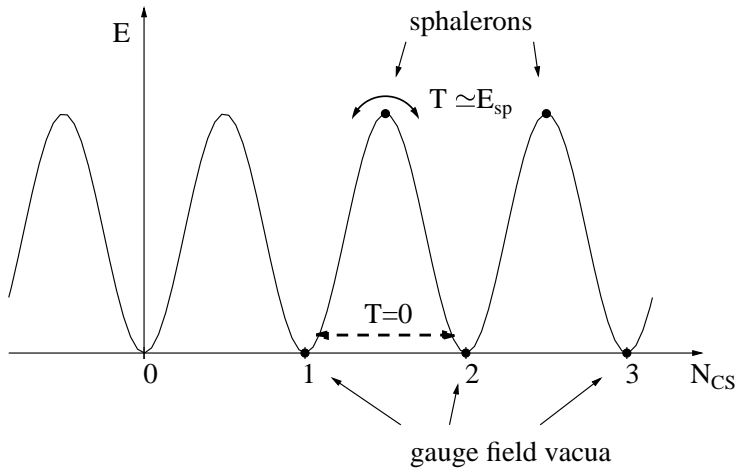}
\vskip-1.2truecm
\caption{The vacuum structure of the $SU(2)_L$ gauge fields. The set
  of degenerate vacua can be labelled with an integer Chern-Simons
  number $N_{CS}$. At zero temperature the transitions
  between different vacua by quantum tunneling through the potential
  barrier are extremely suppressed. At high temperatures the
  transitions are possible by thermal activation via sphaleron configurations.}
\label{gauge_vacua}
\end{figure}
We see that the baryon number changes in integer multiples of $N_f$
in the transitions between different vacua. The structure of the effective
potential for the gauge fields is sketched in Fig.~\ref{gauge_vacua},
where the minima correspond to different vacuum configurations
labelled by the (integer) Chern-Simons number $N_{CS}$. But how could
the transitions between different vacua actually take place? At zero
temperature the only possibility is by quantum tunneling through the potential
barrier. This corresponds to the so called {\em instanton} configuration
\cite{'tHooft76}, but it turns out that the tunneling rate is
negligible: $\Gamma_{\rm in} <
e^{-4\pi/\alpha_W} \sim 10^{-170}$. With this rate not a single proton
could have been produced in the lifetime of the universe! At high
temperature the situation is better. The transitions between the vacua
can take place through a thermal activation over the potential
barriers, via the so called {\em sphaleron} field
configuration \cite{Manton83,KliMan84}. The thermal transition rate
corresponding to this process is shown to be
\cite{KuzRubSha85,ArnMcL88}
\begin{equation}
\Gamma_{\rm sp} \sim T^4 e^{-E_{\rm sp}(T)/T},
\label{sphaleron_rate_mass}
\end{equation}
where $E_{\rm sp}$ is the energy of the sphaleron configuration, which
is related to the vacuum expectation value of the Higgs field
$\langle\phi\rangle$ by $E_{\rm sp}/T \simeq 40 \langle\phi\rangle / T$.
This result is valid only in the broken phase \ie when the gauge
symmetry is spontaneously broken and the gauge bosons are massive. In
the symmetric phase with massless gauge bosons the sphaleron rate is
instead given by (in the minimal standard model) \cite{BodMooRum00}
\begin{equation}
\Gamma_{\rm sp} \simeq (25.4 \pm 2.0)\alpha_W^5 T^4\,,
\label{sphaleron_rate_massless}
\end{equation}
where $\alpha_W \approx 1/30$ is the weak coupling constant. To see if
the sphaleron transitions are fast at the time scales of the expanding
early universe, the sphaleron rate of a unit comoving volume:
$\Gamma_{\rm sp}/T^3$, needs to be compared with the Hubble expansion
rate of the (radiation dominated) universe \cite{KolTur90}
\beq
H = 1.66\,g_*^{1/2} \frac{T^2}{m_{\rm pl}}\,,
\eeq
where $g_*(T)$ counts the total number of effectively massless degrees
of freedom and $m_{\rm Pl} = 1.22 \times 10^{19} \textrm{\ GeV}$ is the
Planck mass. In the minimal standard model at high temperatures of
order $T \gsim 100$ GeV we have $g_* = 106.75$ so that the
symmetric phase sphaleron rate in Eq.~(\ref{sphaleron_rate_massless})
is very large compared to the Hubble rate. Moreover, if $\langle\phi\rangle / T
\lsim {\cal O}(1)$ then also the broken phase sphaleron rate in
Eq.~(\ref{sphaleron_rate_mass}) is large, and the first Sakharov
condition is fulfilled in both of the phases. Actually, it is crucial in
EWBG that the broken phase sphaleron rate is smaller than the Hubble
rate so that the the generated baryon asymmetry is not washed
out. Next we will briefly consider the electroweak phase transition
between the symmetric and broken phases, where the gauge bosons and
fermions become dynamically massive. This transition takes place at
the (electroweak) temperature scale $T \sim 100$ GeV, and it provides
a scheme for the creation of a permanent baryon asymmetry through
EWBG, if the transition is of first order.

\subsubsection{Electroweak phase transition}

In the electroweak theory of the standard model
\cite{Glashow61,Weinberg67,Salam68} the masses of the gauge bosons
$W^{\pm}$ and $Z^0$ and all fermions are generated through spontaneous
symmetry breaking of the gauge symmetry $SU(2)_L\times U(1)_Y$. The
``classical'' potential of the ($SU(2)_L$ doublet) Higgs scalar field
$\Phi$:
\begin{equation}
V(\phi) = -\frac{\mu^2}{2}\phi^2 + \frac{\lambda}{4}\phi^4\,,
\label{Higgs_potential}
\end{equation}
where $\phi \equiv \sqrt{2\,\Phi^\dagger\Phi}$, is minimized for $\phi =
v_0 \equiv \mu/\sqrt{\lambda}$ corresponding to a degenerate set of
vacuum configurations. Choosing a particular vacuum from this set
breaks the gauge symmetry spontaneously giving rise to masses for the
gauge bosons, which are proportional to the vacuum expectation value
$v_0$ of the Higgs field. This is how the Higgs mechanism works at
zero temperature.
\begin{figure}
\centering
\includegraphics[width=0.85\textwidth]{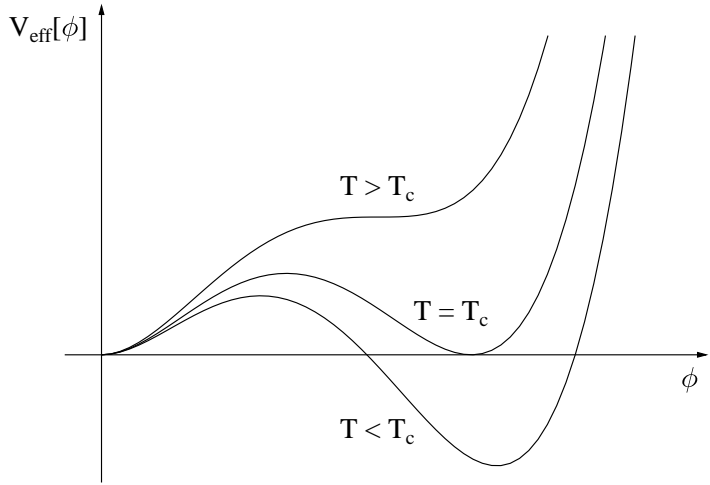}
\caption{The Higgs field effective potential for a first order phase
  transition at different temperatures. Because of the barrier between
  the local minima, the phase transition takes place at a temperature $T < T_c$
  releasing latent heat.}
\label{eff_potential}
\end{figure}

At finite temperatures the ``classical'' Higgs potential
Eq.~(\ref{Higgs_potential}) gets temperature dependent quantum
corrections that will become more and more important when the
temperature increases. These corrections are taken in account by
computing the free energy (or the effective potential) of the Higgs
field \cite{DinKus04}:
\begin{equation}
V_{\textrm{eff}}(\phi, T) = D(T^2 - T_0^2)\phi^2 - ET\phi^3 +
\frac{\lambda}{4}\phi^4 + ...
\label{Higgs_eff_potential}
\end{equation}     
where the values of the parameters $T_0$, $D$ and $E$ depend on the
considered model. We have only given the dominant terms from
perturbative calculations. The temperature dependence of the potential
Eq.~(\ref{Higgs_eff_potential}) gives rise
to an interesting behaviour of the Higgs field. Let us first consider
the case with vanishing third order term $E=0$: when $T > T_0$ the second order
term is positive implying that the minimum of the potential and the
corresponding vacuum expectation value (VEV) of the
Higgs field is zero: $\langle\phi\rangle = 0$, while for $T < T_0$ it is
finite: $\langle \phi \rangle \neq 0$, since the negative second order term
dominates for small $\phi$. At the critical temperature
$T=T_c=T_0$ there will be a phase transition from the former
symmetric phase to the latter broken phase. This is called {\em
electroweak phase transition}, and for the case $E=0$ it is a second
order transition with smoothly increasing VEV
$\langle\phi\rangle$ as the temperature decreases. The case with $E>0$
presented in Fig.~\ref{eff_potential} is more interesting for us. At the
critical temperature $T=T_c$ there is now a potential barrier between
the symmetric and broken minima. Because of this barrier the phase transition
actually takes place at a lower temperature $T < T_c$, and latent heat is
released due to the energy difference between the minima. Hence the
transition is of first order and it happens by nucleation of broken
phase bubbles inside the symmetric phase bulk. These bubbles start to grow
rapidly reaching soon a stationary expansion speed. This scenario with first
order phase transition is of crucial importance for the electroweak
baryogenesis. It is just around the edge (or wall) of these broken phase
bubbles where the last two Sakharov conditions of $CP$ violation and
departure from thermal equilibrium are fulfilled. Unfortunately, it has
been shown by nonperturbative lattice calculations that for the
minimal standard model the phase transition is not of first order
\cite{KLRS96}. For this and other reasons (\eg not enough $CP$
violation) we have to seek new possibilities for electroweak
baryogenesis in extensions of the standard model, such as the minimal
supersymmetric standard model (MSSM). Next we will give a brief
conceptual description of the mechanism of (eletroweak) baryogenesis
for a generic model with a strong first order phase transition.

\subsubsection{Description of the EWBG mechanism}

\begin{figure}
\centering
\includegraphics[width=0.85\textwidth]{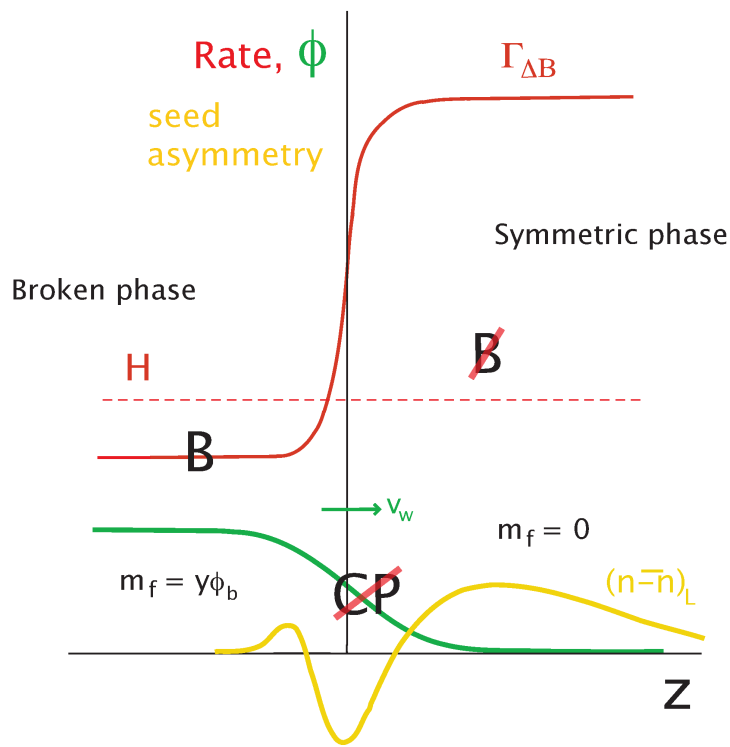}
\vskip-4.2truecm
\caption{Sphaleron rate (red), vacuum expectation value of the Higgs
  field (green), and the source asymmetry (yellow) in the bubble wall
  region as a function of the coordinate normal to the wall front.}
\label{bubblewall}
\end{figure}

In Fig.~\ref{bubblewall} we present a schematic cross-section from the
expanding bubble wall front where baryogenesis takes place. 
In the bubble wall region the VEV of the Higgs field and consequently
the masses of the fermions (green line) are spatially dependent
changing smoothly from zero in the symmetric phase to a finite value
in the broken phase. The varying complex mass of a fermionic
eigenstate gives rise to the
required $CP$ violating effects that will generate an asymmetry in
left chiral number densities between particles and antiparticles
(yellow line). This source asymmetry then creates a pseudo chemical
potential that biases the sphaleron transitions in front
of the wall in the symmetric phase to produce more baryons than
antibaryons. This net baryon number is not washed away in the broken
phase, because the sphaleron rate (red line) is too small there
compared to the Hubble expansion rate $H$. At the end, when the
bubble wall has passed and the plasma is back in thermal
equilibrium, a net baryon number density has been created. 

One of the most difficult problems in the actual calculations of the
baryon asymmetry is to find out the source asymmetry due to the $CP$ violating
effects. An accurate calculation would necessarily involve the use of
nonequilibrium quantum field theory, and this is the background
motivation for our work. In earlier works the problem has been studied
in the semiclassical WKB approach
\cite{JoyProTur95,JoyProTur96a,JoyProTur96b,CliJoyKai98,CliKai00,CliJoyKai00}
and later with the methods of quantum kinetic theory
in~\cite{KPSW01,KPSW02a,KPSW02b,ProSchWei04a,ProSchWei04b},
both approaches using the Boltzmann transport equations for
$CP$-violating phase space densities. These methods should
provide a solid approximation in the semiclassical limit
\ie the case of a thick wall compared to the mean free path of the
interacting fermions. However, in the thin wall limit the dominant
source for the asymmetry comes from the quantum reflection
processes, which are inherently nonlocal and absent in the standard WKB
and kinetic approaches. Attempts have been made to treat 
the reflection phenomena by including collisions in the Dirac equation
\cite{GHOP94,GHOPQ94,HueSat95,HueNel96,RiuSan00}, but no consistent
framework based on quantum field theory has been introduced.
In this work we present an approximation scheme based on the
kinetic approach that enables us
to treat the quantum reflection in a simple but consistent
way in the presence of decohering collisions.

%
%

\section{Preheating}
\label{sec:preheating}

Cosmological inflation
\cite{Kazanas80,Starobinsky80,Guth81,Sato81_1,Sato81_2,MukChi81,GutPi82,Hawking82,Starobinsky82,KolTur90,Linde90}
(for a recent review see \eg \cite{BasTsuWan05}) is a period of rapid
exponential expansion in the very early universe during which the size
of the universe increases by a huge factor (at least $e^{60}$). As
mentioned in the beginning of last section, it provides a natural
explanation for the homogeneity (horizon problem) and flatness of the
universe. It also explains the (almost) scale invariant primordial density
fluctuations that will give rise to large-scale structure
formation and the observed anisotropies in the CMB spectrum
\cite{Bennett_etal03}. Most of the inflationary models are based on the
peculiar dynamics of one or several scalar fields, called inflaton(s),
whose vacuum condensate dominates the evolution of the universe in a
so called ``slow roll'' phase, causing the exponential growth. Because
of the huge and very rapid expansion the universe is
typically\footnote{The warm inflation scenario with particle
  production during inflation is an exception
  \cite{Berera97,BerKep99}.} in a highly non-thermal and very cold
state at the end of inflation. However, the baryogenesis scenarios
require energies greater than the electroweak scale and the primordial
nucleosynthesis requires that the universe is close to thermal
equilibrium at the temperature $\sim 1$ MeV at some stage after
inflation. A mechanism for reheating the universe after
inflation is thus needed to retain the Hot Big Bang scenario. 

The modern scenarios of reheating consist of a {\em preheating} stage
followed by thermalization. In the preheating stage
\cite{DolKir90,TraBra90,KofLinSta94,ShtTraBra95,BVHS96,KofLinSta97}
the inflaton condensate goes through rapid oscillations, so that the
couplings to other matter fields give rise to particle production via
parametric resonance. The basic mechanism is simple: the
coupling to the inflaton(s) gives rise to rapidly oscillating 
  effective masses for the matter fields that will bump up the
particle numbers exponentially for certain momentum modes in close
analogy with the Floquet theory of growing exponents
\cite{WhiWat40,Ince56}. The modes in these
resonance bands will quickly obtain huge occupation numbers (for
scalars with no Fermi blocking). This rapid growth of perturbations is
followed by backreaction and rescatterings. {\em Backreaction} means
the effects of the growing perturbations (particle production) back on the
dynamics of the inflaton condensate. In most scenarios it will rather
quickly shut off the inflaton oscillations (faster than the Hubble
expansion rate of the universe) and consequently terminate the particle
production. Before that the fluctuations of the inflaton field itself
will grow and give rise to {\em rescatterings} \ie couplings between
different momentum modes leading to the growth of occupation numbers for
non-resonance modes as well. When the oscillations of inflaton field
shut off completely the preheating stage ends and the fields start to
thermalize. This thermalization process can be very complex, including
regimes of driven and free turbulence \cite{MicTka04}, which makes it
difficult to estimate the final reheat temperature. Moreover, in
certain cases the universe might enter a ``quasi-thermal'' phase with
a kinetic equilibrium reached much before the full chemical
equilibrium \cite{AllMaz05}. 

In addition to scalar particles, the parametric resonance during
preheating may produce a significant amount of fermionic
particles. This resonant production could possibly lead to dangerous
relic abundances of problematic particles such as gravitinos
\cite{BaaHeiPat98,GreKof99}. Fortunately, extensive studies have shown
that gravitino over-production can be avoided during the preheating in
realistic supersymmetric theories
\cite{GiuRioTka99,KKLV00_1,KKLV00_2,MarMaz00,NilPelSor01_1,NilPelSor01_2,GreKadMur03}. Another
interesting aspect in the fermionic preheating is the possibility to
generate heavy fermions with masses of order $10^{17}$-$10^{18}$ GeV,
that could be important in \eg leptogenesis \cite{GPRT99,PelSor00}. In
chapter \ref{chap:applications} we will apply our approximation scheme
with decohering interactions to study a simple model of fermionic
preheating, during which the fermion is subjected to decays.

%% file: Chapter2.tex
%
%

\chapter{Basic formalism of quantum transport theory}
\label{chap:basic_formalism}

The standard methods of (vacuum or thermal) quantum field theory (QFT) are
not well suited for the study of nonequilibrium quantum fields for
several reasons. First of all, the basic quantities of interest in
nonequilibrium QFT are expectation values of operators in contrast to
transition amplitudes in the standard vacuum QFT, requiring the formal
extension of the time variable into a closed time path with two
different branches. Other specific issues in nonequilibrium QFT are
related to for example secularity\footnote{The standard perturbative
approach suffers from secular terms giving rise to big late-time
contributions in all orders of perturbation expansion, no matter how small the
coupling constant is \cite{BerSer03b,BerSer04,Berges04}. Heuristically,
this can be seen in the sense that $\epsilon\,t \sim 1$ for a big
enough $t$ for every $\epsilon$ (no matter how small).} 
causing the complete failure
of the standard perturbative expansion in many cases of interest
\cite{BerSer03b,BerSer04,Berges04}. In this chapter we introduce the
basic concepts of nonequilibrium quantum field theory in order to
derive the Kadanoff-Baym (KB) transport equations for fermions and scalar
bosons. We start by introducing the closed time path formalism with
four basic propagators. Then, we use the two-particle irreducible
(2PI) effective action methods to derive the self-consistent
(Schwinger-Dyson) equations of motion for the full 2-point correlation
functions of the system, which we then write in the form of
KB-equations. Finally, we list some physical observables that can
be expressed in terms of these 2-point functions.

%
%

\section{Closed time path formalism}

\begin{figure}
\centering
\includegraphics[width=0.65\textwidth]{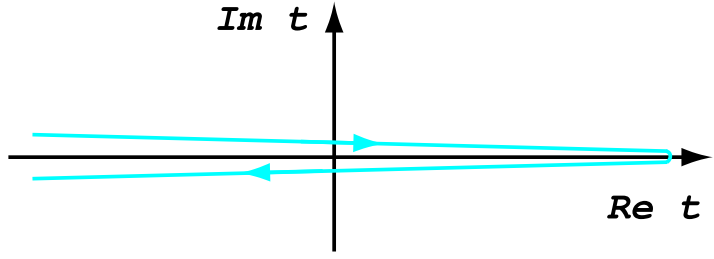}
     \caption{The closed time path of the CTP formalism.}
     \label{fig:KeldyshPath}
\end{figure}

The closed time path (CTP or Schwinger-Keldysh) formalism was
developed by Schwinger \cite{Schwinger61} and Keldysh \cite{Keldysh64}
and further refined by many influential works, including
\cite{BakMah63,ZSHY80,CSHY85,SCYC88,DeWitt84,Jordan86,CalHu87,CalHu88,CalHu89}.
The basic idea of the formalism is simple: In order to study the expectation
values instead of transition amplitudes by the methods of quantum
field theory, the time coordinate must be extended to a {\em closed time path}
from initial time $t_0$ to final time $t_f$ (often taken to be
$\infty$) and then back to $t_0$\footnote{For this reason the CTP
  formalism is often called ``in-in'' formalism on the contrary to the
  traditional ``in-out'' formalism of quantum field theory with
  transition amplitudes between incoming and outgoing states.} 
 (see Fig. \ref{fig:KeldyshPath}). The need for this closed time path can
be demonstrated by writing the expectation value of a real
scalar field $\phi(x)$ in a state defined by arbitrary density
operator $\hat\rho$ in terms of a path integral (we choose here $t_0 = 0$):
\beqa
\langle \hat \phi(x) \rangle &\equiv& {\rm Tr}\{
\hat\phi(x) \hat\rho \} = {\rm
  Tr}\{U(t_f,0)U(0,t)\hat\phi_S(\vec{x})U(t,0)\hat\rho_S(0)U(0,t_f)\}
\nonumber\\[3mm]
 &=& \int d \phi_f \, d \phi_t\, d \phi_0\, d \phi'_0 \Big[\langle 
 \phi_f | U(t_f,t)|\phi_t\rangle\phi(x)\langle\phi_t|U(t,0)|\phi_0\rangle\Big]  
\nonumber\\
&&\qquad\qquad\qquad\quad\;\;\times
\langle\phi_0|\hat\rho_S(0)|\phi'_0\rangle\langle\phi'_0|U(0,t_f)|\phi_f\rangle  
\nonumber\\[4mm]
&=& \int d \phi_f\, d \phi_0\, d \phi'_0
\;\rho\big[\phi_0(\vec{x}),\phi'_0(\vec{x})\big]
\int_{\phi_0(\vec{x})}^{\phi_f(\vec{x})}
D \phi^+ \,\phi^+(x)
\exp\{iS[\phi^+]\}
\nonumber\\[1mm]
&&\qquad\qquad\qquad\qquad\qquad\quad\;\;\;\times
\int_{\phi'_0(\vec{x})}^{\phi_f(\vec{x})}
D \phi^- \exp\{iS[\phi^-]^*\}
\nonumber\\[2mm]
&=& \int D \phi^+ D \phi^-
\,\rho\big[\phi^+(0,\vec{x}),\phi^-(0,\vec{x})\big]\,\phi^+(x)
\exp\{i(S[\phi^+]-S[\phi^-]^*)\}\,,
\nonumber\\
&& 
\label{CTP-integral}
\eeqa
where $\rho\big[\phi^+(0,\vec{x}),\phi^-(0,\vec{x})\big] \equiv
\langle\phi^+|\hat\rho_S(0)|\phi^-\rangle$ is the initial density
matrix and the subscript ${\cal O}_S$ denotes an operator in the
Schr\"odinger picture in contrast to the Heisenberg picture without a
subscript.  We see that the path integral representation involves two
``histories'', for which the evolution is chronological from $0$ to
$t_f$ and antichronological from $t_f$ to $0$, respectively. The field
values in these $+$/$-$ branches are independent except the boundary condition
$\phi^+(t_f,\vec{x})=\phi^-(t_f,\vec{x})$ closing the path at $t=t_f$, which
we did not write explicitly in the last row of Eq.~(\ref{CTP-integral}).   

\subsection{Propagators}

It turns out that the higher $n$-point Green's functions of the system
will automatically become time ordered along the closed time path. Of
special interest in quantum field theory are the 2-point functions or
propagators, which are now defined as 
\beqa
  i\Delta_{\cal C}(u,v) &=& \left\langle T_{\cal C}
  \left[\phi(u)\phi(v)\right] \right\rangle
\,\,\, \equiv {\rm Tr}\left\{\hat\rho \ T_{\cal C} \left[\phi(u) \phi(v)\right]
             \right\}
\label{Gcontour_scalar}
\\
  iG_{\cal C}(u,v) &=& \left\langle T_{\cal C}\left[\psi(u) \bar
      \psi(v)\right] \right\rangle 
\equiv {\rm Tr}\left\{\hat\rho \ T_{\cal C} \left[\psi(u) \bar
    \psi(v)\right] \right\}
\label{Gcontour_fermion}
\eeqa
for a real scalar field $\phi$ and a fermionic field $\psi$, respectively.
$\hat \rho$ is some unknown quantum density operator describing
the statistical properties of the system, and $T_{\cal C}$ defines the
time ordering along the closed time path ${\cal C}$, shown in
Fig. \ref{fig:KeldyshPath}, in the sense that the points
on the lower (negative) branch are ``later'' than those on the
upper (positive) branch. When written in terms of the ordinary
real time variable running from $-\infty$ to $+\infty$, the
closed time path propagators
(\ref{Gcontour_scalar})-(\ref{Gcontour_fermion}) contain four distinct
contributions depending on the time branches of the ``complex''
CTP-coordinates $u$ and $v$. That is, using indices $a,b=\pm$ to label
the positive/negative branches, the scalar propagators are decomposed as:
\begin{eqnarray}
  i\Delta^{+-}(u,v) \equiv i\Delta^<(u,v) &=& 
\langle \phi(v)\phi(u) \rangle 
   \nonumber\\
  i\Delta^{-+}(u,v) \equiv i\Delta^>(u,v) &=& 
\langle \phi(u)\phi(v) \rangle
   \nonumber\\
  i\Delta^{++}(u,v) \equiv i\Delta_F(u,v) &=&   
\theta(u_0-v_0) i\Delta^>(u,v) + \theta(v_0-u_0) i\Delta^<(u,v)
   \nonumber\\
  i\Delta^{--}(u,v) \equiv i\Delta_{\bar F}(u,v) &=&
  \theta(v_0-u_0) i\Delta^>(u,v) + \theta(u_0-v_0)i\Delta^<(u,v)\,,
\nonumber\\
\label{GFs_scalar}
\end{eqnarray}
where now $u^0$ and $v^0$ are ordinary time coordinates. Similarly, for
fermionic propagators we have: 
\begin{eqnarray}
  iG^{+-}(u,v) \equiv -iG^<(u,v)      &=& -\langle \bar\psi(v)\psi(u) \rangle
   \nonumber\\
  iG^{-+}(u,v) \equiv \phm iG^>(u,v)      &=& \langle \psi(u)\bar\psi(v) \rangle
   \nonumber\\
  iG^{++}(u,v) \equiv \phm iG_F(u,v)      &=&   \theta(u_0-v_0) G^>(u,v)
                                               - \theta(v_0-u_0) G^<(u,v)
   \nonumber\\
  iG^{--}(u,v) \equiv \phm iG_{\bar F}(u,v) &=&  \theta(v_0-u_0) G^>(u,v)
                                               -
                                               \theta(u_0-v_0)G^<(u,v)\,.
\nonumber\\
\label{GFs_fermion}
\end{eqnarray}
Using the generic notation $\prop = \{G,\Delta\}$ to denote
fermionic/scalar popagators we see that
$\prop_F$ and $\prop_{\bar F}$ are the chronological (Feynman)
and anti-chronological (anti-Feynman) propagators, respectively,
while $\prop^<$ and $\prop^>$ are the so called \emph{Wightman functions}.
In our further analysis we are especially interested in the dynamics of
these Wightman functions, which contain the essential thermal or
out-of-equilibrium statistical information of the quantum system under
study, in order to compute for example the expectation values of the
number current $j^\mu$ and the energy momentum tensor $T^{\mu\nu}$.

Before we start to build the calculational scheme of the CTP formalism,
let us introduce a few more Green's functions, which are useful in the
following analysis, and list some of their properties. First, we
define the retarted and advanced propagators: 
\begin{eqnarray}
  \prop^r(u,v)  &\equiv& \prop_F \mp \prop^< = \phantom{-}
  \theta(u^0-v^0) (\prop^> \mp
  \prop^<) \nonumber\\ 
  \prop^a(u,v)  &\equiv& \prop_F - \prop^> = -\theta(v^0-u^0) (\prop^>
  \mp \prop^<),
\label{raGFs}
\end{eqnarray}
where now $\mp$ refers to bosons/fermions. The definitions
(\ref{GFs_scalar}), (\ref{GFs_fermion}) and (\ref{raGFs}) then
imply that the propagators have the following hermiticity properties:
\begin{eqnarray}
  \left[i\Delta^{<,>}(u,v)\right]^\dagger &=& i\Delta^{<,>}(v,u)
  \nonumber\\
  \left[iG^{<,>}(u,v)\gamma^0\right]^\dagger &=& iG^{<,>}(v,u)\gamma^0\,,
\label{CEq1}
\end{eqnarray}
and further $\left[i\Delta^{r}(u,v)\right]^\dagger = -i\Delta^{a}(v,u)$
and $\left[iG^{r}(u,v)\gamma^0\right]^\dagger =
-iG^{a}(v,u)\gamma^0$. These latter identities for retarted and
advanced propagators suggests to decompose them into hermitian and
antihermitian parts:
\begin{eqnarray}
  \prop_H   &\equiv&  \frac{1}{2}\left(\prop^a + \prop^r\right) \nonumber\\ 
  {\cal A}  &\equiv&  \frac{1}{2i}\left(\prop^a - \prop^r\right) 
                     = \frac{i}{2}\left(\prop^> \mp \prop^<\right).  
\label{Hdecomposition}
\end{eqnarray}
The antihermitian part ${\cal A}$ is called the \textit{spectral
  function}. Based on Eqs. (\ref{raGFs}) it is easy to show that
$\prop_H$ and ${\cal A}$ obey the spectral relation:
\beq 
\prop_H(u,v) = -i {\rm sgn}(u^0-v^0) {\cal A} (u,v)\,. 
\label{spec_rel}
\eeq
%

%
%

\section{Two-particle irreducible effective action\\ and Schwinger-Dyson equations}

In a nonlinear quantum field theory, including the majority of
the interacting theories, the 2-point correlation functions necessarily
couple to higher order correlators and so on, to form an infinite
Schwinger-Dyson hierarchy of equations analogous to
Bogoliubov-Born-Green-Kirkwood-Yvon (BBGKY) hierarchy in classical
statistical mechanics \cite{CalHu08}. For this reason it is an
impossible task to solve the 2-point correlators exactly; that would correspond
to a full solution of the nonlinear QFT. A major paradigm in practical
applications thus is to truncate this hierarchy (slave the
higher order correlators) in an appropriate way. The truncation can
be done in many ways, for example by a brute use of perturbation theory or a
loop expansion. However, these standard methods do not generally
provide a good approximation for out-of-equilibrium dynamics, because
of several problems \eg with secularity \cite{Berges04}.   

A way to evade these problems is a method of obtaining the (truncated)
equations of motion from variational principles of increasing
complexity. On the first level we obtain an equation of motion for
the field expectation value $\langle \phi \rangle$ only, in the next
level to $\langle \phi \rangle$ and 2-point function $\prop \sim
\langle(\phi - \langle \phi \rangle)^2\rangle$, and so forth. The
effective action corresponding to the $n$-th level of this hierarchy
is called {\em n-particle irreducible} (nPI) effective action
$\Gamma_{\rm nPI}$. The heuristic difference between the nPI-method
and the standard perturbation theory is that in the latter the
{\em solutions} are written as an expansion in a small parameter, say coupling
constant, while in the former the {\em equations} themselves are expanded. This
difference is of crucial importance in nonequilibrium quantum field
theory; it is the very reason for the problems of \eg secularity with
the standard perturbation method. The truncation procedure using the
nPI-method is practically feasible as it turns out that the
higher order nPI effective actions become redundant, once the order of
expansion is fixed. That is, for
example in a loop expansion\footnote{A loop expansion in the nPI
  effective action corresponds to an expansion of the {\em equations},
  not to the perturbative expansion of the {\em solutions}.} at
$m$-loop order all nPI effective actions with $n \geq m$ are
equivalent. In addition to these reductions there may be further
simplifications depending on special conditions, such as vanishing of
the average field \cite{Berges04}.

\subsection{From a generating functional to 2PI effective action}

In this work we will concentrate on the two-particle irreducible (2PI)
effective action
\cite{LutWar60,DomMar64a,DomMar64b,DahLas67,CorJacTom74,CSHY85,CalHu88},
which will lead to a self-consistent dynamics for the 2-point correlation
function $\prop$ as well as the 1-point function $\langle \phi
\rangle$. We show, following ref.~\cite{CalHu08}, how the 2PI
effective action and the corresponding equations of motion are derived
for a real scalar field. For fermions we will only give the
appropriate results. We start by defining 2PI {\em generating
  functional} on the closed time path:
\beq
Z[J,K] = e^{iW[J,K]} = \int D \phi^a \,\rho\big[\phi^a(0,\vec{x})\big]\,
\exp\Big\{i\big(S[\phi^a] + J_a \phi^a + \frac12 K_{ab}\phi^a
\phi^b\big)\Big\}\,,
\label{2PI_gen}
\eeq
where we use notation with a branch doublet $\phi^a = (\phi^+,\phi^-)$, so that
\eg $ D \phi^a = D \phi^+ D \phi^-$, and define a ``metric''
$c_{ab}={\rm diag}(1,-1)$, so that $J_1(x)=J^1(x)$ and
$J_2(x)=-J^2(x)$ and repeated indices are summed over. The CTP
action is defined as $S[\phi^a] = S[\phi^+]-S[\phi^-]^*$, and $J_a(x)$
and $K_{ab}(x,x')$ are local and nonlocal Gaussian sources,
respectively. We also use de Witt summation convention to
leave out integrals in the notation for the source terms:
$K_{ab}\phi^a \phi^b \equiv \int d^4 x\,d^4
x'\,K_{ab}(x,x')\phi^a(x)\phi^b(x')$ and similarly for
$J_a\phi^a$. All $n$-point Green's functions are obtained from
$Z[J,K]$ through functional differentiation with respect to sources
$J^a$, while $W[J,K]$ generates the connected $n$-point
functions. Especially, the {\em average field}\footnote{This is
  usually called mean field, but in this work we have a different
  notion for mean field, explained later in chapter
  \ref{chap:scheme}.} is defined as
\beq
\bar{\phi}^a(x) = \frac{\delta W[J,K]}{\delta J_a(x)}\,.
\label{average_field}
\eeq
If we set $J_a = K_{ab} = 0$ after the variation, then
$\bar{\phi}^+=\bar{\phi}^-=\langle\phi\rangle$ is the physical expectation value
without sources. The 2-point functions can be obtained form $W[J,K]$ either
through a double derivative with respect to $J_a$ or through a
derivative with respect to the nonlocal source $K_{ab}$. For later
purposes we use the latter way, and define the full propagators
$\Delta^{ab}$ as
\beq
\frac{\delta W[J,K]}{\delta K_{ab}(x,x')} = \frac12
\big[\bar{\phi}^a(x)\bar{\phi}^b(x') + \Delta^{ab}(x,x')\big]\,.
\label{2-point_correlator}
\eeq
From the definition we see that $\Delta^{ab} = \langle T_{\cal C}
\left[(\phi^a - \bar{\phi}^a)(\phi^a - \bar{\phi}^a)\right]\rangle$ \ie it
  corresponds to fluctuations with respect to the average field, and
  thus it actually reduces to Eq.~(\ref{GFs_scalar})
  only for vanishing average field $\bar{\phi}$ (and vanishing
  sources). To proceed, we define the 2PI {\em
  effective action} as a double Legendre transformation of $W[J,K]$:
\beq
\Gamma_{\rm 2PI}[\bar{\phi},\Delta] = W[J,K] - J_a \bar{\phi}^a - \frac12
K_{ab}\big[\bar{\phi}^a\bar{\phi}^b + \Delta^{ab}\big]\,,
\label{2PI_action}
\eeq
where it is understood that the sources $J_a(x)$ and $K_{ab}$ are
eliminated through the relations between them and the correlators
$\bar{\phi}^a$ and $\Delta^{ab}$ arising from
Eqs.~(\ref{average_field})-(\ref{2-point_correlator}). These relations are
always invertible following from the general properties of Legendre
transformation. The desired equations of motion for the correlators
$\bar{\phi}^a$ and $\Delta^{ab}$ are now obtained by functional differentiation:
\beq 
\frac{\delta\Gamma_{\rm 2PI}}{\delta \bar{\phi}^a} = -J_a -
K_{ab}\bar{\phi}^b\,, \qquad\quad \frac{\delta\Gamma_{\rm 2PI}}{\delta
  \Delta^{ab}} = -\frac12 K_{ab}\,,
\label{2PI_equations}
\eeq
so that in the case of physical (sourceless) dynamics we get the equations:
$\delta\Gamma_{\rm 2PI}/\delta \bar{\phi}^a = 0$ and 
$\delta\Gamma_{\rm 2PI}/\delta \Delta^{ab} = 0$, which corresponds to
finding the extremum for the effective action
$\Gamma_{\rm 2PI}[\bar{\phi},\Delta]$.

\subsection{Formula for the 2PI effective action in terms of the
  fluctuation field}

In order to use the equations of motion (\ref{2PI_equations}), we want to find
a practical method to compute the 2PI effective action $\Gamma_{\rm
  2PI}$. To implement the so called background field method, we write
the effective action in the form:
\beq
e^{i\Gamma_{\rm 2PI}} = \int D \phi^a \,\exp\Big\{i\big[S[\phi^a] +
J_a (\phi^a - \bar{\phi}^a) + \frac12 K_{ab}(\phi^a \phi^b -
\bar{\phi}^a \bar{\phi}^b - \Delta^{ab})\big]\Big\}\,,
\label{Gamma_2PI_eq}
\eeq
which follows directly from Eqs.~(\ref{2PI_gen}) and
(\ref{2PI_action}). Here we have left out the initial density matrix
contribution $\rho\big[\phi^a(0,\vec{x})\big]$. The justification for
this can be seen in two ways. First, if the initial state is Gaussian,
the initial density matrix can be written as
$\rho\big[\phi^a(0,\vec{x})\big] = \exp(-\frac12 R_{ab}\phi^a\phi^b)$,
and the new initial-time Kernel $R_{ab}$ can be absorbed in the source
$K_{ab}$. But this inclusion seems to ruin the desired condition that the
physical evolution is given by vanishing sources $J_a$ and $K_{ab}$.
However, since $R_{ab}$ vanishes for all but initial
time, we see that it affects only the initial conditions for
the 1- and 2-point functions $\bar{\phi}^a$ and $\Delta^{ab}$, and
hence can be neglected in the dynamical equations if we do adjust these
initial conditions correctly \cite{Berges04}. The
other possibility when the neglection of $\rho\big[\phi^a(0,\vec{x})\big]$
is justified is to consider such initial conditions, where the initial
state in the distant past $t_0 \rightarrow -\infty$ is in the {\em in}
vacuum. This condition is implemented by just shifting the mass $m^2$
to $m^2 - i \epsilon$ in the first first branch and to $m^2 + i
\epsilon$ in the second branch\footnote{This is why the complex
  conjugate is explicitly written in the second branch action
  $S[\phi^-]^*$, even though the classical action is always real.} \ie
``tilting'' the time path in
the complex plane in the same way as in the standard vacuum quantum field
theory \cite{CalHu08}. For an initial state that is neither of these
cases the omitting of $\rho\big[\phi^a(0,\vec{x})\big]$ is not
strictly justified and the following developments in this
section provide only an approximation. If one wishes to consider those
non-Gaussian initial states more accurately one needs to use higher nPI
effective actions. 

\begin{figure}
\center
\includegraphics[width=0.45\textwidth]{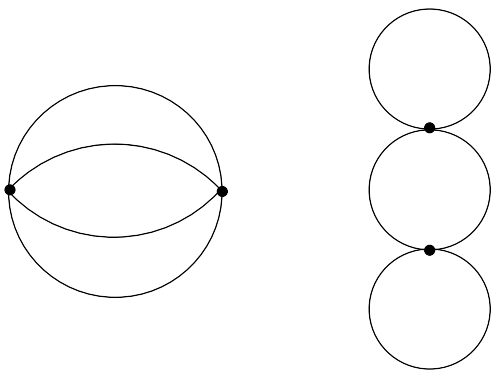}
\caption{Three-loop vacuum graphs in a theory with quartic interaction. Only the
  left one is two-particle irreducible.}
\label{fig:2PI_graphs}
\end{figure}

To come back to equation (\ref{Gamma_2PI_eq}), we see that using
Eqs.~(\ref{2PI_equations}) and the symmetry of the source $K_{ab}$ the
exponent becomes
\beq
S[\phi^a] - \frac{\delta\Gamma_{\rm 2PI}}{\delta \bar{\phi}^a}(\phi^a
- \bar{\phi}^a) - \frac{\delta\Gamma_{\rm 2PI}}{\delta
  \Delta^{ab}}\big[(\phi^a - \bar{\phi}^a)(\phi^b - \bar{\phi}^b) -
\Delta^{ab}\big]\,. 
\label{exponent}
\eeq
Next we shift the integration variable in Eq.~(\ref{Gamma_2PI_eq}) by the
average fields: $\phi^a = \bar{\phi}^a + \varphi^a$ and expand the
classical action in powers of the new {\em fluctuation field} $\varphi^a$:
\beq
S[\bar{\phi}^a + \varphi^a] = S[\bar{\phi}^a] + S_{,a}\varphi^a +
\frac12 S_{,a,b} \varphi^a \varphi^b + S_2\,,
\label{S-expansion}
\eeq
where $S_{,a}$ means functional derivatives of $S$ with respect to
$\phi^a$ evaluated at $\phi^a = \bar{\phi}^a$ and similarly for the
second derivative $S_{,a,b}$, and $S_2$ denotes the collection of the
higher order terms (cubic and so forth) in the fluctuation field
$\varphi^a$. Furthermore, we separate the trivial (lowest orders) and
nontrivial parts of $\Gamma_{\rm 2PI}$ by writing it in the
form 
\beq
\Gamma_{\rm 2PI}[\bar{\phi},\Delta] = S[\bar{\phi}^a] + \frac12 i
\Delta_{0,ab}^{-1}(\bar{\phi})\Delta^{ab} - \frac12 i {\rm Tr}[\ln \Delta] + \Gamma_2[\bar{\phi},\Delta] + {\rm const} \,,   
\label{2PI_ansatz}
\eeq
where the infinite constant (often discarded since it does not affect the
equations of motion) is $-\frac12 \int d^4 x\, \delta(0)$ and we
denote $i\Delta_{0,ab}^{-1}(\bar{\phi}) \equiv S_{,a,b}$\footnote{The
motivation for this notation is that $\Delta_{0,ab}^{-1}(\bar{\phi})$ becomes
the inverse free propagator in the limit of vanishing $\bar{\phi}^a$.}.
Plugging these expressions in Eq.~(\ref{Gamma_2PI_eq}) we find that
the nontrivial part $\Gamma_2$ can be expressed as:
\beqa
e^{i\Gamma_2} &=& [\det \Delta]^{-1/2} \int D \varphi^a
\nonumber\\
&&\times\exp\Big\{-\frac12
\Delta_{ab}^{-1}\varphi^a \varphi^b + i\big[S_2[\varphi^a] - \tilde{J}_a\varphi^a - \tilde{K}_{ab}\big(\varphi^a \varphi^b - \Delta^{ab}\big)
\big]\Big\}\,,
\nonumber\\[2mm]
\label{Gamma_2_eq}
\eeqa
where
\beq
\tilde{J}_a = \frac12 S_{,a,b,c}\Delta^{bc} +
\frac{\delta\Gamma_2}{\delta \bar{\phi}^a}\,, \qquad\quad
\tilde{K}_{ab} = \frac{\delta\Gamma_2}{\delta
  \Delta^{ab}}\,.
\eeq
We see that despite the $\tilde{K}_{ab} \Delta^{ab}$-term $\Gamma_2$
has the form of a generating functional for a new theory with
classical action $\frac{i}{2}
\Delta_{ab}^{-1}\varphi^a \varphi^b + S_2[\varphi^a]$ and sources
$\tilde{J}_a$ and $\tilde{K}_{ab}$. Next we show that these sources
with the additional term $\tilde{K}_{ab} \Delta^{ab}$ will just fix the 1-
and 2-point functions of this new theory. To show that, we start by
considering the matrix
\beq
\left(\begin{array}{cc}
\frac{\delta J_a}{\delta \phi^c} & \frac{\delta J_a}{\delta \Delta^{cd}}  \\
\frac{\delta K_{ab}}{\delta \phi^c} & \frac{\delta K_{ab}}{\delta \Delta^{cd}}
\end{array} \right)\,,
\eeq    
which we know is invertible, because of the general invertibility of Legendre
transformations. In terms of the derivatives of $\Gamma_2$ this
becomes (also subtracting a singular matrix which does not affect invertibility)
\beq
\left(\begin{array}{cc}
\frac{\delta^2 \Gamma_2}{\delta \phi^a \delta \phi^c} - 2\frac{\delta
  \Gamma_2}{\delta \Delta^{ac}} & \frac{\delta^2 \Gamma_2}{\delta
  \phi^a \delta \Delta^{cd}} \\ 2\frac{\delta^2 \Gamma_2}{\delta
  \Delta^{ab} \delta \phi^c} & \frac{\delta^2 \Gamma_2}{\delta
  \Delta^{ab} \delta \Delta^{cd}} 
\end{array} \right)\,.
\eeq    
On the other hand, by taking variational derivatives of
Eq.~(\ref{Gamma_2_eq}) we get the equations
\beq
\left(\begin{array}{cc}
\frac{\delta^2 \Gamma_2}{\delta \phi^a \delta \phi^c} - 2\frac{\delta
  \Gamma_2}{\delta \Delta^{ac}} & \frac{\delta^2 \Gamma_2}{\delta
  \phi^a \delta \Delta^{cd}} \\ 2\frac{\delta^2 \Gamma_2}{\delta
  \Delta^{ab} \delta \phi^c} & \frac{\delta^2 \Gamma_2}{\delta
  \Delta^{ab} \delta \Delta^{cd}} 
\end{array} \right)
\left(\begin{array}{c}
\langle \varphi^c \rangle  \\
\langle \varphi^c \varphi^d \rangle - \Delta^{cd}
\end{array} \right) = 0\,.
\eeq
So, since the coefficient matrix is invertible, we conclude that 
\beq
\langle \varphi^c \rangle = 0\,, \qquad\qquad \langle \varphi^c
\varphi^d \rangle = \Delta^{cd}\,.
\eeq
This result has tremendous implications: As the sources $\tilde{J}_a$
and $\tilde{K}_{ab}$ just kill the 1-point function and fix
the 2-point function to $\Delta^{cd}$, it follows that we can
neglect these sources in the practical calculations, if we include
only the {\em vacuum} contribution\footnote{In non-vacuum graphs the
  external legs are connected
  to the average field of $\varphi$-theory that vanishes} to the
effective action $\Gamma_2$ using $\Delta^{cd}$ as the full propagator of this
$\varphi$-theory. Furthermore, because the {\em full} propagator is fixed to
$\Delta^{cd}$, it follows that in the diagrammatic calculations we need to
consider only {\em two-particle irreducible} (2PI) graphs \ie graphs
that do not become disconnected while cutting two internal lines (see
Fig.~\ref{fig:2PI_graphs}),
hence the name for the 2PI action. So, we conclude that the 2PI
effective action for the original theory is given, besides the
terms explicit in equation (\ref{2PI_ansatz}) (classical and one-loop
contribution), by the sum of all 2PI vacuum graphs in a theory with
action $\frac{i}{2}\Delta_{ab}^{-1}\varphi^a \varphi^b +
S_2[\varphi^a]$. For example, for the real scalar field with quartic
interaction:
\beq 
S[\phi^a] = S[\phi^+] - S[\phi^-] = \int d^4 x
\Big[\frac{1}{2}c_{ab}\big(\partial_\mu \phi^a \partial^\mu \phi^b - m^2
\phi^a \phi^b \big) -
\frac{\lambda}{4!}h_{abcd}\phi^a\phi^b\phi^c\phi^d \Big]\,,
\eeq
where $h_{1111}=-h_{2222}=1$ and the other components are zero, we
find by performing the shift $\phi^a = \bar{\phi}^a + \varphi^a$ that
the interaction part for the $\varphi$-theory is given by 
\beq
S_2[\varphi^a] =  \int d^4 x
\Big[-\frac{\lambda}{6}h_{abcd}\bar{\phi}^a\varphi^b\varphi^c\varphi^d -
\frac{\lambda}{4!}h_{abcd}\varphi^a\varphi^b\varphi^c\varphi^d\Big]\,.
\eeq
Note that a cubic interaction with an effective vertex depending on
the average field $\bar{\phi}^a$ is generated.

\subsection{Schwinger-Dyson equations for the propagators}
 
\begin{figure}
\centering
\includegraphics[width=0.7\textwidth]{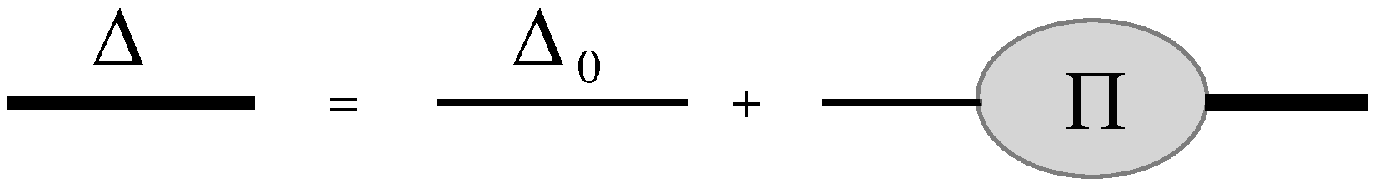}
     \caption{A generic form of a Schwinger-Dyson equation for the
    full 2-point function $\Delta^{ab}$.}
     \label{fig:SchDysGen}
\end{figure}

Using the expression (\ref{2PI_ansatz}) for $\Gamma_{\rm 2PI}$ in the
equations of motion (\ref{2PI_equations}) we get the following equation
for the propagator $\Delta^{ab}$ in the (physical) case of vanishing sources:
\beq
\frac{\delta \Gamma_{\rm 2PI}[\Delta]}{\delta \Delta^{ab}(x,y)} =
\frac12 i\Delta_{0,ab}^{-1}(x,y) - \frac12 i\Delta_{ab}^{-1}(x,y) +
\frac{\delta \Gamma_2[\Delta]}{\delta \Delta^{ab}(x,y)} = 0\,,
\label{SD1}
\eeq
where the second term follows from variation: $\delta {\rm Tr}[\ln
\Delta]/\delta \Delta^{ab} = \Delta_{ab}^{-1}$. By defining the {\em
  self energy}:
\beq
\Pi_{ab}(x,y) \equiv 2 i\frac{\delta \Gamma_2[\Delta]}{\delta
  \Delta^{ab}(x,y)}\,, 
\eeq
we see that this equation (\ref{SD1}) is of the form of famous {\em
  Schwinger-Dyson equation} for the full propagator:
\beq 
\Delta_{ab}^{-1}(x,y) = \Delta_{0,ab}^{-1}(x,y) + \Pi_{ab}(x,y)\,, 
\label{SD2}
\eeq
which (upon inverting) is presented graphically in Fig.~\ref{fig:SchDysGen}.
To get this equation in a form that is feasible for practical
calculations we multiply it from the right by $\Delta^{ab}$ to obtain
\beq
\int d^4 z \Delta_{0,ac}^{-1}(x,z) \Delta^{cb}(z,y) = \delta_{ab}
\delta^4(x-y) + \int d^4 z \Pi_{ac}(x,z) \Delta^{cb}(z,y)\,.
\label{SD_scalar}
\eeq
This is an integro-differential equation for the full propagator
$\Delta^{ab}$, because the inverse free propagator
$\Delta_0^{-1}$ on the LHS contains explicit spacetime derivatives acting on
$\Delta^{ab}$, while on the other hand the full propagator appears
inside the integral on the RHS, and also the self energy $\Pi$ is
(typically) a nontrivial functional of $\Delta^{ab}$. One should stress
that this Schwinger-Dyson equation is formally exact in the case of
Gaussian (or distant past vacuum) initial conditions. However, the
actual computation of the self energy via the 2PI action $\Gamma_2$
is a nontrivial task and in most cases of interest necessarily
involves some truncation, such as the loop expansion or a large $N$
expansion for ${\cal O}(N)$-invariant theories.

For fermions the construction of the 2PI effective action is very similar to
scalar fields. However, due to the fact that fermionic fields are Grassmann
numbers, there are some differences in the final formula, which for
vanishing fermionic average field is given by \cite{Berges04}:
\beq
\Gamma_{\rm 2PI}[G] = - i G_{0,ab}^{-1} G^{ba} + i {\rm Tr}[\ln G] +
\Gamma_2[G] + {\rm const} \,, 
\label{2PI_ferm} 
\eeq
where $\Gamma_2$ is again the sum of two-particle irreducible vacuum
graphs in the theory with propagator $G^{ab}$ and the vertices are the
same as in the original action, because of the vanishing average field. 
By comparing this with the scalar expression (\ref{2PI_ansatz}) we see a
difference of a factor $-1/2$ in the first two terms. This is due to different
values of the functional determinant arising from the scalar and
fermionic path integrals. In the same way as for scalar fields we
obtain a completely analogous Schwinger-Dyson equation for the
propagator $G^{ab}$:
\beq
\int d^4 z G_{0,ac}^{-1}(x,z) G^{cb}(z,y) = \delta_{ab}
\delta^4(x-y) + \int d^4 z \Sigma_{ac}(x,z) G^{cb}(z,y)\,,
\label{SD_ferm}
\eeq
where the self energy $\Sigma$ is now defined as:
\beq
\Sigma_{ab}(x,y) \equiv -i \frac{\delta \Gamma_2[G]}{\delta G^{ba}(y,x)}\,.
\label{self_ferm}
\eeq
Note that for a {\em complex} scalar field the 2PI effective action would
be similar to the fermionic expression (\ref{2PI_ferm}), without
$1/2$ factors, but with different signs in the first two
terms (see \eg \cite{ProSchWei04a}). Consequently the definition
of the self energy would be the same as in Eq.~(\ref{self_ferm})
except for the sign difference. The reason why the complex scalar field
differs from the real one is basically just the doubling of the degrees of
freedom for a complex field. Finally, we note
that for an interacting theory that combines fermionic and scalar fields
 \eg by a Yukawa interaction, we would just need to combine the nontrivial
$\Gamma_2$-parts of the 2PI effective actions and include contribution
from the Yukawa interaction. This would naturally lead to
cross-couplings in the fermionic and scalar Schwinger-Dyson equations.

%
%

\section{Kadanoff-Baym transport equations}

Let us use again the generic notation $\prop = \{G,\Delta\}$ for fermionic and
scalar propagators and denote both self energies simply by $\Sigma$, and
also adopt the same notation: $\Sigma^{++} \equiv \Sigma_F$
etc., as for scalar and fermionic propagators
Eqs.~(\ref{GFs_scalar})-(\ref{GFs_fermion}). It follows that the
different components of Schwinger-Dyson equations 
(\ref{SD_scalar}) and (\ref{SD_ferm}) are consistent provided that the
self energies can be divided in local (singular) and nonlocal
parts
\beq
\Sigma^{ab}(u,v) = c^{ab}\delta^4(u-v)\Sigma_{\rm sg}(u) +
\bar{\Sigma}^{ab}(u,v)\,,
\label{self_rel1}
\eeq
where the nonlocal part obeys similar relations as the propagators:
\beqa
\bar{\Sigma}_F(u,v) &=& \theta(u_0-v_0) \bar{\Sigma}^>(u,v) \pm
\theta(v_0-u_0) \bar{\Sigma}^<(u,v) 
\nonumber\\
\bar{\Sigma}_{\bar F}(u,v) &=& \theta(v_0-u_0) \bar{\Sigma}^>(u,v) \pm
\theta(u_0-v_0)\bar{\Sigma}^<(u,v)
\label{self_rel2}
\eeqa
for scalars and fermions, respectively. 
These relations (\ref{self_rel1})-(\ref{self_rel2}) should hold
generally for any reasonable approximation of the self energy
\cite{ProSchWei04a}. The singular term $\Sigma_{\rm sg}$ can be
absorbed into the
inverse propagator $\prop_{0,ab}^{-1}$ on the LHS of the
Schwinger-Dyson equations either to the mass renormalization
or to a classical background field (for example for gauge
interactions). From now on we assume that this absorption is made
and denote simply: $\Sigma^{ab} = \bar{\Sigma}^{ab}$. We further
define the retarted and advanced self energies and their
(anti)hermitian parts analogously to the propagators in Eqs.~(\ref{raGFs}) and
(\ref{Hdecomposition})\footnote{The defined self energies
clearly follow the hermiticity properties of the propagators,
Eq.~(\ref{CEq1}) and below}. The antihermitian part is denoted as $\Gamma$:
\beq
\Gamma \equiv \frac{1}{2i}\left(\Sigma^a - \Sigma^r\right) 
                     = \frac{i}{2}\left(\Sigma^> \mp \Sigma^<\right)\,,
\label{Gamma_def}
\eeq
corresponding to the scattering width of the field
excitations. For scalar fields Eq.~(\ref{Gamma_def}) is conventionally
defined as $\omega \tilde\Gamma$, when $\tilde\Gamma$ is directly the
scattering width with the correct dimension. For fermions $\Gamma$ is a
$4 \times 4$ (spinor) matrix and the physical meaning of various
elements is more obscure and will be discussed later in section
\ref{sec:ferm_preheating} in the case of interaction with a thermal
background.
   
Next we want to write the Schwinger-Dyson equations in a different form
to make a separation between the dynamical and spectral properties of
the system more evident. Using the above definitions and the corresponding
ones for the propagators in Eqs.~(\ref{raGFs}) and (\ref{Hdecomposition}),
and the fact that for vanishing average fields the inverse free
propagator obeys $\prop_{0,ab}^{-1} = c_{ab}\prop_{0,F}^{-1}$, 
it is a matter of simple algebra to show that the Schwinger-Dyson
equations (\ref{SD_scalar}) and (\ref{SD_ferm}) can be written in the
form:
\begin{eqnarray}
   (\prop_{0,F}^{-1}-\self_H) \otimes {\cal A} - \Gamma
   \otimes \prop_H = 0
\nonumber \\
    (\prop_{0,F}^{-1}-\self_H) \otimes \prop_H + \Gamma
   \otimes {\cal A} = \delta 
\label{SpecEq1}
\end{eqnarray}
and
\begin{equation}
   (\prop_{0,F}^{-1}-\self_H) \otimes \prop^< - \self^< \otimes \prop_H
   = \frac{1}{2}\left( \self^> \otimes \prop^< - \self^< \otimes \prop^>  
\right) \,,
\label{Dyneq}
\end{equation}
where, as stated before, we have assumed that the singular self energy
$\Sigma_{\rm sg}(u)$ is absorbed into $\prop_{0,F}^{-1}$, and we use the
notation $\otimes$ for the convolution integral:
\begin{equation}
    f \otimes g \equiv \int {\rm d}^4z f(u,z)g(z,v)\,.
\label{otimes}
\end{equation}
The equations (\ref{SpecEq1}) are called \emph{pole equations},  while
Eq.~(\ref{Dyneq}) is one of the two \emph{Kadanoff-Baym (KB)
  equations} \cite{KadBay62}. The similar KB-equation for the other
Wightman function
$\prop^>$ needs not to be considered, since from the definition
(\ref{Hdecomposition}) it immediately follows that $\prop^> =
\pm\prop^< - 2i {\cal A}$. In general, the pole equations will fix
the spectral properties of the theory, while the KB-equations will
give the dynamical evolution, \ie the quantum transport effects. Indeed,
in the classical limit the KB-equations (\ref{Dyneq}) for fermions and
scalars will reduce to well known quantum Boltzmann transport
equations for the phase space number densities (see \eg
\cite{CalHu08,KPSW01,KPSW02b,ProSchWei04a,ProSchWei04b}).

\subsection{Mixed representation and gradient expansion}

If there is a clear separation between internal (microscopic) and
external (macroscopic) scales in the system, it is appropriate to
analyze the pole- and KB-equations (\ref{SpecEq1})-(\ref{Dyneq}) in so
called mixed or Wigner representation, where a partial Fourier
transformation with respect to the internal coordinate $r=u-v$ is
performed. This transformation leads to a gradient expansion in
derivatives of the external (average) coordinate $x=(u+v)/2$, which
contains, in general, infinitely many terms. However, if the separation of the
scales is manifest, this expansion can be truncated (or
resummed) to a good approximation.

To begin with, let us define the Wigner transformation for an arbitrary
2-point function $F(x,y)$: 
\begin{equation}
F(k,x) \equiv \int d^{\,4} r \, e^{ik\cdot r} F(x + r/2,x-r/2) \,,
\label{wigner1}
\end{equation}
where $x\equiv (u+v)/2$ is the average coordinate, and $k$ is the
internal momentum conjugate to the relative coordinate $r\equiv
u-v$. Using this definition it is easy to transform the equations
(\ref{SpecEq1}) and (\ref{Dyneq}) into the mixed representation to get
\begin{eqnarray}
  \tilde\prop_{0,F}^{-1} {\cal A}
  -  e^{-i\Diamond}\{ \self_H \}\{ {\cal A}\}
  -  e^{-i\Diamond}\{ \Gamma \}\{\prop_H\} &=& 0
\label{SpecEqMix1} \\
  \tilde\prop_{0,F}^{-1} \prop_H
  -  e^{-i\Diamond}\{ \self_H \}\{ \prop_H \}
  +  e^{-i\Diamond}\{ \Gamma \}\{{\cal A}\} &=& 1
\label{SpecEqMix2}
\end{eqnarray}
and
\begin{equation}
 \tilde\prop_{0,F}^{-1}\prop^<
  -  e^{-i\Diamond}\{ \self_H \}\{ \prop^< \}
  -  e^{-i\Diamond}\{ \self^< \}\{ \prop_H \}
  = {\cal C}_{\rm coll} \,,
\label{DynEqMix}
\end{equation}
where the collision term in Eq.~(\ref{DynEqMix}) is given by
\begin{equation}
{\cal C}_{\rm coll} = -i e^{-i\Diamond}
                             \left( \{\Gamma\}\{\prop^<\} -
                                    \{\self^<\}\{{\cal A}\}\right)\,,
\label{collintegral}
\end{equation}
and the $\Diamond$-operator is the following generalization of the
Poisson brackets:
\begin{equation}
\Diamond\{f\}\{g\} = \frac{1}{2}\left[
                   \partial_x f \cdot \partial_k g
                 - \partial_k f \cdot \partial_x g \right]\,.
\label{diamond}
\end{equation}
The differential operators $\tilde\prop_{0,F}^{-1}$ are related to the inverse
propagators:
\begin{eqnarray}
\tilde\Delta_{0,F}^{-1} &\equiv& k^2 - \frac14 \partial^2 +
ik\cdot\partial - m^2 e^{-\frac{i}{2}\partial^m_x \cdot \partial_k}
\\
\tilde G_{0,F}^{-1} &\equiv& \kdag + \frac{i}{2} \deldag_x - m_R
e^{-\frac{i}{2}\partial^m_x \cdot \partial_k} - i\gamma^5 m_I
e^{-\frac{i}{2}\partial^m_x \cdot \partial_k} \,,
\label{inv_freeprop}
\end{eqnarray}
for bosons/fermions respectively, where $\partial^m_x$ means that the
derivative is acting on the left to $m^2, m_{R,I}$. Note that
these operators are not the transformed inverse propagators, but
follow from the identity: $e^{-i\Diamond}\{\prop_{0,F}^{-1}\}\{ F \} =
\tilde\prop_{0,F}^{-1} F$.

Eqs. (\ref{SpecEqMix1})-(\ref{DynEqMix}) are the desired quantum
transport equations for the 2-point correlation functions $\prop^<$,
$\prop_H$ and ${\cal A}$. We are primarily interested in solving the
Kadanoff-Baym equation (\ref{DynEqMix}) for the Wightman function
$\prop^<$, but in general, also the pole equations
(\ref{SpecEqMix1})-(\ref{SpecEqMix2}) have to be considered because of
the cross-couplings between the equations. One can see that these
equations indeed contain the derivatives of the masses $m$ and the
self-energies $\Sigma$ up to infinite order, restraining their use in
practical applications unless this gradient
expansion can be truncated (or resummed) in some reasonable way. 

In the standard approach to {\em quantum kinetic theory} the
conditions of so called quasiparticle or on-shell approximation
are assumed, including slowly varying (\ie ``nearly'' translation invariant)
background fields {\em and} correlators, as well as weak interactions
\cite{CalHu08}. This approach provide a consistent approximation to
truncate the gradient expansion in KB-equations (\ref{DynEqMix}) to
leading order, culminating in the derivation of the famous quantum Boltzmann
equations. However, because of the assumptions made, it follows
that the correlators whose dynamics we are studying, are ``close'' to
local thermal equilibrium throughout the evolution. Especially the
information on quantum coherence will be irrevocably lost. In the
next chapter we start to build an extended approximation scheme that
incorporates the good features of the standard kinetic approach with
easy-to-use Boltzmannian-type transport equations, yet including the
effects of nonlocal quantum coherence.

%
%

\section{Physical quantities from 2-point correlation functions}
\label{sec:physical1}

We conclude this chapter by writing down some familiar physical
observables, including the number currents and the energy momentum
tensors for fermionic and scalar fields, in terms of the 2-point
correlation functions $G^<$ and $\Delta^<$. These expressions follow
directly from the definitions of the correlators in
Eqs.(\ref{GFs_scalar}-\ref{GFs_fermion}) written in the mixed
representation, so we simply list the results here. The
expectation value of the fermionic number current is given by
\beq
\langle j_F^\mu(x)\rangle = \langle \bar\psi\gamma^\mu\psi\rangle =
\int \frac{{\rm d}^4 k}{(2\pi)^4}\,{\rm Tr}\left[\gamma^\mu
    iG^<(k,x)\right]\,,
\label{ferm_current1}
\eeq
and for {\em complex}\footnote{No conserved Noether current
  can be defined for a real scalar field} scalar bosons we have:
\beq
\langle j_B^\mu(x)\rangle = -i\big\langle
\partial^\mu\phi^\dagger\phi
  - \phi^\dagger\partial^\mu\phi\big\rangle =
\int \frac{{\rm d}^4 k}{(2\pi)^4}\,2 k^\mu i\Delta^<(k,x)\,.
\label{bos_current1}
\eeq
The symmetric (Belinfante) energy momentum tensor (see \eg
\cite{Weinberg95}) for fermions is
\beqa
\langle\theta^{\mu\nu}(x)\rangle &=&
\frac{i}{4}\Big\langle\bar\psi\gamma^\mu\partial^\nu\psi
- \partial^\nu\bar\psi\gamma^\mu\psi\Big\rangle + \mu
\leftrightarrow \nu
\nonumber\\[1mm]
&=& \int \frac{{\rm d}^4 k}{(2\pi)^4}\,{\rm
  Tr}\Big[\frac12(\gamma^\mu k^\nu + \gamma^\nu k^\mu) iG^<(k,x)\Big]\,,
\label{ferm_energy_momentum}
\eeqa
while the bosonic tensor for real scalar field is given by
\beqa
&&\langle T^{\mu\nu}(x)\rangle =
\Big\langle \partial^\mu\phi\,\partial^\nu \phi -
\frac12 g^{\mu\nu}\big[(\partial \phi)^2 - m^2
\phi^2 \big] \Big\rangle
\nonumber\\[1mm]
&&= \int \frac{{\rm d}^4 k}{(2\pi)^4}\,\Big[k^\mu k^\nu +
\frac14 \partial_x^\mu \partial_x^\nu - \frac12 g^{\mu\nu}\big(k^2 - m^2 +
\frac14 \partial_x^2 \big)\Big]i\Delta^<(k,x)\,,
\label{bos_energy_momentum}
\eeqa
where $g^{\mu\nu}$ is the standard Minkowskian metric with signature
$(+,-,-,-)$. For a complex scalar field we get a result with the last row
of Eq.~(\ref{bos_energy_momentum}) multiplied by two. These relations
demonstrate the importance of the 2-point Wightman functions in
nonequilibrium quantum field theory. Later, in section
\ref{sec:physical2} we will use these results to express the
observables in terms of the on-shell distribution functions.

%% file: Chapter3.tex
%
%

\chapter{Quasiparticle picture including nonlocal quantum coherence}
\label{chap:scheme}

%
%

\section{Extended quasiparticle approximation}
\label{sec:quasi_approx}

In the literature (see \eg \cite{LeBellac00}) a {\em quasiparticle
approximation} (QPA)
usually refers to a set of approximations leading to a spectral phase
space structure for the 2-point correlation functions $\prop^{<,>}$,
 composed of sharp singular shells with definite
energy momentum dispersion relations, such as the standard free particle mass
shell. An alternative, stronger definition for QPA \cite{CalHu08}
requires that the functional forms of the free theory propagators
are preserved in (QPA) interacting theory, with only the mass
parameters replaced by effective masses. For both of these definitions
the necessary conditions for the quasiparticle approximation to be
justified include weak interactions and slowly varying background
fields. Moreover, the standard treatment of QPA relies on the
assumption that system be close to thermal equilibrium and
consequently nearly translation invariant, the famous example being
the derivation of the quantum Boltzmann transport equation from KB-equations
\cite{CalHu08}.

Our {\em extended quasiparticle approximation} (eQPA) scheme
relinquishes the assumption that the system needs to be close to
thermal equilibrium. The key observation in our approach is that under the
(otherwise) same conditions of QPA with weak interactions and slowly
varying background fields, the phase space of the 2-point correlators
$\prop^{<,>}$ contains novel and completely different singular shell
solutions, in addition to the standard (quasi)particle
mass-shell solutions. These new $k_{0,z}=0$-shell solutions are unavoidably
absent if we demand that the system is near thermal equilibrium, hence their
lacking in the standard quasiparticle treatments. We will examine
these new solutions in detail in section \ref{sec:shell}, where we
interpret them as describing {\em the nonlocal quantum coherence
  between ``opposite'' (quasi)particle excitations}. After the complete spectral
structure of the correlators is discovered, we feed it as an
ansatz to the dynamical equations to find out the equations of
motion for the corresponding on-shell distribution functions $f$, including
the novel coherence shells. In this way we get an extension to the
quantum Boltzmann transport equation to include the effects of
nonlocal quantum coherence.

First, we examine the necessary conditions for the extended
quasiparticle approximation to be justified. These conditions include
weak interactions, slowly varying background fields, and existence of
particular spacetime symmetries. If some of these conditions are not
fulfilled, the spectral approximation for the phase space breaks down. 
At the end of this section we outline the procedure of using the
dynamical equations to find the desired equations of motion for
the on-shell functions. The actual derivation of the appropriate equations for
fermions and scalar bosons is presented in section \ref{sec:eq_motion}.

\subsection{Weak interactions}
\label{sec:weak_int}

The limit of weak interactions in the context of quasiparticle
approximation means $\Gamma\rightarrow 0$ in
Eqs.~(\ref{SpecEqMix1})-(\ref{DynEqMix}), where $\Gamma$ is the
interaction width defined in Eq~(\ref{Gamma_def}). This limit is taken
strictly whenever the {\em phase space properties} of the correlators
are studied. However, when studying the {\em dynamical properties}, one
has to include $\Gamma$ to at least leading order to get any
thermalization effects. This is precisely the way how the Boltzmann
transport equation is obtained in the classical limit.  

Neglecting the terms proportional to $\Gamma$ in
Eqs.~(\ref{SpecEqMix1})-(\ref{DynEqMix}), including the collision term
${\cal C}_{\rm coll}$ (in general $\Sigma^{<,>}\sim \Gamma$), leads to
the free field equations except for the self energies $\self_H$. It is not
completely obvious how these self energies should be handled, however. For a
controlled expansion in the coupling constant, all the self energies need
to be treated in an equal footing; if we neglect $\Gamma \sim g^n$,
we have to neglect 
also $\self_H$ at the same order. So, it would be justified to retain
$\self_H$ only if it is of lower order in coupling constant than
$\Gamma$. This is indeed the case for gauge interactions for example,
with $\Sigma_H \sim g^2$ and $\Gamma \sim g^4$ in the lowest
order. Another approach is to treat $\self_H$ and
$\Gamma$ completely independently, so that even if there is no
coupling hierarchy, $\self_H$ will be retained in the
equations when the phase space properties of the correlators are
examined. The motivation for this non-controlled approximation is that
retaining $\self_H$ will not ruin 
the spectral structure of the correlators; it merely modifies the
dispersion relations of the excitations, so that the standard free
particles become {\em quasiparticles}.

\subsection{Slowly varying background, mean field limit}

In general, it is not enough to neglect the terms proportional to
$\Gamma$ to obtain a spectral phase space structure for the 2-point
correlators. This can be demonstrated for a scalar field with
nonvanishing constant $\partial_t m^2 \equiv \epsilon > 0$, while
other derivatives of the mass $m$ are vanishing. The spatially
homogeneous solution for the free field ($\self_H$ is also neglected)
KB-equation (\ref{DynEqMix}) is then:
\beq
i\Delta^<(k_0,|\vec k|; t) \propto
{\rm Ai}\left(\frac{4^{1/3}(k^2-m^2(t))}{\epsilon^{2/3}} \right)\,,
\label{Airy}
\eeq
where ${\rm Ai}(x)$ is the Airy-function. This is not a singular
distribution in momentum for nonvanishing $\epsilon$, but indeed in the limit
$\epsilon\rightarrow 0$ it reduces to $i\Delta^< \propto \delta(k^2 -
m^2(t))$. This example illustrates that in order to obtain spectral
phase space structure one needs to consider slowly varying background
fields. Moreover, to actually get singular shell solutions {\em all}
derivatives of the background field have to be neglected in the equations
(\ref{SpecEqMix1})-(\ref{DynEqMix}). This approximation of including
only the zeroth order gradients of the background is called {\em mean
  field} (or adiabatic) limit. The resulting KB-equations for the
study of the phase space properties of the correlators in combined weakly
interacting and mean field limits are then
\beqa
  \big(\kdag - \frac{i}{2} \deldag_x - m_R
       - im_I \gamma^5 - \Sigma_H\big) \, G^< &=& 0
\label{DynEqMix_QPA}
\\
 \big(k^2 - \frac14 \partial^2 + ik\cdot\partial 
      - m^2 - \Pi_H\big) \Delta^< &=& 0
\label{DynEqMix_QPA_sca}
\eeqa
for fermions and scalars, respectively. The corresponding pole
equations are completely the same with $0$ replaced by $1$ on the RHS
of the equations for $G_H$.

\subsection{Special spacetime symmetries}

It turns out that even the mean field equations
(\ref{DynEqMix_QPA})-(\ref{DynEqMix_QPA_sca}) do not in general have
spectral solutions for the correlators. We see this by considering a
$1+1$ dimensional free scalar field, for which the solution for
Eq.~(\ref{DynEqMix_QPA_sca}) reads (for $k_0 \neq 0$) \cite{Mrowczynski97}:
\beq
i\Delta^<(k,x) = \left[\theta(-k^2) +
  \theta(k^2-m^2)\right]\big[A(k)\cos(q\cdot x) + B(k) \sin(q\cdot x) \big]\,,
\eeq
where $A(k)$ and $B(k)$ are real functions of $k$ determined by the
initial conditions, and $q$ is defined as:
\beq
q \equiv 2|k_0|\sqrt{(k^2-m^2)/k^2}\left(\frac{k_1}{k_0},1\right)\,.
\eeq
This solution is not restricted to spectral form with support only on
singular shells; instead it potentially has support everywhere inside the
mass shell $k^2 \geq m^2$, or outside the light cone $k^2 < 0$,
depending on the unspecified functions $A(k)$ and $B(k)$. For
an arbitrary $x$-dependence this is in conflict with the quasiparticle
approximation, which requires that the phase space structure is
spectral. However, if we demand for example the complete translational
invariance: $\partial_{x_0,x_1}\Delta^<(k,x)=0$, then $q$ must be zero,
implying that $k^2-m^2=0$. We conclude that there can be support only on the
mass shell, and the solution is indeed spectral. In the same way, if we
demand only time translational invariance,
$\partial_{x_0}\Delta^<(k,x)=0$, we find two different spectral solutions: the
former mass-shell solution, but also a nonconstant solution in $x_1$
with $k_1 =0$. Later, in section \ref{sec:shell}, we examine the
spectral phase space solutions in more detail with a different
approach by subjecting the solutions of
Eqs.~(\ref{DynEqMix_QPA})-(\ref{DynEqMix_QPA_sca}) on certain
spacetime symmetries in the first place. At least for fermions, this seems to be
the only reasonable method, because of the complex spinor
structure. In section \ref{sec:shell} we will find that the phase
space structure of fermionic and scalar correlators is indeed spectral for two
particular spacetime symmetries of interest: {\em spatial homogeneity}
  and {\em static planar symmetry}.

\subsection{Spectral ansatz for dynamical equations}

The next step in our approximation scheme is to insert the
spectral solutions as an {\em ansatz} in the full KB-equations (\ref{DynEqMix}).
Now we are interested in the dynamics of the spectral
solutions, so we will consider only those of the resulting component
equations that include direct spacetime derivatives. We will resort to some
approximations also in this step. As stated before, we are assuming
the limit of weak interactions. However, now we do not want to neglect the terms
proportional to $\Gamma$ completely, since that would lead to
collisionless plasma dynamics. Instead we will include only the
leading order terms in $\Gamma$. To see which terms are actually
leading order can be
somewhat difficult in practice. It has been shown that for a scalar
field close to thermal equilibrium, the term proportional to $\prop_H$
on the LHS of Eq.~(\ref{DynEqMix}) is of higher order in $\Gamma$ than
the dominant contribution from the collision term \cite{ProSchWei04a}, so in
this limit it is justified to neglect that term. For more general
situations, however, it is not evident that neglecting this term would be
justified by any simple arguments. For fermions this hierarchy has not
been shown even for systems close to thermal equilibrium, as far as we know.
Nevertheless, neglecting the term $\propto\prop_H$ may anyway be a good first
approximation; for example the well-known Boltzmann transport equation is
derived in this limit. Further investigations on the role of this term
(in dynamical equations) are definitely needed to make any conclusion
on its importance.

The role of the term proportional to $\Sigma_H$ on the LHS of
Eq.~(\ref{DynEqMix}) poses another question. Clearly, if the mean field
part has been included in the study of spectral properties in equations
(\ref{DynEqMix_QPA})-(\ref{DynEqMix_QPA_sca}), the same term should be
included in dynamical equations as well. The gradient
corrections\footnote{The higher gradient corrections in $\Sigma\,
  \prop$-terms do {\em not} necessarily correspond to gradients of the
  background field, and thus are not on the same footing as the
  overall mean field approximation} to
this term are yet another question, and in the sense of a controlled expansion
in coupling constant, they should also be included. However, the role of
these higher gradient contributions would probably be the same
as the mean field contribution to $\Sigma_H$ \ie to affect the
spectral properties by ``modifying'' parameters like masses and
momenta in the equations. By these arguments, neglecting higher
gradient terms of $\Sigma_H$ when the dynamics of the spectral
solutions are studied, seems to be quite well justified.

On the other hand, it is of course possible to include all of these
interaction dependent terms in Eq.~(\ref{DynEqMix}). Based on the
above discussion, it is appropriate to do this
by including everything else than the mean field contribution of
$\Sigma_H$ formally into the collision term. The KB-equations for fermions
and scalars then become         
\beqa
  \Big(\kdag - \frac{i}{2} \deldag_x - m_R
e^{-\frac{i}{2}\partial^m_x \cdot \partial_k} - i\gamma^5 m_I
e^{-\frac{i}{2}\partial^m_x \cdot \partial_k} - \Sigma_H\Big)\, G^<
&=& \tilde{\cal C}_{\rm coll}^\psi \,.
\label{DynEqMix_QPA2}
\\
 \Big(k^2 - \frac14 \partial^2 + ik\cdot\partial 
      - m^2 e^{-\frac{i}{2}\partial^m_x \cdot \partial^\Delta_k} -
      \Pi_H\Big) \Delta^< 
&=& \tilde{\cal C}_{\rm coll}^\phi \,,
\label{DynEqMix_QPA_sca2}
\eeqa
where the ``extended'' collision terms are defined as 
\beqa
\tilde{\cal C}_{\rm coll}^\psi &=& e^{-i\Diamond}
                             \left( \{\Sigma_H-i\Gamma\}\{G^<\} +
                                 \{\Sigma^<\}\{G_H+i{\cal A}\}\right)
                               - \Sigma_H G^<\,. 
\\
\tilde{\cal C}_{\rm coll}^\phi &=& e^{-i\Diamond}
                             \left( \{\Pi_H-i\Gamma\}\{\Delta^<\} +
                                    \{\Pi^<\}\{\Delta_H + i{\cal A}\}\right)
                                     - \Pi_H \Delta^<\,. 
\label{collintegral_mod}
\eeqa
The higher order gradients in the collision term are a
delicate issue in our approximation scheme. Usually, when slowly
varying backgrounds are studied and the solutions are close to
thermal equilibrium, it is justified to neglect consistently all
higher (than zeroth) order derivatives in collision term, since those will
necessarily correspond to higher order gradients in the background
field. However, in our scheme the coherence shell solutions are
rapidly oscillating even in constant backgrounds, and consequently the
higher derivatives in the collision terms {\em do not} necessarily
correspond to higher order gradients in the background field and thus
it is not \apriori justified to neglect them. In practical
calculations this gradient expansion has to be truncated, however,
unless the different order gradients of the self energy terms can be
resummed in some useful way. Later on in chapter \ref{chap:applications},
we show that this resummation is indeed possible for a scalar field
interacting with a thermal background.
   
To summarize, our approximation scheme works as follows:
First, we find out the spectral properties of the 2-point functions by
using the weakly interacting mean field equations
(\ref{DynEqMix_QPA})-(\ref{DynEqMix_QPA_sca}). Then, we substitute the
obtained spectral solutions as an ansatz in the full interacting equations
(\ref{DynEqMix_QPA2})-(\ref{DynEqMix_QPA_sca2}) (with the
extended collision terms $\tilde{\cal C}_{\rm coll}$ or the
standard ones of Eq.~(\ref{collintegral})) to find out the
equations of motion for the on-shell functions in the presence of collisions.
The crucial difference compared with the standard treatment based on the
quasiparticle
approximation is that we do not assume that the system is close to
thermal equilibrium at any moment, yet we are using the spectral
solutions for the correlators. This apparent paradox will be settled in
section \ref{sec:shell}, where we find the novel coherence shell solutions
that have completely different properties than the standard
(quasi)particle mass-shell solutions. In what follows, we 
will neglect the $\Sigma_H$ and $\Pi_H$ terms on the LHS of
Eqs.~(\ref{DynEqMix_QPA2})-(\ref{DynEqMix_QPA_sca2}), since we are
primarily interested in the general structure of the phase space and
the dynamics of the on-shell functions, and not so much in the
specific modifications of dispersion relations caused by
interactions. Thus, while studying the phase space properties, the
corresponding equations are effectively reduced to the {\em noninteracting
  mean field limit}.

%
%

\section{Reduction of the spin structure in fermionic equations}

Before we enter the study of the phase space shell structure, let us
first simplify the fermionic KB-equation (\ref{DynEqMix_QPA2}) further
in the cases of two spacetime symmetries of interest: the spatial
homogeneity and the static planar symmetry. To begin with, we write
the equation (\ref{DynEqMix_QPA2}) for the hermitian Wightman
function, defined as
\beq
\bar G^<(k,x) \equiv i G^<(k,x)\gamma^0\,.
\label{herm_gless}
\eeq
By multiplying both sides of Eq.~(\ref{DynEqMix_QPA2}) by $\gamma^0$ we get
\beq
\Big( k_0 + \frac{i}{2}\partial_t
    - \vec \alpha \cdot (\vec k - \frac{i}{2}\vec \nabla)
    - \gamma^0 \hat m_0 - i\gamma^0 \gamma^5 \hat m_5
\Big) \bar G^<
      = \gamma^0 i{\cal C}_{\rm coll}^\psi\gamma^0 \,,
\label{DynEqMixQPAFinal}
\eeq
where we use the notation:
\beq
\hat m_{\rm 0,5}(x) \equiv m_{R,I}(x) e^{-\frac{i}{2}
       \partial_x^m \cdot \partial_k}\,,
\label{massoperators}
\eeq
and we have dropped the tilde in the collision term to denote either
the extended collision term in Eq.~(\ref{collintegral_mod}) or the
standard one in Eq.~(\ref{collintegral}).

\subsection{Spatially homogeneous case, helicity diagonal\\ correlator}
\label{sec:heli_diag}

In a spatially homogeneous case the spatial gradients of $G^<$ and $m$
in Eq.~(\ref{DynEqMixQPAFinal}) vanish, and consequently the
helicity operator $\hat h= \hat k \cdot \vec S = \hat k \cdot \gamma^0
\vec \gamma \gamma^5$, where $\hat k \equiv \vec k/|\vec k|$, commutes
with the differential operator on the LHS of
Eq.~(\ref{DynEqMixQPAFinal}). This implies that different helicity projections
\beq
\bar G_{hh'}^<(k,x) \equiv P_h \bar G^<(k,x) P_{h'}\,,  
\eeq
where $P_h$ denotes the helicity projector:
\beq
P_h = \frac12(1 + h \hat h),\qquad P_h P_{h'} = \delta_{hh'},\qquad h =
\pm 1\,,
\eeq
do not mix in a noninteracting theory\footnote{By noninteracting theory
  we
  mean here that the self-energies $\Sigma^{ab}$ and consequently
  the collision term are vanishing. The system still interacts with the
  varying classical background (giving rise to varying mass).}, so
that helicity is a good quantum number \ie a conserved quantity. It
follows that the helicity off-diagonals couple to the dynamics of the
diagonal part only through the collision term. In this work we will
not consider these cross-couplings, but use the helicity diagonal part
of the correlator: $G^< = \sum_{h=\pm 1} G_{hh}^<$, as an ansatz for
the interacting theory. In the Weyl basis, where the gamma-matrices
are given by the following direct product expressions (both $\rho^i$
and $\sigma^i$ are the usual Pauli matrices referring here to chiral /
spin d.o.f. respectively):
\beq
\gamma^0 = \rho^1 \otimes 1 \,,
\qquad
\vec \alpha = -\rho^3 \otimes \vec \sigma
\quad {\rm and} \quad \gamma^5 = -\rho^3 \otimes 1 \,,
\label{gammamatrices}
\eeq
the helicity diagonal correlator can be written as: 
\beq
  \bar G_{hh}^<  \equiv g_h^< \otimes
  \frac12(1 + h \hat k\cdot \vec \sigma),
\label{connectionHOMOG}
\eeq
where $g_h^<$ are (unknown) hermitian $2\times 2$ matrices (for $h = \pm
1$) in chiral indices. By taking a helicity diagonal projection of
Eq.~(\ref{DynEqMixQPAFinal}) we get then an equation for $g_h^<$:
\beq
\Big( k_0 + \frac{i}{2}\partial_t
       + h |\vec k| \rho^3
       - {\hat m}_0 \rho^1 + {\hat m}_5 \rho^2
\Big) g_h^< = {\cal C}_h\,,
\label{Gs-eomHOMOG}
\eeq
where the $2\times 2$ matrix ${\cal C}_h$ is the chiral part of the
helicity diagonal projection of the collision term:
\beq
 P_h\left(\gamma^0 i{\cal C}_{\rm coll}^\psi \gamma^0\right)P_h \equiv {\cal
   C}_h \otimes \frac12(1 + h \hat k\cdot \vec \sigma)\,.  
\label{coll_chiral_HOM}
\eeq
Given that $g_h^<$ is a hermitian matrix, it is useful to decompose the
equation (\ref{Gs-eomHOMOG}) into hermitian (H) and antihermitian (AH) parts:
\begin{eqnarray}
{\rm (H)}:\quad 2k_0 g^<_h &=& \hat H g_h^< + g_h^< \hat H^\dagger + {\cal
   C}_h^+
\label{Hermitian22}
\\ 
{\rm (AH)}:\quad i \partial_t g^<_h &=& \hat H g_h^< - g_h^< \hat
H^\dagger  + {\cal C}_h^- \,,
\label{AntiHermitian22}
\end{eqnarray}
where  
\begin{equation}
\hat H \equiv - h|\vec k|\rho^3 + \hat m_0 \rho^1 - \hat m_5\rho^2
\label{GenHamiltonian}
\end{equation}
can be interpreted as a local free field Hamiltonian operator, and
${\cal C}_h^\pm \equiv {\cal C}_h \pm {\cal C}_h^\dagger$ are the
hermitian and antihermitian parts of ${\cal C}_h$.
We see that in the noninteracting mean field limit with ${\cal
  C}_h=0$ and $\hat m_{\rm 0,5} = m_{R,I}$ only the
AH-equation contains an explicit time derivative of $g^<_h$, while the
H-equation is a purely algebraic matrix equation. For this reason the
AH-equation is called a {\em ``kinetic equation}'' describing the
dynamical evolution of the helicity diagonal Wightman function in a
varying background. The H-equation on the other hand is called a {\em
  ``constraint equation}'', which will constrain the solutions of the
kinetic equation in the 4-dimensional phase space and thus is the one
to be used in determining the phase space properties of the (eQPA)
interacting theory.

\subsection{Static planar symmetric case, spin-$z$ diagonal\\ correlator}
\label{sec:spinz}

In a static planar symmetric case all but one spatial derivative
(chosen to be $z$ here) of $G^<$ and $m$ vanish. Then, apart from the
$\vec \alpha_{\pp } \cdot \vec k_\pp$-term, the
differential operator on the LHS of Eq.~(\ref{DynEqMixQPAFinal})
commutes with the spin in $z$-direction: $\hat S^3 =
\gamma^0\gamma^3\gamma^5$. We can proceed analogously to the previous
section \ref{sec:heli_diag}, by first performing a Lorentz
boost to a frame where this term vanishes
\cite{KPSW01,KPSW02b,ProSchWei04a}. The Dirac representation of 
the desired boost is found to be 
\beq
  S_\pp = {\rm sgn(k_0)} \frac{k_0 + \tilde{k}_0
               - \vec{\alpha}\cdot\vec{k}_\pp}
                {\sqrt{2\tilde{k}_0(k_0+\tilde{k}_0)}}\,.
\label{boostperp}
\eeq
The boosted correlator\footnote{Whereas $G^<$ transforms conventionally:
  $G^< \rightarrow S\,G^< S^{-1}$, the hermitian correlator $\bar G^<
  = iG^< \gamma^0$ obeys a peculiar transformation law: $\bar G^<
  \rightarrow S\,\bar G^< S$.}
\beq
  \bar G^<_\pp(\tilde k_0, k_z; z)
    \equiv S_\pp  \bar G^<(k; z) S_\pp
\eeq
then obeys an equation
\beq
\Big( \,  \tilde k_0
-  \alpha^3 \, (k_z - \frac{i}{2}\partial_{z_w})
        - \gamma^0  \hat m_{0}  - i\gamma^0\gamma^5 \hat m_{5}
        \, \Big) \, \bar G_\pp^<(\tilde k_0, k_z; z) = S_\pp^{-1}\gamma^0 i{\cal
          C}_{\rm coll}^\psi \gamma^0 S_\pp\,,
\label{G-lessEq5_noperb}
\eeq
where $\tilde k_0 = {\rm sgn}(k_0) (k_0^2-k_\pp^2)^{1/2}$. After this
boost the differential operator on the LHS of Eq.~(\ref{G-lessEq5_noperb})
indeed commutes with the spin $\hat S^3$, as expected. Analogously to
the spatially homogeneous case this implies that different spin projections  
\beq
\bar G_{\pp ss'}^< \equiv P_s \bar G_\pp^< P_{s'}\,,  
\eeq
where $P_s$ denotes the spin-$z$ projector:
\beq
P_s = \frac12(1 + s \hat S^3),\qquad P_s P_{s'} = \delta_{ss'},\qquad s =
\pm 1\,,
\eeq
do not mix in a noninteracting theory, so that spin-$z$ is a good
quantum number. We again neglect the effects of spin off-diagonals and
consider only the spin diagonal correlator, which is written in the Weyl
basis as:
\beq
\bar G_{\pp s}^<  \equiv g_s^< \otimes \frac12(1 + s \sigma^3)
\,,
\label{connection}
\eeq
where $g_s^<$ are hermitian $2\times 2$ matrices (for $s = \pm 1$) in
chiral indices. By taking the spin-$z$ diagonal projection of
Eq.~(\ref{G-lessEq5_noperb}) we get now the following equation for
$g_s^<$: 
\beq
\Big( \tilde k_0
       + s (k_z - \frac{i}{2}\partial_z) \rho^3
       - {\hat m}_0 \rho^1 + {\hat m}_5 \rho^2
\Big) g_s^< = {\cal C}_s \,,
\label{Gs-eom}
\eeq
where the $2\times 2$ matrix ${\cal C}_s$ is the chiral part of the
spin-$z$ diagonal projection of the boosted collision term:
\beq
 P_s\left(S_\pp^{-1} \gamma^0 i{\cal C}_{\rm coll}^\psi\gamma^0
   S_\pp\right)P_s \equiv {\cal C}_s \otimes \frac12(1 + s \sigma^3)\,. 
\eeq
Now, because the $\rho^3$ matrix is multiplying the derivative
$\partial_z$ in Eq.~(\ref{Gs-eom}), the straightforward division into hermitian
and antihermitian parts does not lead to a convenient separation of the
derivatives. However, by first multiplying Eq.~(\ref{Gs-eom}) from the
left by $\rho^3$ and only then taking the hermitian and antihermitian
parts gives the desired division:
\begin{eqnarray}
{\rm (H)}:\quad -2 s k_z g^<_s &=& \hat P g_s^< + g_s^< \hat P^\dagger
- (\rho^3{\cal C}_s)^+ 
\label{Hermitian22z}
\\ 
{\rm (AH)}:\quad\ \  is \partial_z g^<_s &=& \hat P g_s^< - g_s^< \hat
P^\dagger - (\rho^3{\cal C}_s)^-\,,
\label{AntiHermitian22z}
\end{eqnarray}
where
\begin{equation}
\hat P \equiv k_0 \rho^3  + i(\hat m_0 \rho^2 + \hat m_5\rho^1)\,, 
\label{Pamilton}
\end{equation}
and $(\rho^3{\cal C}_s)^\pm \equiv \rho^3{\cal C}_s \pm {\cal
  C}_s^\dagger \rho^3$ are the hermitian and antihermitian parts of
$\rho^3{\cal C}_s$. These equations are analogous to
Eqs.~(\ref{Hermitian22})-(\ref{AntiHermitian22}) of the spatially
homogeneous case. Again, the AH-equation is called a ``kinetic equation''
describing the dynamical $z$-evolution of the spin diagonal Wightman
function, while the H-equation is called a ``constraint equation''
determining the phase space properties of the correlator.

%
%

\section{Phase space shell structure}
\label{sec:shell}

We now begin to examine the phase space structure of the fermionic and
scalar Wightman functions $iG^<$ and $i\Delta^<$ in the (extended) quasiparticle
limit discussed in section \ref{sec:quasi_approx} for the case of
$\Sigma_H=\Pi_H=0$. We find out that in
addition to the standard mass-shell excitations, with the dispersion
relation $k^2 - |m|^2$, the phase space consists of novel
singular shell solutions that are located at $k_0 = 0$ for a spatially
homogeneous case and at $k_z=0$ for a static planar symmetric case.

\subsection{Fermions}

As discussed above, the relevant equations that describe the phase space
properties of the fermionic Wightman function are the constraint (H)
equations Eqs.~(\ref{Hermitian22}) and (\ref{Hermitian22z}), that in
the noninteracting mean field limit reduce to
\beqa
2k_0 g^<_h &=& \{H, g^<_h\}\,, \qquad\quad\ \ \, H \equiv - h|\vec
k|\rho^3 + m_R \rho^1 -  m_I \rho^2
\label{constraint_HOM}
\\[2mm]
-2 s k_z g^<_s &=& P g_s^< + g_s^< P^\dagger\,, \qquad P \equiv \ k_0
\rho^3  + i(m_R \rho^2 + m_I\rho^1)
\label{constraint_STA}
\eeqa
for the spatially homogeneous and the static planar symmetric cases,
respectively.

\subsubsection{Spatially homogeneous case}

\begin{figure}
\centering
\hskip -3truecm
\includegraphics[width=0.90\textwidth]{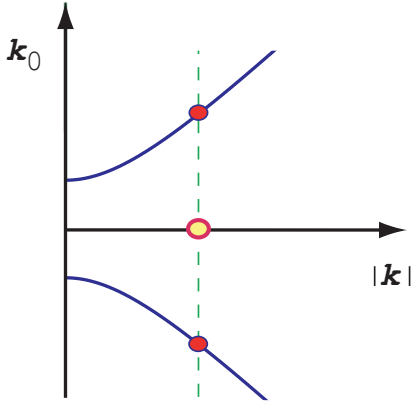}
\vskip -1.2cm 
    \caption{Shell structure of the correlators $G^<$ and $\Delta^<$ in the
      spatially homogeneous case. The dark filled (red) blobs show the
      mass-shell contributions for a given $|\vec k|$, and the light
      (yellow) blob shows the corresponding coherence contribution
      from the new $k_0=0$-shell.}
    \label{fig:DR-homog}
\end{figure}

To further analyze the constraint equation (\ref{constraint_HOM}) for the
spatially homogeneous case, it is convenient to introduce the
so called Bloch-representation for the chiral matrix $g_h^<$: 
\beq
g^<_h \equiv \frac12 \left( g^h_0 +  g^h_i \rho^i \right)\,,
\label{bloch_HOM}
\eeq
where $\rho^i$ are the (chiral) Pauli matrices and $g^h_\alpha(k,t)$ are
real functions, because of the
hermiticity of $g^<_h$. It is easy to see that in the
Bloch-representation the constraint equation
(\ref{constraint_HOM}) decomposes into a simple homogeneous
matrix equation
\beq
B_\alpha^{\ \beta} g^h_\beta = 0\,,
\label{matrix_eq}
\eeq
where the $4\times4$ coefficient matrix is 
(index ordering is here defined as $\alpha = 0,3,1,2$):

\beq
B = \left( \begin{array}{cccc}
    k_0  &  h|\vec{k}|  &  -m_R  &  m_I \\
    h|\vec{k}| &  k_0   &  0     &  0   \\
    -m_R &  0     &  k_0   &  0   \\
    m_I  &  0     &  0     &  k_0 
    \end{array} \right)\,.
\eeq
A homogeneous matrix equation, such as Eq.~(\ref{matrix_eq}), may have nonzero
solutions only when the determinant of the matrix vanishes. Here the
determinant is simply:
\beq
\det(B) = \left( k^2 - |m|^2 \right)k_0^2 \,,
\label{constraint_det}
\eeq
which implies that the nonzero solutions are possible only when
\beq
k^2 - |m|^2 = 0 \qquad {\it or} \qquad k_0 = 0\,.
\eeq
These constraints give rise to a {\em singular shell structure} for the
solutions, since they need to be proportional to $\delta(k_0^2 - \vec{k}^2 -
|m|^2)$ or $\delta(k_0)$. The former class is identified as the
standard one particle mass-shell solutions, with the dispersion
relation $k_0 = \pm \omega_k \equiv \pm (\vec{k}^2 + |m|^2)^{1/2}$,
while the latter class of solutions with $k_0=0$ are completely novel
in the context of quantum field theory (see
Fig.~\ref{fig:DR-homog}). Based on the observation that the quantum
interference of the plane waves $\sim e^{\pm i\omega t}$ contains a
contribution with $k_0=0$, we make an interpretation that
{\em these additional ($k_0=0$)-shell solutions describe the quantum
  coherence between the particles and antiparticles (positive and
  negative energy states) with opposite momenta and spin.}

The explicit matrix structure of these solutions is found easily by
setting $k_0 \neq 0$ and $k_0 = 0$ in the matrix equation
(\ref{matrix_eq}) for the mass-shell and the
coherence shell solutions, respectively \cite{paper1}. The full
chiral matrix corresponding to the mass-shell solution is given by: 
\beq
g^<_{h,{\rm m-s}}(k_0,|\vec{k}|;t)  =
2 \pi f^h_{s_{k_0}}(|\vec{k}|,t) |k_0|
                                  \left(\begin{array}{cc}
                                  1 - h|\vec{k}|/k_0  &  m/k_0 \\
                                  m^*/k_0 &  1 + h|\vec{k}|/k_0
                                  \end{array} \right)
                                 \delta(k^2 - |m|^2)\,,
\label{m-s_matrix_HOM}
\eeq
where $s_{k_0} \equiv {\rm sgn}(k_0)$ and $f^h_{s_{k_0}}(|\vec{k}|,t)$
are real functions parametrizing this solution. The on-shell
distribution functions $f^h_\pm$ are called the phase space densities for
positive and negative energy modes, respectively. Indeed, in the thermal
limit they are related to the number densities of physical
particles, as we will see later. For $k_0=0$-shell solution we
get on the other hand

\begin{eqnarray}
 g^<_{h,{\rm 0-s}}(k_0,|\vec{k}|;t) &=&  \pi \left[
                      f^h_{1}(|\vec{k}|,t) \left( \begin{array}{cc}
                                  h\, m_R/|\vec{k}| &  1 \\
                                  1       & - h\, m_R/|\vec{k}|
                                 \end{array} \right)
                             \right.
  \nonumber \\    && + \left.
                   f^h_{2}(|\vec{k}|,t) \left( \begin{array}{cc}
                                  -h\, m_I/|\vec{k}| &  -i \\
                                  i       &  h\, m_I/|\vec{k}|
                                 \end{array} \right)
                    \right] \,
                    \delta (k_0) \,,
\label{k0zero_matrix}
\end{eqnarray}
where $f^h_{1}(|\vec{k}|,t)$ and $f^h_{2}(|\vec{k}|,t)$ are new
undetermined real functions corresponding to the degrees of freedom of
this coherence solution. The most general solution satisfying the
constraint equation (\ref{constraint_HOM}) (or equivalently the matrix
equation (\ref{matrix_eq})) for a spatially
homogeneous case is the linear combination of Eqs.~(\ref{m-s_matrix_HOM}) and
(\ref{k0zero_matrix}):
\beq
g^<_h = g^<_{h,{\rm m-s}} + g^<_{h,{\rm 0-s}}\,. 
\label{fullchiral_HOM}
\eeq
This general solution contains four independent on-shell distribution
functions $f^h_{\pm,1,2}$ (for both $h = \pm 1$), which is just the
number of independent components in a hermitian $2\times2$ matrix,
such as the chiral matrix $g^<_h$. Indeed, in section
\ref{sec:eq_motion} we find that there is a {\em one-to-one mapping}
between these on-shell functions and the components of the $k_0$-integrated
chiral matrix $\langle g^<_h \rangle$.

\subsubsection{Static planar symmetric case}

\begin{figure}
\centering
\includegraphics[width=0.65\textwidth]{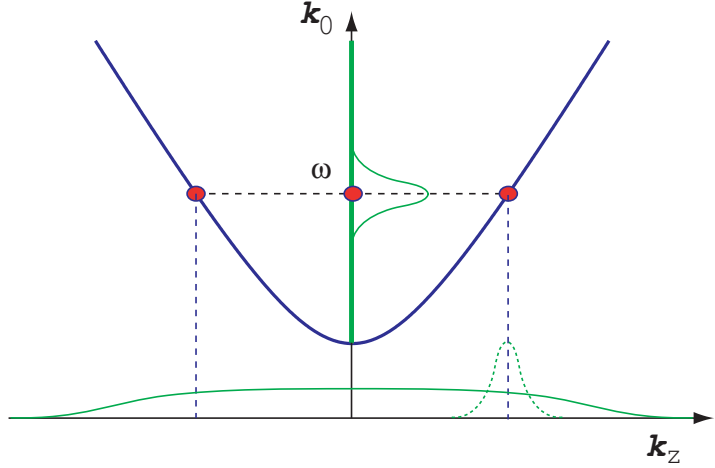}
    \caption{Shell structure of the correlators $G^<$ and $\Delta^<$ in the
      static planar symmetric case. The filled blobs show the
      contributions from the mass-shell and the new $k_z=0$-shell for
      a given $k_0=\omega$. Peaked functions on each axis represent
      the accuracy of possible exterior 
    knowledge or ``measurement'' on $k_0$ and $k_z$ allowing coherent mixing
    (wide solid peak in $k_z$-axis) or reducing the system to
    a noncoherent evolution (dashed peak).}
    \label{fig:DR-planar}
\end{figure}

The analysis of the static planar symmetric case proceeds in
complete analogy. By introducing a Bloch representation for
$g^<_s$:
\beq
g^<_s \equiv \frac12 \left( g^s_0 +  g^s_i \rho^i \right)
\label{bloch_STA}
\eeq
with real $g^h_\alpha$, the constraint equation 
(\ref{constraint_STA}) decomposes again into a homogeneous
matrix equation
\beq
\tilde B_\alpha^{\ \beta} g^s_\beta = 0\,,
\label{matrix_eq_STA}
\eeq
where the $4\times4$ coefficient matrix is now 
(index ordering is again $\alpha = 0,3,1,2$):
\beq
\tilde B = \left( \begin{array}{cccc}
    k_0  &  s k_z  &  -m_R  &  m_I \\
    s k_z &  k_0   &  0     &  0   \\
    0 &  m_R     &  s k_z   &  0   \\
    0  & -m_I     &  0     &  s k_z 
    \end{array} \right)\,.
\eeq
The nonzero solutions of this matrix equation are found at the
zeros of the determinant
\begin{equation}
\det(\tilde B) = k_z^2 (k^2 - |m|^2) \,,
\end{equation}
which are now:
\beq
k^2 - |m|^2 = 0 \qquad {\it or} \qquad k_z = 0\,.
\eeq
The former condition gives again the standard one particle mass-shell
solutions, with $k_z = \pm k_m \equiv \pm(k_0^2 - |m|^2)^{1/2}$, while
the latter condition $k_z=0$ gives now a different
novel class of solutions (see Fig.~\ref{fig:DR-planar}). By the same
argument as in the spatially homogeneous case, we
interpret that {\em these additional
($k_z=0$)-shell solutions describe the quantum coherence between the states of
same spin and energy travelling in opposite z-directions.}

The matrix structure of these solutions is found by setting $k_z \neq
0$ (mass-shell) and $k_z=0$ (coherence shell) in the matrix equation
(\ref{matrix_eq_STA}). For the mass-shell solution we get 
\beq
  g^<_{s,\rm m-s}(k_0,k_z;z) = 2 \pi |k_0| \, f^s_{s_{k_z}}(k_0,z) 
                            \left( \begin{array}{cc}
                              1 - sk_z/k_0  &  m/k_0 \\
                              m^*/k_0       &  1 + sk_z/k_0
                            \end{array} \right)
                             \delta(k^2 - |m|^2)\,,
\label{m-s_matrix_STA}
\eeq
where $s_{k_z} \equiv {\rm sgn}(k_z)$ and $f^s_{s_{k_z}}(k_0,z)$ are
real on-shell distribution functions parametrizing this solution. The
coherence shell solution is given by
\begin{eqnarray}
 g^<_{s,{\rm 0-s}}(k_0,k_z;z) &=&  \pi \left[
                      f^s_{1}(k_0,z) \left( \begin{array}{cc}
                                  m_R/k_0 &  1 \\
                                  1       &  m_R/k_0
                                 \end{array} \right)
                             \right.
  \nonumber \\    && + \left.
                   f^s_{2}(k_0,z) \left( \begin{array}{cc}
                                  -m_I/k_0 &  -i \\
                                  i       &  -m_I/k_0
                                 \end{array} \right)
                    \right] \,
                    \delta (k_z) \,,
\label{kzzero_matrix}
\end{eqnarray}
where $f^s_{1}(k_0,z)$ and $f^s_{2}(k_0,z)$ are new real distribution
functions. The most general solution satisfying the constraint
equation (\ref{constraint_STA}) for a static planar symmetric case is
now the linear combination of Eqs.~(\ref{m-s_matrix_STA}) and
(\ref{kzzero_matrix}):
\beq
  g^<_s = g^<_{s,{\rm m-s}} + g^<_{s,{\rm 0-s}} \,.
\label{fullchiral}
\eeq
In this case also, we later find that the four independent
on-shell functions $f^s_{\pm,1,2}$ will uniquely correspond to the
components of the integrated chiral matrix $\langle g^<_s \rangle$,
where the integration is now over $k_z$ instead of $k_0$.

We later see that even in the simplest possible example of
constant mass and no interactions, the new $k_{0,z}$-shell solutions
for the spatially homogeneous and static planar symmetric cases
are not constant, but oscillate rapidly with the frequencies $2
\omega_{\vec{k}}$ and $2 k_m$, respectively. Thus they break the translational
invariance of the correlator $G^<$ badly even in this trivial limit. This is
the very reason why these coherence solutions have not been found
(used) in the standard treatments of quasiparticle approximation,
where it is assumed that the correlator is close to thermal
equilibrium.

\subsection{Scalar bosons}

Next, we discuss the singular shell structure for scalar
fields. It appears that the method of finding these solutions is not so
straightforward, since there is no purely algebraic equation in this case.
To analyze the phase space properties of the scalar Wightman function
$i\Delta^<$, we consider the KB-equation (\ref{DynEqMix}) in the
noninteracting mean field limit:
\begin{equation}
 \Big(k^2 - \frac14 \partial^2 + ik\cdot\partial 
      - m^2 \Big) i\Delta^< = 0\,.
\label{DynEqMix2}
\end{equation}
Like for fermions, we first consider the spatially homogeneous case
and then the static planar symmetric case.

\subsubsection{Spatially homogeneous case}

In the spatially homogeneous case the spatial gradients of the
correlator $\Delta^<$ and the mass $m$ vanish. Splitting the equation
(\ref{DynEqMix2}) into real and imaginary parts gives then
\begin{eqnarray}
 \Big(k^2 - m^2 - \frac14 \partial_t^2 \Big)i\Delta^<(k,t)
      &=& 0
\label{KG_Eq_HOM2}\\
 k_0\partial_t i\Delta^<(k,t) &=& 0\,.
\label{KG_Eq_HOM3}
\end{eqnarray}
In contrast to the fermionic case we cannot divide these equations 
into ``kinetic'' and ``constraint'' equations, since both of them
contain time derivatives. However, we can now determine the phase
space structure indirectly by using both of the equations (\ref{KG_Eq_HOM2}) and
(\ref{KG_Eq_HOM3}) in an appropriate way.

We begin by setting $k_0 \neq 0$. Then Eq.~(\ref{KG_Eq_HOM3}) requires
that $\partial_t i\Delta^< = 0$ at all times implying that also
$\partial_t^2 i\Delta^< = 0$. Substituting this to
Eq.~(\ref{KG_Eq_HOM2}) now leads to an algebraic equation
\begin{equation}
\left(k^2  -m^2 \right)i\Delta^<_{\rm m-s} = 0\,,
\label{alkepraeq1}
\end{equation}
which has the spectral solution:
\begin{equation}
i\Delta^<_{\rm m-s}(k_0,|\vec{k}|,t) = 
    2\pi\,{\rm sgn}(k_0)f_{s_{k_0}}(|\vec{k}|,t)
    \delta\big(k^2 -m^2\big)\,, 
\label{SpecSol_HOM}
\end{equation}
corresponding to the standard one particle mass-shell solution, with
the dispersion relation $k_0 = \pm \omega_k \equiv \pm (\vec{k}^2 +
m^2)^{1/2}$. Note that this solution satisfies the equation
(\ref{KG_Eq_HOM3}) {\em only} in the mean field limit \ie neglecting all the
time derivatives of the mass $m$. This is expected, since based on the
discussion in section \ref{sec:quasi_approx}, the solutions will
spread in phase space when the gradients are taken in account (see
Eq.~\ref{Airy}). 

On the other hand, by setting $k_0 = 0$ in the first place,
Eq.~(\ref{KG_Eq_HOM3}) is identically satisfied, and no extra
constraint for the time derivatives of $i\Delta^<$ follows. Then
Eq.~(\ref{KG_Eq_HOM2}) simply becomes
\begin{equation}
(\partial_t^2 + 4 \omega_{\vec k}^2)\Delta^<_{\rm 0-s} = 0 \,,
\label{q-cohHOM}
\end{equation}
which has the (mean field) solution,
\beq
i\Delta^<_{\rm 0-s}(k_0,|\vec{k}|,t) = 2\pi
\big[A(|\vec{k}|,t)\cos(2\omega_{\vec k} t) +
B(|\vec{k}|,t)\sin(2\omega_{\vec k} t) \big]\, \delta(k_0)
\label{k0zerospec_OLD}
\eeq
where $A(|\vec{k}|,t)$ and $B(|\vec{k}|,t)$ are real functions
that become constants when the mass $m$ (and $\omega_{\vec k}$) is a constant. 
The $\delta(k_0)$-factor explicitly fixes the restriction to 
the shell $k_0=0$. From now on, we will call (parametrize) the factor in the
square brackets in Eq.~(\ref{k0zerospec_OLD}) as 
$f_c(|\vec{k}|,t)$\footnote{Note that $f_c$ is not dimensionless here,
but has the dimension of $1/M$}, so that the solution
(\ref{k0zerospec_OLD}) is written simply as
\begin{equation}
i\Delta^<_{\rm 0-s}(k_0,|\vec{k}|,t) = 2\pi\,f_c(|\vec{k}|,t)\delta(k_0)\,.
\label{k0zerospec}
\end{equation}
The full spectral solution satisfying the (mean
field) equations (\ref{KG_Eq_HOM2}) and (\ref{KG_Eq_HOM3}) is then the
combination of Eqs.~(\ref{SpecSol_HOM}) and (\ref{k0zerospec}) (see
Fig.~\ref{fig:DR-homog}):
\begin{equation}
i\Delta^< = i\Delta^<_{\rm m-s} + i\Delta^<_{\rm 0-s}\,.
\label{fullspec_HOM}
\end{equation}
This complete solution has three independent on-shell distribution functions
$f_{\pm,c}$, which are now one-to-one related to the three lowest
$k_0$-moments of $i\Delta^<$, as we will see in section \ref{sec:eq_motion}.

\subsubsection{Static planar symmetric case}

For the static planar symmetric case with $\partial_{t,x,y} i\Delta^<
= \partial_{t,x,y} m = 0$ the real and imaginary parts of
Eq.~(\ref{DynEqMix2}) are 
\begin{eqnarray}
 \Big(k^2 - m^2 + \frac14 \partial_z^2 \Big)i\Delta^<(k,z)
      &=& 0
\label{KG_Eq_STA2}\\
 k_z\partial_z i\Delta^<(k,z) &=& 0\,.
\label{KG_Eq_STA3}
\end{eqnarray}
The analysis proceeds now in complete analogy with the spatially
homogeneous case. For $k_z \neq 0$ we have $\partial_z
i\Delta^< = \partial_z^2 i\Delta^< \equiv 0 $, so that we find the same
mass-shell solution as before:
\begin{equation}
i\Delta^<_{\rm m-s}(k_0,|\vec{k}_\pp|,k_z,z) = 2\pi\,{\rm
  sgn}(k_0)f_{s_{k_z}}(k_0,|\vec{k}_\pp|,z)\delta\big(k^2 -
  m^2\big)\,.
\label{SpecSol_STA}
\end{equation}
A new solution is found by setting $k_z=0$ in the first
place, leading to (the other equation is identically satisfied)
\begin{equation}
 \left(\partial_z^2 + 4 k_m^2(z) \right)i\Delta^<_{\rm 0-s} = 0\,,
\label{q-cohSTA}
\end{equation}
where $k_m \equiv (k_0^2 -\vec{k}_\pp^2 - m^2)^{1/2}$. Analogously to
the spatially homogeneous case, this
equation has a (mean field) solution that can be parametrized as
\begin{equation}
i\Delta^<_{\rm 0-s}(k_0,|\vec{k}_\pp|,k_z,z) =
2\pi\,f_c(k_0,|\vec{k}_\pp|,z)\delta(k_z)\,,
\label{kzzerospec}
\end{equation}
where $f_c(k_0,|\vec{k}_\pp|,z)$ is a real on-shell distribution
function corresponding to this coherence solution.
The most general solution satisfying the (mean field) equations
(\ref{KG_Eq_STA2}) and (\ref{KG_Eq_STA3}) is once again the combination of
Eqs.~(\ref{SpecSol_STA}) and (\ref{kzzerospec}) (see Fig.~\ref{fig:DR-planar}):
\begin{equation}
i\Delta^< = i\Delta^<_{\rm m-s} + i\Delta^<_{\rm 0-s}\,. 
\label{fullspec_STA}
\end{equation}
In this case the three independent on-shell distribution functions
$f_{\pm,c}$ are one-to-one related to the three lowest
$k_z$-moments of $i\Delta^<$. 

Like for fermions, we interpret the new
$k_{0,z}=0$-shell solutions for scalar fields as describing the
nonlocal quantum coherence between the states with
opposite momenta and $z$-momenta in the spatially homogeneous and the
static planar symmetric cases, respectively. For a complex scalar field
the $k_0=0$-coherence is between particles and antiparticles, while for
a real scalar field it is between the different modes of the same
particle\footnote{For a real scalar field particles are their own
  antiparticles}. The oscillatory
behaviour of the coherence shell solution with
the frequency $2 \omega_{\vec{k}}$ in the constant mass limit is now 
directly seen from Eq.~(\ref{k0zerospec_OLD}).

%
%

\section{Equations of motion with collisions}
\label{sec:eq_motion}

Having determined the singular phase space shell structure in the
(extended) quasiparticle limit, we now want to use the corresponding
spectral correlators as an ansatz for the dynamical equations
including the gradients and collisions. Our goal is to find a closed set
of equations of motion for the on-shell functions $f$ (or some generic
quantities related to these). We will only consider the spatially homogeneous
case here, but the static planar symmetric case would be highly
analogous to this\footnote{For the free theory the static planar
  symmetric case will be considered
  in connection with the Klein problem in section \ref{sec:klein}}. To
proceed, the singular shell structure of the correlator naturally
suggests to integrate the dynamical equations over the momentum. This
procedure is supported also in the sense that (local) physical
quantities, like currents, energy density and pressure, are
obtained from the correlator $\prop^<$ by a full 4-momentum
integration (see section \ref{sec:physical1}). There is one profound
difference between the procedures for fermions and
scalar bosons, however. For fermions, it is enough to consider the bare
integral (zeroth moment) of the correlator to find out a
closed set of equations for the on-shell functions $f_\alpha$. For scalars, on
the other hand, the different moment integrals couple to each other,
and due to our decomposition with three on-shell functions $f_{\pm,c}$   
we need to consider the three lowest moment integrals to get a closure.

\subsection{Integrated matrix equations for fermions}

To derive the equations of motion for the on-shell functions 
$f_{\pm,1,2}$ for fermions in the spatially homogeneous case, we start with the
kinetic (AH) equation Eq.~(\ref{AntiHermitian22}) for the chiral
$2\times2$ matrix $g^<_h$: 
\begin{equation}
i \partial_t g^<_h = \hat H g_h^< - g_h^< \hat H^\dagger  + {\cal C}_h^-\,,
\label{AntiHermitian22_2}
\end{equation}
where the operator $\hat H$ is given by Eq.~(\ref{GenHamiltonian})
and ${\cal C}_h^-$ is the antihermitian chiral part of the
collision term: $P_h\left(\gamma^0 i{\cal C}_{\rm
    coll}^\psi\gamma^0\right)P_h$, defined in
Eq.~(\ref{coll_chiral_HOM}). Integration over $k_0$ gives now the
following equation:
\beq
i\partial_t \rho_h = [H, \rho_h] + \langle{\cal C}_h^-\rangle\,,
\label{rho_coll_intHOMOG}
\eeq
where we use notation $\langle\ldots\rangle \equiv \int \frac{{\rm
    d}k_0}{2\pi}(\ldots)$ for the $k_0$-integral, $\rho_h \equiv
\langle g^<_h \rangle$ denotes the (zeroth moment) integral of the
correlator $g^<_h$, and the local Hamiltonian operator is given by
\begin{equation}
H \equiv - h|\vec k|\rho^3 + m_R \rho^1 - m_I\rho^2\,.
\label{Hamiltonian}
\end{equation}
The $k_0$-derivatives appearing in Eq.~(\ref{GenHamiltonian}) have
disappeared as total derivatives due to the integration.
When we substitute the spectral correlator Eq.~(\ref{fullchiral_HOM})
as an ansatz for $g^<_{h}$, the (chiral) components of
the integrated correlator $\rho_h$ are simply related to the on-shell
functions $f_\alpha$, $\alpha = \pm,1,2$:
\begin{eqnarray}
\rho_{h,LL} &=&  \frac{1}{2}\Big(1-h\frac{|\vec{k}|}{\omega_{\vec
    k}}\Big)f^h_{+}
             + \frac{1}{2}\Big(1+h\frac{|\vec{k}|}{\omega_{\vec
                 k}}\Big)f^h_{-}
             + h\Big(\frac{m_R}{2|\vec{k}|} f^h_{1} -
               \frac{m_I}{2|\vec{k}|} f^h_{2}\Big)
\nonumber \\
\rho_{h,RR} &=&  \frac{1}{2}\Big(1+h\frac{|\vec{k}|}{\omega_{\vec
    k}}\Big)f^h_{+}
             + \frac{1}{2}\Big(1-h\frac{|\vec{k}|}{\omega_{\vec
                 k}}\Big)f^h_{-}
             - h\Big(\frac{m_R}{2|\vec{k}|} f^h_{1} -
               \frac{m_I}{2|\vec{k}|} f^h_{2}\Big)
\nonumber \\
\rho_{h,LR} &=&  \frac{m}{2\omega_{\vec k}}(f^h_{+} - f^h_{-})
             + \frac{1}{2}(f^h_{1} - if^h_{2})
\nonumber \\
\rho_{h,RL} &=&  \frac{m^*}{2\omega_{\vec k}}(f^h_{+} - f^h_{-})
             + \frac{1}{2}(f^h_{1} + if^h_{2}) \,.
\label{rhocomp_HOM}
\end{eqnarray}
It is easy to show that these linear relations can be inverted whenever
$|\vec k| \neq 0$. Moreover, because of the singular shell structure
of the spectral correlator Eq.~(\ref{fullchiral_HOM}), the collision
integral $\langle{\cal C}_h^-\rangle$ is composed of projections of
the self energies $\Sigma^{<,>}$ on these singular shells, and is thus
depending on the correlator $g^<_h$ only through the on-shell
functions $f_\alpha$. Based on these observations, we conclude that
the matrix equation (\ref{rho_coll_intHOMOG}) can be used to find the
desired closed set of equations of motion for the on-shell functions
$f_\alpha$. We will not give the resulting equations explicitly here,
since they would be more messy than Eq.~(\ref{rho_coll_intHOMOG}), and
in any case the specifications of the self energies in the collision
term are needed to get the equations in the final form. 
Instead, we refer to the matrix
equation (\ref{rho_coll_intHOMOG}) together with the relations
(\ref{rhocomp_HOM}) as our {\em master equations} to study the
dynamics of either the on-shell functions $f_\alpha$ or the
components of the chiral matrix $\rho_h$. When the self energy
functionals $\Sigma^{<,>}$ are specified, these equations can be used
in practical calculations. In chapter \ref{chap:applications} we use
them to study coherent particle production in an
oscillating background field in the presence of decohering collisions.

\subsection{Moment equations for scalar bosons}
\label{sec:moment_eq}

For scalar bosons the relevant dynamical equation is
Eq.~(\ref{DynEqMix}), which in the spatially homogeneous case reduces
to the following coupled equations (real and imaginary parts):
\begin{eqnarray}
 \Big(k^2 - \frac14 \partial_t^2 - m^2 \cos(\sfrac{1}{2}
      \partial_t^m \partial_{k_0}^\Delta)\Big)i\Delta^<
      &=& {\cal C}^+
\label{KG_Eq_Coll1}
\\
 \Big(k_0 \partial_t + m^2 \sin(\sfrac{1}{2}
      \partial_t^m \partial_{k_0}^\Delta)\Big)i\Delta^<
      &=& {\cal C}^-\,,
\label{KG_Eq_Coll2}
\end{eqnarray}
where ${\cal C}^\pm$ are the real and imaginary parts of $i{\cal
  C}_{\rm coll}^\phi$, respectively. We now see that because of the
explicit $k_0$-factors in Eqs.~(\ref{KG_Eq_Coll1})-(\ref{KG_Eq_Coll2})
the $k_0$-integration of these equations will couple different moments
\begin{equation}
\rho_n(|\vec{k}|,t) \equiv \langle k_0^n\, i\Delta^<
\rangle = \int \frac{{\rm d} k_0}{2\pi}\;k_0^n\, i\Delta^<(k_0,|\vec{k}|,t)\,
\label{n-moment}
\end{equation}
in the same equation. That is, by taking the zeroth moment of
Eq.~(\ref{KG_Eq_Coll1}) and the zeroth and first moments of
Eq.~(\ref{KG_Eq_Coll2}) we find the following coupled moment equations:
\begin{eqnarray}
\frac14 \partial_t^2 \rho_0 + \omega_{\vec k}^2 \rho_0 - \rho_2 &=&
\langle{\cal C}^+\rangle
\nonumber\\
\partial_t\rho_1 &=& \langle{\cal C}^-\rangle
\nonumber\\
\partial_t\rho_2 - \frac12 \partial_t(m^2) \rho_0 &=& \langle k_0{\cal
  C}^-\rangle \,.
\label{rho_Eq_Coll1}
\end{eqnarray}
Moreover, when the spectral correlator
Eq.~(\ref{fullspec_STA}) is used as an ansatz for $\Delta^<$,
the moments $\rho_{0,1,2}$ are related to the on-shell functions
$f_{\pm,c}$ in a simple way:
\begin{eqnarray}
\rho_0 &=& \frac{1}{2\omega_{\vec k}}(f_+ - f_-) + f_c 
\nonumber\\
\rho_1 &=& \frac{1}{2}(f_+ + f_-)
\nonumber\\[1mm]
\rho_2 &=& \frac{\omega_{\vec k}}{2}(f_+ - f_-)\,.
\label{rho-f_HOM}
\end{eqnarray}
We see that this linear set of equations is invertible whenever $|\vec k|
\neq 0$, and also the collision integrals $\langle{\cal
  C}^\pm\rangle$ and $\langle k_0{\cal C}^-\rangle$ are depending on
the correlator $\Delta^<$ only through the on-shell functions
$f_{\pm,c}$, in complete analogy with fermions. So, we conclude now
that the moment equations (\ref{rho_Eq_Coll1}) can be used to obtain the desired
closed set of equations of motion for the on-shell functions
$f_{\pm,c}$. Again, we will not give these equations here explicitly,
but refer to Eqs.~(\ref{rho_Eq_Coll1})-(\ref{rho-f_HOM}) as our {\em master
  equations} for the study of the dynamics of either the on-shell
functions $f_{\pm,c}$ or the three lowest moments $\rho_{0,1,2}$. In
chapter \ref{chap:applications} we use these equations to study the
coherent production of unstable particles in an oscillating background.

\section{Spectral function and thermal limit}

Let us next consider the fermionic and scalar spectral
functions ${\cal A}$ in terms of our eQPA scheme. In our approach
these are needed in the evaluation of the collision terms in
Eqs.~(\ref{rho_coll_intHOMOG}) and (\ref{rho_Eq_Coll1}). We consider
here only the spatially homogeneous case, since all the problems
we are going to study with collisions in chapter \ref{chap:applications}
have this particular symmetry. 
The relevant equations for the study of the phase space properties of
spectral functions (in the noninteracting mean field limit)
are the same as the corresponding ones for the Wightman
functions $G^<$ and $\Delta^<$,
Eqs.~(\ref{DynEqMix_QPA})-(\ref{DynEqMix_QPA_sca}). Consequently, the spectral
functions have spectral solutions given by
Eqs.~(\ref{fullchiral_HOM}) and (\ref{fullspec_HOM}), with yet
undefined on-shell functions $f^{h{\cal A}}_{\pm,1,2}$ for fermions
and $f^{\cal A}_{\pm,c}$ for scalars. {\em In addition},
however, the spectral functions must obey the sum rules that
can be derived from the pole equations
(\ref{SpecEqMix1})-(\ref{SpecEqMix2}) using the spectral relation
(\ref{spec_rel}) (or alternatively from the canonical equal time
(anti)commutation relations of the fields):
\beq
  \int \frac{{\rm d}k_0}{\pi}
                  {\cal A}(k,x) \gamma^0 = 1
\label{sumrule_ferm}
\eeq
for fermions, and 
\beq
\int \frac{{\rm d}k_0}{\pi}\big(k_0 + \frac{i}{2}\partial_t\big){\cal
  A}(k,x) = 1
\label{sumrule_sca}
\eeq
for scalars. It is easy to show that for the fermionic spectral function
the conditions from the sum rule (\ref{sumrule_ferm}) completely fix
the values of all the on-shell functions 
\beq 
f^{h\,\cal A}_\pm = \frac12 \quad\quad {\textrm{and}} \quad\quad f^{h\,\cal
  A}_{1,2} = 0
\label{specAconstrHOMOG}
\eeq
for both helicities, so that the spectral function reduces to the familiar local thermal
equilibrium form:
\beq
 {\cal A} = \pi {\rm sgn}(k_0) (\kdag +  m_R - i\gamma^5 m_I)
            \delta(k^2-|m|^2) \,.
\label{specA3}
\eeq
For the scalar fields, on the other hand, the sum rule
(\ref{sumrule_ferm}) does not immediately fix the values of the
on-shell functions $f^{\cal A}_{\pm,c}$. However, by use of the
dynamical equations for the moment functions $\rho^{\cal A}_n\equiv
\int \frac{{\rm d}
  k_0}{2\pi}\;k_0^n\, {\cal A}$ that are identical to
(\ref{rho_Eq_Coll1}) with vanishing collision terms, the sum rule gives
that $\rho^{\cal A}_0=\rho^{\cal A}_2=0$ and $\rho^{\cal A}_1 = 1/2$
in order to get a continuous constant mass limit. The connection
relations identical to Eq.~(\ref{rho-f_HOM}) between $\rho^{\cal A}_n$
and $f^{\cal A}_{\pm,c}$ then give:
\begin{equation}
f^{\cal A}_\pm = \frac12\,, \qquad f^{\cal A}_c = 0\,,
\end{equation}
so that the spectral function ${\cal A}$ again reduces to its standard
thermal form
\begin{equation}
{\cal A} = \pi {\rm sgn}(k_0)\, \delta \left(k^2 - m^2\right)\,.
\label{spectral_function_sca}
\end{equation}
Note that the contributions from the coherence shell $k_0=0$ are
completely absent in the (eQPA) spectral functions (\ref{specA3}) and
(\ref{spectral_function_sca}), because of the strong constraints
imposed by the spectral sum rule.

Let us conclude this section by giving the Wightman functions
$G^<$ and $\Delta^<$ in the thermal limit. It is easy to
show that the full translation invariance in thermal
equilibrium kills the coherence shell contributions
completely, and furthermore implies the Kubo-Martin-Schwinger (KMS) conditions
for the Wightman functions \cite{Kubo57,KubYokNak57,MarSch59}:
\beq
\prop^>_{\rm eq}(k_0) = e^{\beta k_0}\prop^<_{\rm eq}(k_0)\,.
\eeq
The KMS-conditions and the relation $\prop^> =
\pm\prop^< - 2i {\cal A}$ as well as the thermal spectral 
functions Eqs.~(\ref{specA3}) and (\ref{spectral_function_sca}) will then
fix the mass-shell distribution functions to Fermi-Dirac and
Bose-Einstein distributions for fermions and bosons, respectively:
\beqa
&&f^h_{s_{k_0}}(|\vec{k}|) \rightarrow n^F_{\rm eq}(k_0) \equiv \frac{1}{e^{\beta k_0}
  + 1} 
\nonumber\\
&&f_{s_{k_0}}(|\vec{k}|) \rightarrow n^B_{\rm eq}(k_0) \equiv \frac{1}{e^{\beta k_0}
  - 1}\,,
\eeqa
leading to the following thermal equilibrium correlators \cite{LeBellac00}:  
\beqa
iG^<_{\rm eq} &=& 
       2\pi\,{\rm sgn}(k_0) (\kdag +  m_R - i\gamma^5 m_I) \,n^F_{\rm eq}(k_0) \,
            \delta(k^2-|m|^2) 
\nonumber\\
i\Delta^<_{\rm eq} &=& 2\pi\,{\rm sgn}(k_0) n^B_{\rm eq}(k_0) \delta\left(k^2
  - m^2\right)\,.
\eeqa
These expressions will be used in chapter \ref{chap:applications} when
we compute the collision terms in the case of interaction with a thermal
background.

%
%

\section{Physical quantities in terms of the on-shell functions}
\label{sec:physical2}

Having found the spectral phase space structure of the correlators
$G^<$ and $\Delta^<$, we want to write some
of the physical observables considered in section \ref{sec:physical1}
in terms of the on-shell functions $f$. In addition, we introduce
the concept of phase space particle number.

\subsection{Particle number and fluxes}

In contrast to standard vacuum QFT, the concept of particles is not
very well defined in nonequilibrium quantum field theory
\cite{BirDav82,CalHu08}. The problem is that different observers will
see different vacuum states, and on the other hand vacuum states are
not remaining ``empty'' even in the free field evolution, because of
the nontrivial background. That said, in the case of clear
asymptotic regions in the background field, the problem of particle
production from the classical background is well understood.
Most definitions for particle number are relying on some sort of
diagonalization of the Hamiltonian
\cite{BirDav82,CalHu08,GarProSch04,AaBer01}, the idea being that the
energy of the $\vec{k}$-mode would entirely consists of a vacuum part
plus (anti)particle excitations: ${\cal E}_{\vec{k}} \propto
\omega_{\vec{k}}(n_{\vec{k}} + \bar{n}_{\vec{k}} \mp 1)$ for fermions and
bosons, respectively. This idea is of course supported by thermal
equilibrium, where this relation is satisfied. 

Here we adopt a somewhat different concept for the particle number. At
the end of the previous section we saw
that in thermal equilibrium the mass-shell distribution functions of
$G^<$ and $\Delta^<$ are just the Fermi-Dirac and Bose-Einstein
distributions, which correspond to particle
number densities for positive energies: $n^{F,B}_{\vec
  k} = n^{F,B}_{\rm eq}(k_0=\omega_{\vec k})$, and to antiparticle
number densities for negative energies according to the
Feynman-Stuckelberg interpretation: $\bar n^{F,B}_{\vec k} = \pm 1 -
n^{F,B}_{\rm eq}(k_0=-\omega_{\vec k})$. Within our eQPA scheme the
Wightman functions $G^<$ and $\Delta^<$ always have a spectral phase
space structure including the singular mass-shell, even in
out-of-equilibrium conditions. This suggests us to generalize these
relations and use them as definitions for the {\em phase space
  particle and antiparticle numbers}, \ie for a spatially homogeneous
system we define:
\beqa
n_{\vec{k}h}(t) &\equiv& f^h_+(|\vec k|,t) \qquad {\rm and} \qquad \bar
n_{\vec{k}h}(t) \equiv 1 - f^h_-(|\vec k|,t)
\nonumber\\
n_{\vec{k}}(t) &\equiv& f_+(|\vec k|,t) \qquad {\rm and} \qquad\;\, \bar
n_{\vec{k}}(t) \equiv - 1 - f_-(|\vec k|,t)
\eeqa
for fermions and scalars, respectively. The soundness of these
definitions is consolidated by expressing the number current densities
in Eqs.~(\ref{ferm_current1})-(\ref{bos_current1}) in terms of the
on-shell functions. We get
\beqa
\langle j_F^0(t)\rangle &=& \sum_h \int \frac{{\rm d}^3k}{(2\pi )^3}
(f^h_{+} + f^h_{-}) = \sum_h \int \frac{{\rm d}^3k}{(2\pi )^3}
(n_{\vec{k}h} - \bar n_{\vec{k}h} +1)
\nonumber\\
\langle j_B^0(t)\rangle &=& \int\frac{{\rm d}^3k}{(2\pi )^3}
(f_{+} + f_{-}) =  \int\frac{{\rm d}^3k}{(2\pi )^3}
(n_{\vec{k}} - \bar n_{\vec{k}} - 1)\,,
\label{currentdensity}
\eeqa
which are just the expected results with the correct vacuum energy
contributions. Now, using the inverse relations
of Eq.~(\ref{rhocomp_HOM}) we can express the fermionic (anti)particle
numbers in
terms of the Bloch components of the integrated chiral matrix $\rho_h
= \frac{1}{2}(\langle g^h_0 \rangle + \langle \vec{g}^h \rangle \cdot
\vec{\sigma})$:
\beqa
n_{\vec{k}h} &=& \frac{1}{2 \omega}\left(-h|\vec{k}| \langle
  g^h_3 \rangle + m_R \langle g^h_1 \rangle - m_I \langle g^h_2
  \rangle \right) + \frac{1}{2} \langle g^h_0 \rangle 
\nonumber \\ 
{\bar n}_{\vec{k}h} &=& \frac{1}{2 \omega}\left(-h|\vec{k}| \langle
  g^h_3 \rangle + m_R \langle g^h_1 \rangle - m_I \langle g^h_2
  \rangle \right) - \frac{1}{2} \langle g^h_0 \rangle + 1 \,.
\label{partnumber_ferm}
\eeqa
For scalar fields, on the other hand, the (anti)particle numbers can
be related to the three lowest moments $\rho_{0,1,2}$ by the inverse
relations of Eq.~(\ref{rho-f_HOM}):
\beqa
n_{\vec{k}} &=& \frac{1}{\omega_{\vec k}}\rho_2 + \rho_1
\nonumber\\
\bar{n}_{\vec{k}} &=& \frac{1}{\omega_{\vec k}}\rho_2 - \rho_1 - 1\,. 
\label{partnumber_sca}
\eeqa
We see that the particle and antiparticle numbers coincide if
$\langle g^h_0 \rangle = 1$ for fermions and $\rho_1 = -1/2$ for
scalars\footnote{For a real scalar field particle is
  its own antiparticle, thus $\rho_1 = -1/2$ is the only sensible
  value}. These values thus correspond to zero ``chemical
potential'', for which the number current densities (\ref{currentdensity})
have only vacuum contributions.    

Let us now compare our phase space particle number to other
definitions in literature. Apart from some sign conventions our
fermionic particle number density in Eq.~(\ref{partnumber_ferm}) with zero
chemical potential reduces to the one obtained in
ref.~\cite{GarProSch04} by using the Bogoliubov transformation to
diagonalize the Hamiltonian. Moreover, using the free field equations
of motion (\ref{rho_Eq_Coll1}), we see that our scalar particle number in
Eq.~(\ref{partnumber_sca}) agrees with the one in ref.~\cite{AaBer01}, namely
\begin{equation}
\Big(n_{\vec{k}} + \frac12\Big)^2 = \rho_0\Big(\omega_{\vec k}^2\rho_0 +
\frac12 \partial_t^2 \rho_0 \Big)\,,
\label{Berges}
\end{equation}
if we take the adiabatic limit: $\partial_t^2 \rho_0 \ll \omega_{\vec
  k}^2\rho_0$. These comparisons further strengthen the soundness of
our definitions.

For spatially dependent problems one is often more interested in
fluxes, \ie the spatial components of the number currents, than
densities. Using Eqs.~(\ref{ferm_current1})-(\ref{bos_current1}) we
find that the $z$-fluxes in a {\em static planar symmetric case} are given by:
\beqa
\langle j_F^3(z)\rangle &=& \sum_s \int \frac{{\rm d}k_0\,{\rm
    d}^2k_\pp}{(2\pi )^3}\,{\rm sgn}(k_0)\,s\,(f^s_{+} - f^s_{-})
\nonumber\\
\langle j_B^3(z)\rangle &=& \int \frac{{\rm d}k_0\,{\rm
    d}^2k_\pp}{(2\pi )^3}\,{\rm sgn}(k_0)(f_{+} - f_{-})\,.
\label{flux}
\eeqa
In this case it is natural to interpret or define $f_\pm$ as
{\em right/left moving particle fluxes} per unit volume in
$(k_0,\vec{k}_\pp)$-phase space. We use these definitions in
sections \ref{sec:klein} and \ref{sec:mass_wall} for solving quantum
reflection problems with our methods.

\subsection{Energy density and pressure for spatially\\ homogeneous
  systems}
 
Let us proceed by considering the energy density and pressure in a
spatially homogeneous case. These are by definition the $00$- and
$ii$-components of the energy momentum tensor, respectively. Using
Eqs.~(\ref{ferm_energy_momentum})-(\ref{bos_energy_momentum}) we get
for the energy density (in terms of $n$ and $\bar n$ instead of
$f_\pm$):
\beqa
\langle{\cal E}_F(t)\rangle = \langle\theta^{00}(t)\rangle &=& \sum_h
\int \frac{{\rm d}^3
  k}{(2\pi)^3}\,\omega_{\vec k}\big(n_{\vec{k}h}+{\bar n}_{\vec{k}h}
  -1\big)
\nonumber\\
\langle{\cal E}_B(t)\rangle = \langle T^{00}(t)\rangle &=& \int \frac{{\rm d}^3
  k}{(2\pi)^3}\,\frac12\omega_{\vec k}\big(n_{\vec{k}} +
\bar{n}_{\vec{k}}+1\big)\,,
\label{energy_dens}
\eeqa
for fermions and scalars, respectively. For the pressure we get instead:
\beqa
\langle P_F(t)\rangle = \langle\theta^{ii}(t)\rangle 
&=& \sum_h \int \frac{{\rm d}^3 k}{(2\pi)^3}
\,\frac{1}{3}\bigg[\frac{{\vec{k}}^2}{\omega_{\vec{k}}}\big(n_{\vec{k}h}+{\bar
      n}_{\vec{k}h}  - 1\big) - m_R f^h_1 + m_I f^h_2 \bigg]
\nonumber\\
\langle P_B(t)\rangle = \langle T^{ii}(t)\rangle 
&=& \int \frac{{\rm d}^3 k}{(2\pi)^3}\,\frac13\bigg[\frac{\vec{k}^2}{\omega_{\vec k}}
\frac12\big(n_{\vec{k}} + \bar{n}_{\vec{k}}+1\big) - \big(3\omega_{\vec k}^2 -
\vec{k}^2\big) f_c \bigg]\,.
\label{pressure}
\eeqa
The expressions for energy densities in Eq.~(\ref{energy_dens}) are, as
expected, consisting of particle and antiparticle contributions and
the vacuum energies, which are opposite for fermions and scalars. The
expressions for pressure in Eq.~(\ref{pressure}), however, involve
explicit contributions from the coherence shell as well. Thus, in the
presence of coherence the ``quantum pressure'' may be drastically
different from the standard classical one. However, because of
the typical oscillatory behavior of the coherence solutions at
microscopic time-scales $\Delta t_{\rm osc} \sim 1/\omega$, this
deviation would average out in most cases of interest at time-scales much
larger than $\Delta t_{\rm osc}$.

%
%

\section{On the validity of eQPA approximation}
\label{sec:validity}

We conclude this chapter by discussing the main concerns regarding the
validity of our eQPA approximation scheme, the breakdown
(spreading) of the singular shell picture and the handling of the
gradient expansion in the collision term. To begin with, we recall that
the basic requirements of the quasiparticle approximation, weak
(quantum) interactions and a slowly varying background field,
should be fulfilled for the singular shell picture to be a good
approximation. However, in our eQPA scheme we relax the assumption
that the system is close to a local thermal equilibrium, and hence
nearly translation invariant. This allows us to include the solutions
describing nonlocal quantum coherence, which are oscillatory (and thus
translationally non-invariant) on quantum scales with frequencies of
order $2 \omega_{\vec{k}}$ (or $2 k_m$ in a static planar symmetric
case). To that end, it is worth noticing that in this work we have
considered the coherence only in the cases of two particular spacetime
symmetries: ($k_0=0$-shell) in a spatially homogeneous case and
($k_z=0$-shell) in a static (or stationary) planar symmetric case. For
other spacetime symmetries, it is expected that similar coherence
solutions would exist, living in different singular shells in the
phase space. One interesting symmetry, worth studying in the future,
is the rotationally symmetric geometry.

\begin{figure}
\centering
\includegraphics[width=1.00\textwidth]{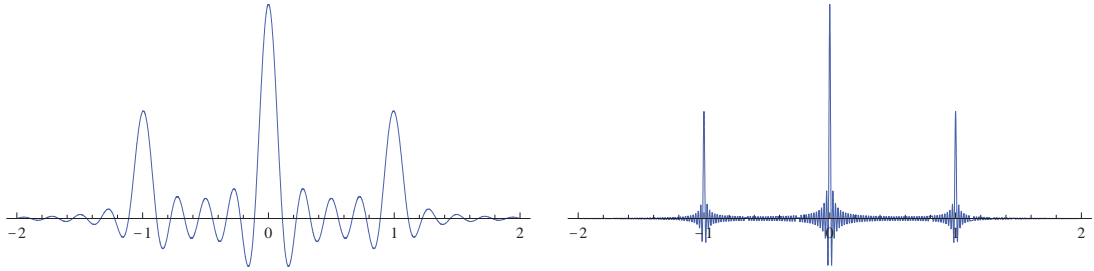}
     \vskip-0.3truecm
     \caption{The phase space structure of the 2-point Wightman
       $i\Delta^<$ in the case of a 1-dimensional infinite square well
       potential. Shown is the phase space density Eq.~(\ref{spread_shells})
       as a function of $k_z/k_n$ at the middle of the square well at
       $z=0$, for odd bound states with $n=9$ (left) and $n=99$
       (right). For more details, see ref.~\cite{paper3}.}
     \label{fig:shells}
\end{figure}

Of course, the singular shell picture is not exact for any (nontrivial)
practical application, as the shells will spread out in the phase
space when the interaction width $\Gamma$ and the gradients of the
background field are nonvanishing. Thus, an important question
concerning the eQPA scheme is whether the new $k_{0,z}=0$-shells are
spreading similarly (in magnitude) to the well-known mass-shell
solutions, when the strict quasiparticle limit is lifted.
To elaborate on this question, we review a simple
nonrelativistic example from ref.~\cite{paper3}, where we computed the exact
free-field Wightman function from the one-particle wave functions in
the case of an infinite square well potential. In this case, the
standing waves of bound states are highly coherent superpositions of the
opposite travelling plane waves. The essential phase space
structure is plotted in Fig.~\ref{fig:shells}. We clearly see that
the spreads of the mass-shell contributions (at $k_z = \pm k_n$) and the
$k_z=0$-shell are of the same order. This observation is confirmed by
an analytical calculation, which shows that the functional forms of
individual peaks are actually the same in this specific case. That is, the
plotted phase space density in the middle of the square well
is given by 
\beq
\sum_\pm f(k_z \mp k_n) + 2 f(k_z)\,,
\label{spread_shells}
\eeq
where the density corresponding to each ``shell'' is $f(q) \equiv
\sin(qL)/q$. Although this is just one simple example, it
supports the expectation that the different phase space shells are
 spreading in an equal manner also more generally. Indeed,
 the spreads for all shells should be controlled by the same parameters, the
 interaction width $\Gamma$ and the gradients of the background.  

Another important issue in our eQPA scheme is the treatment of the gradient
expansion in the collision term. For the flow term this expansion is
trivial as the $k$-derivatives can be handled easily in the
$k$-integration (partial integrations are straightforward). For the
collision term, on the other hand, nontrivial
contributions arise that are proportional to spacetime derivatives of
the correlator $\partial_x^n \prop^<$. In the standard kinetic
approach these terms would necessarily be proportional to $\Gamma$ or
the gradients of the background field $\partial^n m$\footnote{This
  notation represents the $n$-th gradient order, where it is assumed that
  the $n$-th derivative and the first derivative to the power $n$ are
  of the same gradient order}, as the correlator is adiabatic. In our eQPA
approach, in contrast, this is not the case: because of the
oscillatory coherence solutions, one typically has (in a spatially
homogeneous case): $\partial_t^2 \prop^< \sim - 4 \omega_{\vec{k}}^2
\prop^<$, such that all orders of the expansion involve terms that are
not controlled by any small parameter. This severe-looking problem can
be cured, however, by a recursive use of the equations of motion and a
resummation of these coherence contributions, to get a controlled
expansion in powers of $\Gamma$ and $\partial^n m$. In an example of
coherent production of decaying scalar particles in section
\ref{part_production_sca}, this expansion is worked out for a thermal
interaction to the leading order, neglecting terms of order ${\cal
  O}(\Gamma^2, \Gamma\,\partial m)$ and higher.

%% file: Chapter4.tex
%
%

\chapter{Applications}
\label{chap:applications}

%
%

\section{Klein problem}
\label{sec:klein}

As our first application for the eQPA scheme introduced in chapter
\ref{chap:scheme} we consider the famous Klein problem \ie quantum reflection
from a step potential. The problem with noninteracting quantum fields
can be solved conveniently using the Dirac and
Klein-Gordon equations for fermions and scalars,
respectively. However, here we want to
illustrate the necessity to include the coherence shell solutions when the
problem is solved with the methods of quantum transport
theory. Moreover, when the interacting fields are considered, the wave
equation approach is not possible anymore, and one needs to use some
more advanced methods, like our approximation scheme, to solve the problem.
 
\begin{figure}
\centering
\includegraphics[width=0.65\textwidth]{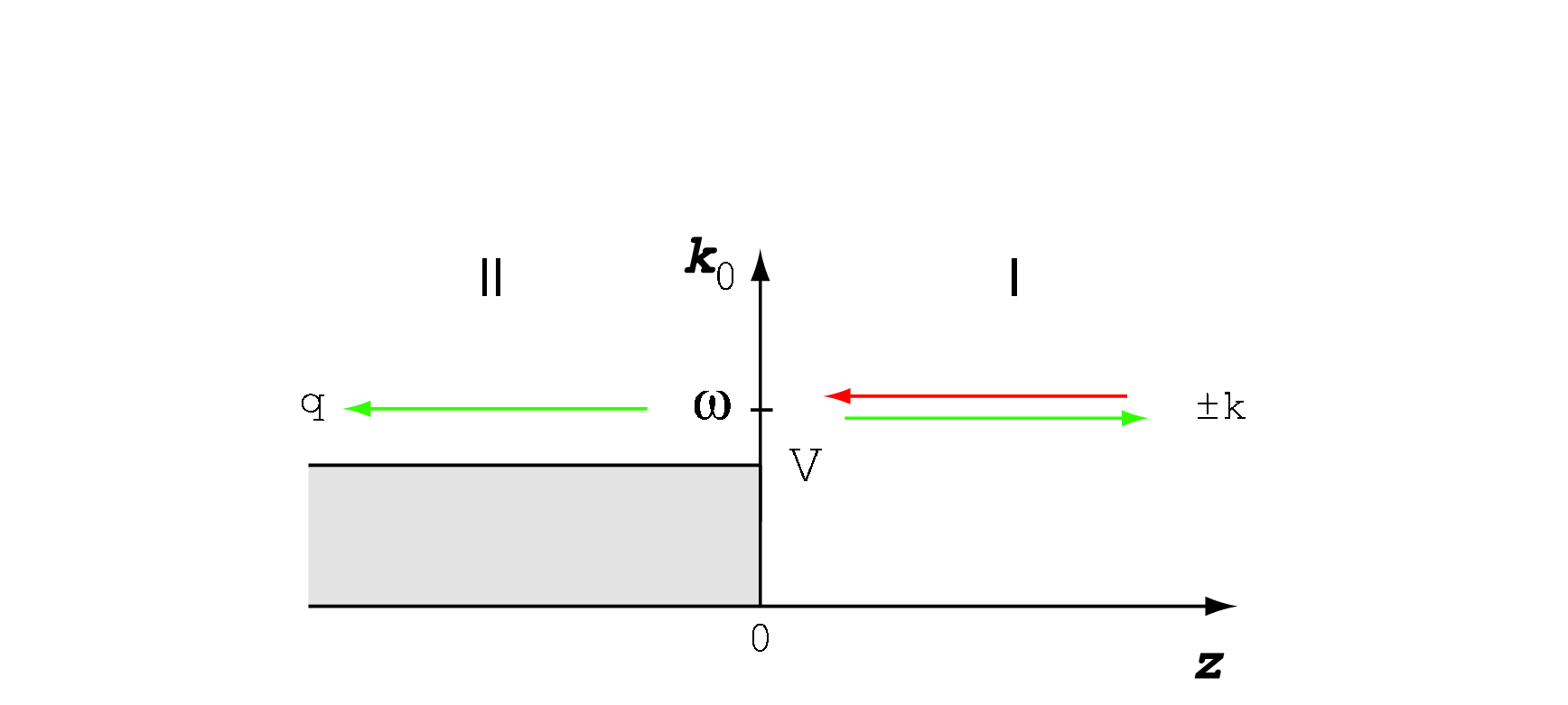}
    \caption{Reflection from a step potential $V(z) =
      V\theta(-z)$ with the momenta of incoming (red) and
      outgoing (green) particles described by arrows.}
    \label{fig:KleinWall}
\end{figure}

The setup for the problem is illustrated in figure
\ref{fig:KleinWall}. We consider a step potential
\beq
V(z) = V\theta(-z)\,,
\eeq
such that the mass-shell momenta are 
$k \equiv k_m^{\rm I} = (k_0^2-|m|^2)^{1/2}$ and $q \equiv k_m^{\rm
  II} = ((k_0-V)^2-|m|^2)^{1/2}$ in regions I and II,
respectively. The momentum $k$ is always real, while $q$ can be
either real or imaginary, depending on the values of energy $k_0$ and potential
$V$. The latter case describes the quantum tunneling inside the potential
barrier. Without collisions energy is a conserved quantity, and we
fix now: $k_0 \equiv \omega > 0$.  In our approach the particle fluxes
are just the mass-shell distribution functions $f_\pm$. The {\em
  asymptotic} boundary conditions for this setup are the following:
We normalize the incoming flux from the right to unity $f^{\rm
  I}_-=1$, and we have no incoming flux form the left $f^{\rm
  II}_+=0$. Also, since there is no incoming flux from the left, we
set the coherence asymptotically to zero in region II. In addition, in
the case of imaginary mass-shell momentum $q$ in region II, we cannot
have any asymptotic mass-shell solution there, since no
propagating wave would penetrate a potential barrier to infinite
distance. The reflected and transmitted fluxes we want to compute are
$f^{\rm I}_+$ and $f^{\rm II}_-$, respectively. We will first
consider fermions and then scalar bosons. For both, we analyze
the cases with real and imaginary $q$ separately.

\subsection{Fermions}

For a varying potential instead of the mass, the fermionic free field
KB-equation becomes in the mixed representation
\beq
   \big(\kdag + \frac{i}{2} \deldag_x  - m_R - i\gamma^5 m_I
    - \gamma_0 V(x) e^{-\frac{i}{2}{\partial}^V_x \cdot \partial_k}
   \big) G^<(k,x) = 0\,.
\label{MspaceKleineq}
\eeq
In this case the potential is planar symmetric, and the reduction of the spin
structure is completely analogous to section \ref{sec:spinz}
leading to the collisionless equations
\begin{eqnarray}
{\rm (H)}:\quad -2 s k_z g^<_s &=& \hat P_V g_s^< + g_s^< \hat P_V^\dagger 
\label{Hermitian22z_V}
\\ 
{\rm (AH)}:\quad\ \  is \partial_z g^<_s &=& \hat P_V g_s^< - g_s^< \hat
P_V^\dagger
\label{AntiHermitian22z_V}
\end{eqnarray}
with
\begin{equation}
\hat P_V \equiv (k_0 - \hat V) \rho^3  + i ( m_R \rho^2 + m_I \rho^1)\,,
\label{PamiltonV}
\end{equation}
where we denote $\hat V(z) \equiv V(z) 
e^{\,\frac{i}{2}\partial^V_z \partial_{k_z}}$. The spectral phase
space structure of the correlator $g^<_s$ that is obtained by
analyzing the H-equation (\ref{Hermitian22z_V}) in the mean field
limit, is now identical to the case of a varying mass given
by Eq.~(\ref{fullchiral}), except for the replacement $k_0 \rightarrow k_0
- V$. By integrating the kinetic (AH) equation
(\ref{AntiHermitian22z_V}) over $k_z$ we get
\beq
is \partial_z \rho_s = P_V \rho_s - \rho_s 
P_V^\dagger\,,
\label{rho_eq_z}
\eeq
where we denote $\rho_s \equiv \int \frac{{\rm
    d}k_z}{2\pi}\,g^<_s$. The $k_z$-derivatives in $\hat V$ have disappeared
due to the integration reducing $\hat V$ and $\hat P_V$ to their
corresponding mean field forms $V$ and $P_V$. When the spectral
correlator is used as an ansatz for $g^<_s$, we get the following
linear relations between the components of the chiral matrix $\rho_s$
and the on-shell functions $f_\alpha$, $\alpha = \pm,1,2$:
\begin{eqnarray}
\rho^s_{LL} &=&  \frac{1}{2}(\frac{\tilde k_0}{k_m}+s)f^s_{-}
             + \frac{1}{2}(\frac{\tilde k_0}{k_m}-s)f^s_{+}
             + \frac{m_R}{2\tilde k_0} f^s_{1} - \frac{m_I}{2\tilde k_0} f^s_{2}
\nonumber \\
\rho^s_{RR} &=&  \frac{1}{2}(\frac{\tilde k_0}{k_m}-s)f^s_{-}
             + \frac{1}{2}(\frac{\tilde k_0}{k_m}+s)f^s_{+}
             + \frac{m_R}{2\tilde k_0} f^s_{1} - \frac{m_I}{2\tilde k_0} f^s_{2}
\nonumber \\
\rho^s_{LR} &=&  \frac{m}{2k_m }(f^s_{-} + f^s_{+})
             + \frac{1}{2}(f^s_{1} - if^s_{2})
\nonumber \\
\rho^s_{RL} &=&  \frac{m^*}{2k_m }(f^s_{-} + f^s_{+})
             + \frac{1}{2}(f^s_{1} + if^s_{2}) \,,
\label{rhocomp_STA}
\end{eqnarray}
where $\tilde k_0 \equiv k_0 - V(z)$ and $k_m \equiv ({\tilde
  k_0}^2-m^2)^{1/2}$. Formally these linear relations are invertible whenever
$k_m \neq 0$. However, the on-shell momentum $k_m$ may become
imaginary, depending on the value of $k_0$ with respect to $V$. In
this case, one needs to set $f_\pm = 0$, since it actually corresponds
to a situation where the mass-shell functions do not contribute
in $k_z$-integration at all due to imaginary roots in the
$\delta(k^2-m^2)$-factor. In general, the spectral approximation for
the correlator $g^<_s$, based on the mean field limit with vanishing
gradients, is expected to become better when the distance from the
potential wall increases. {\em Asymptotically} far away from the wall
the spectral decomposition and thus the relations (\ref{rhocomp_STA})
should be {\em exact}. Now, if we formally replace $k_m
\rightarrow |k_m|$ in Eq.~(\ref{rhocomp_STA}), we see that the
resulting relations are invertible for all $k_m \neq 0$ and there are
no problems with imaginary $k_m$. Thus, we
can always parametrize $\rho_s$ in terms of effective
$f$-functions via these connection relations, which will reduce to
the actual on-shell functions only asymptotically when $k_m$ is
real (or $f_\pm=0$). The use of such a parametrization is that the
equations of motion for the effective $f$-functions, resulting from the
matrix equation (\ref{rho_eq_z}), become very simple inside the regions
I and II (we set here $m_I=0$ for simplicity): 
\begin{eqnarray}
s\partial_z f^s_\pm &=& 0
\nonumber \\[1mm]
s\partial_z f^s_1 &=& -2\tilde k_0 f^s_2 
\nonumber \\
s\partial_z f^s_2 &=& \frac{2k_m^2}{\tilde k_0}f^s_1 \,.
\label{kivaeqn}
\end{eqnarray}
Moreover, the structure of the dynamical matrix equation
(\ref{rho_eq_z}) readily implies that the only consistent condition
for matching $\rho_s$ at $z=0$ is that all of its components are
continuous, while the off-diagonals have kinks \ie their
derivatives have finite discontinuities.

\subsubsection{Real $q$, partial reflection}

Let us first consider the case with real $q$. The equations
(\ref{kivaeqn}) now imply that inside the regions I and II the functions
$f^s_{\pm}$ are constants, so the asymptotic boundary conditions for
the incoming fluxes fix $f^{s({\rm I})}_- = 1$ and $f^{s({\rm II})}_+ =
0$ throughout the regions. Moreover, the functions $f^{s({\rm
    I,II})}_{1,2}$ are oscillatory in both regions. Hence, the
asymptotic boundary condition for the coherence to vanish as $z
\rightarrow -\infty$, kills it completely in region II: $f^{s({\rm
    II})}_{1,2} = 0$. Using these conditions and the four matching
conditions for the components of $\rho_s^{\rm I}$ and $\rho_s^{\rm
  II}$ at $z = 0$, we can fix all the remaining constants: $f^{s({\rm
    I})}_+$, $f^{s({\rm II})}_-$, as well as two constants related to
the oscillatory coherence solutions in region I. We find that the
reflected and transmitted fluxes are related by flux conservation:
\beq
f^{s({\rm I})}_+ = \frac{1-x}{1+x} \qquad\; {\rm and} \qquad
f^{s({\rm II})}_- = 1 - f^{s({\rm I})}_+ = \frac{2x}{1+x}\,,
\label{KleinSolution1}
\eeq
while the oscillatory coherence solution in region I is given by
\begin{equation}
f^{s({\rm I})}_1(z) =  \frac{m\omega V}{k^2q}\frac{2x}{1+x} \cos(2kz) 
\qquad {\rm and} \qquad 
f^{s({\rm I})}_2(z) = -\frac{1}{2\omega}s\partial_z f^{s({\rm I})}_1(z) \,, 
\label{KleinSolution2}
\end{equation}
where we denote
\beq
x \equiv \frac{qk}{\omega (\omega - V) - m^2}\,.
\eeq

\subsubsection{Imaginary $q$, total reflection}

In the case of imaginary $q$ we have an additional boundary
condition. That is, we cannot have any asymptotic mass-shell
solution in region II: $f^{s({\rm II})}_\pm = 0$ throughout (since
they are constants). However, we see from
Eq.~(\ref{kivaeqn}) that the coherence solutions
are exponentials in region II for imaginary $q$, so that the asymptotic
boundary condition at $z\rightarrow-\infty$ does not kill the exponentially
decaying mode. An otherwise similar analysis gives now the result with a
complete reflection:
\beq
f^{s({\rm I})}_+ = 1 \qquad {\rm and} \qquad
f^{s({\rm II})}_- = 0\,,  
\eeq
while the coherence solutions are
\beq
f^{s({\rm I})}_1(z) = \left(\frac{k(\omega-V)}{mV} -
  \frac{2m}{k}\right)\cos(2kz) + \frac{|q|\omega}{mV}\sin(2kz)
\eeq
with $f^{s({\rm I})}_2(z) = -s\partial_z f^{s({\rm I})}_1(z)/{2\omega}$ in region
I (there was a misprint in Eq.~(118) in ref.~\cite{paper1}), and
\begin{equation}
f^{s({\rm II})}_1(z) = \frac{k(\omega-V)}{mV} e^{2|q|z}
\label{KleinSolution4}
\end{equation}
with $f^{s({\rm II})}_2(z) = -s\partial_z f^{s({\rm
    II})}_1(z)/{2(\omega-V)}$ in region II. Note that the results in
Eqs.~(\ref{KleinSolution1}-\ref{KleinSolution4}), which agree with the
standard Dirac equation approach \cite{ItZu}, would not have been
obtained, should we have dropped the coherence shell solutions
$f_{1,2}^s$ from our analysis.

\subsection{Scalar bosons}

For scalar fields the free field KB-equation with a $z$-dependent
potential becomes
\begin{equation}
 \Big([k_0 - V(z)]^2
 e^{-\frac{i}{2}{\partial}^V_z \partial_{k_z}} - k_z^2 -
   \frac14 \partial_z^2 + ik_z\partial_z - m^2 \Big)i\Delta^< = 0\,,
\label{Klein1}
\end{equation}
where we have taken ${\vec k}_\pp=0$. Taking the real and imaginary
parts of Eq.~(\ref{Klein1}) we find: 
\begin{eqnarray}
 \Big(k_z^2 - \frac14 \partial_z^2 - k_m^2 \cos(\sfrac12
      \partial^V_z \partial_{k_z})\Big)i\Delta^<
      &=& 0
\label{Klein2} \\
 \Big(k_z \partial_z + k_m^2 \sin(\sfrac12
      \partial^V_z \partial_{k_z})\Big)i\Delta^< &=& 0\,,
\label{Klein3}
\end{eqnarray}
where $k_m(z) \equiv ((k_0-V(z))^2 - m^2)^{1/2}$. We
find again that the spectral phase space structure of the correlator
$\Delta^<$, obtained by analyzing these equations in the mean field
limit, is identical to the case of a spatially varying mass
given by Eq.~(\ref{fullspec_STA}), except for the replacement $k_0
\rightarrow k_0 - V$. We proceed now analogously to the spatially
homogeneous case in section \ref{sec:moment_eq}. The $n$-th moments of
the correlator $\Delta^<$ are defined as integrals over $k_z$:
\begin{equation}
\rho_n(k_0,z)\equiv \int \frac{{\rm d} k_z}{2\pi}\;k_z^n\,
i\Delta^<(k_0,\vec{k}_\pp=0,k_z,z)\,.
\label{n-moment_Klein}
\end{equation}
By taking the 0th moment of Eq.~(\ref{Klein2}) and
the 0th and 1st moments of Eq.~(\ref{Klein3}) we get the following
closed set of equations for the three lowest moments $\rho_{0,1,2}$:
\begin{eqnarray}
\frac14 \partial^2_z \rho_0 + k_m^2 \rho_0 - \rho_2 &=& 0
\nonumber\\
\partial_z\rho_1 &=& 0
\nonumber\\
\partial_z\rho_2 - \frac12 (\partial_z k_m^2) \rho_0 &=& 0\,.
\label{rho_Eq1_STA}
\end{eqnarray}
When the spectral correlator Eq.~(\ref{fullspec_STA}) is used as an
ansatz for $\Delta^<$, we get the following relations between the lowest
moments $\rho_{0,1,2}$ and the on-shell functions $f_{\pm,c}$:  
\begin{eqnarray}
\rho_0 &=& {\rm sgn}(k_0-V(z))\frac{1}{2k_m}(f_+ + f_-) + f_c 
\nonumber\\
\rho_1 &=& {\rm sgn}(k_0-V(z))\frac{1}{2}(f_+ - f_-)
\nonumber\\[1mm]
\rho_2 &=& {\rm sgn}(k_0-V(z))\frac{k_m}{2}(f_+ + f_-)\,.
\label{rho-f_STA}
\end{eqnarray}
These linear relations are invertible whenever $k_m \neq 0$, but
for the case of imaginary $k_m$ we have to set $f_\pm = 0$, as for
fermions. Again, the spectral approximation for the correlator, and
consequently the relations (\ref{rho-f_STA}), become exact
asymptotically far away from the wall. The matching
conditions at $z=0$ induced by the moment equations 
(\ref{rho_Eq1_STA}) are now more complicated than with fermions. First,
we note that $\partial_z\rho_1$ vanishes everywhere, so $\rho_1$ is a
constant throughout. The last Eq.~(\ref{rho_Eq1_STA}) implies that $\rho_2$ must
have a finite discontinuity over the barrier, which can be
computed by integrating it over a step from $z=-\epsilon$ to
$z=\epsilon$ to give $\rho^{\rm I}_2-\rho^{\rm II}_2 =
\shalf[k^2-q^2]\rho_0(z=0)$. Using this in the first
Eq.~(\ref{rho_Eq1_STA}), we see that also $\partial^2_z \rho_0$ has at
most a finite discontinuity over the step, implying finally that $\rho_0$ and
its derivative $\partial_z\rho_0$ are continuous at $z=0$.

\subsubsection{Real $q$, partial reflection}

We consider first the case with real $q$.
The two latter Eqs.~(\ref{rho_Eq1_STA}) imply that $\partial_z\rho_{1,2} = 0$,
\ie $\rho_{1,2}$ are constants inside the regions I and II. These
constants are partially fixed by the asymptotic boundary conditions
$f^{\rm I}_-=1$ and $f^{\rm II}_+=0$ through the relations (\ref{rho-f_STA}).
Moreover, from the first Eq.~(\ref{rho_Eq1_STA}) we now find that
$\rho_0$ is oscillatory with a constant shift of amount
$\rho_2/k_m^2$ in both regions I and II, respectively. Combining
Eqs.~(\ref{rho-f_STA}) and (\ref{rho_Eq1_STA}) we then find
that asymptotically the coherence solutions
are also oscillatory: $f_c = -\partial_z^2\rho_0/(4 k_m^2)$, so
the asymptotic vanishing of the coherence as $z \rightarrow -\infty$
kills the oscillatory part of $\rho_0$ completely in region II. Using
these conditions and the four matching conditions for $\rho_{0,1,2}$
and $\partial_z\rho_0$ at $z=0$ described above, we can fix all the
remaining constants to obtain the solution
\begin{equation}
f^{\rm I}_+ =\frac{(k-q)^2}{(k+q)^2} \;\qquad {\rm and} \qquad
f^{\rm II}_- = 1 - f^{\rm I}_+ = \frac{4 k q}{(k+q)^2} \,,
\label{Klein_sca_real_mass-shell}
\end{equation}
while the coherence solution in the region I is given by
\begin{equation}
f^{\rm I}_c = \frac{1}{k}\sqrt{f^{\rm I}_+} \cos(2 k z)\,.
\label{Klein_sca_real_coh}
\end{equation}
Note that these solutions describe the shell structure of the
correlator $\Delta^<$ exactly only asymptotically as
$z\rightarrow\pm\infty$. However, as in the case of fermions, replacing
$k_m \rightarrow |k_m|$ the relations are invertible for all $k_m \neq
0$, so that the moments $\rho_{0,1,2}$ can be completely parametrized with
these $f$-functions. In this (effective) sense the solutions
(\ref{Klein_sca_real_mass-shell}) and (\ref{Klein_sca_real_coh}) are
exact throughout the regions I and II.

\subsubsection{Imaginary $q$, total reflection}

As for fermions, in the case of imaginary $q$ we cannot have any
mass-shell solutions in region II, so that asymptotically both $f^{\rm
  II}_\pm=0$. Now the coherence solution $f_c=-\partial_z^2\rho_0/(4
k_m^2)$ is a superposition of decaying and increasing
exponentials in region II, so that the asymptotic boundary condition at
$z\rightarrow-\infty$ does not kill the decaying mode. Because of
these differences we get the solution with total reflection:
\beq
f^{\rm I}_+=1 \qquad {\rm and} \qquad f_\pm^{\rm II} = 0\,,
\eeq
while the coherence solution is given by
\begin{equation}
f^{\rm I}_c(z) = \frac{1}{k}\frac{k^2-|q|^2}{k^2+|q|^2}\cos(2 k z) -
\frac{2|q|}{k^2+|q|^2}\sin(2 k z)
\end{equation}
in region I, and
\begin{equation}
f^{\rm II}_c(z) = \frac{2 k}{k^2 + |q|^2} e^{2|q|z}
\label{KleinSolution4_sca}
\end{equation}
in region II. Again, we would not have obtained the
results in Eqs.~(\ref{Klein_sca_real_mass-shell}-\ref{KleinSolution4_sca}) that
agree with the standard Klein-Gordon approach, should we have dropped
the coherence shell solutions $f_c$ from our analysis.

%
%

\section{Quantum reflection from a CP-violating\\ mass wall}
\label{sec:mass_wall}

\begin{figure}
\centering
\includegraphics[width=0.68\textwidth]{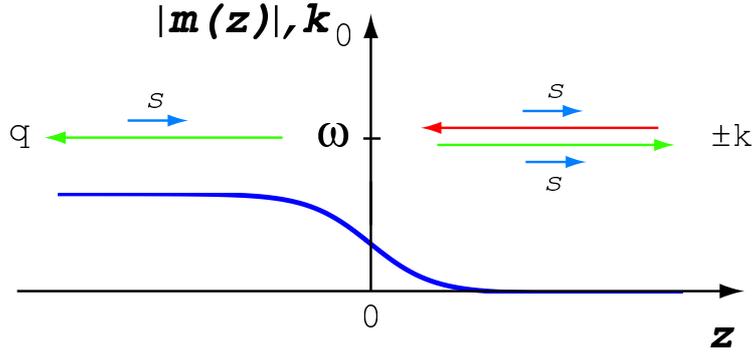}
    \caption{Reflection from a spatially varying mass wall, with the
      momenta of incoming (red) and outgoing (green) particles
      described by arrows. Spin $s$ is conserved in the collisionless case.}
    \label{fig:CPwall}
\end{figure}

\begin{figure}
\centering
\includegraphics[width=1.0\textwidth]{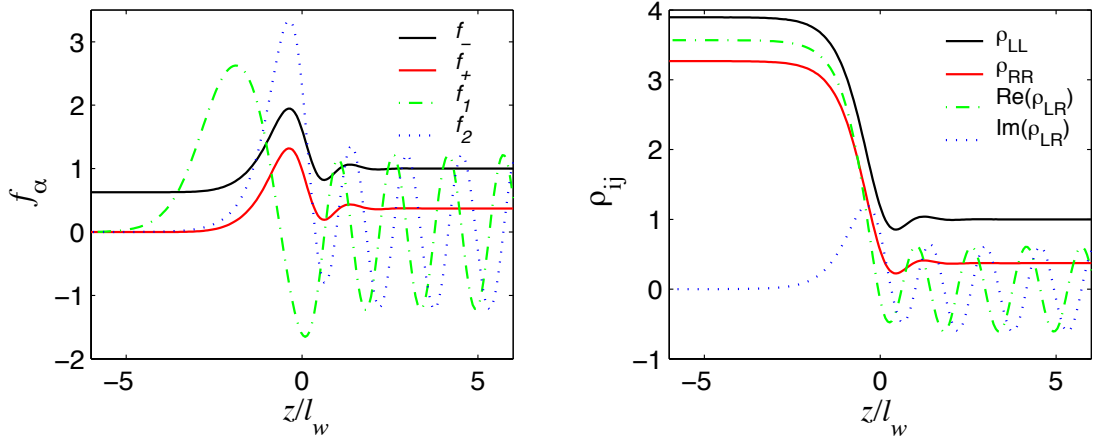}
    \caption{a) Shown are the mass-shell functions $f_\pm$ corresponding to the
             left and right moving fluxes and functions $f_{1,2}$, which 
             encode the quantum coherence. We have taken $s=1$ and 	 	
             $q/|m_{-\infty}|=0.088$.
             b) The chiral density matrix components for
             the same solution. For the wall width and the total
             change of the phase we used $\ell_w = 2$ and $\Delta\theta=-1$.}
    \label{fig:fprofs}
\end{figure}

Let us next consider the quantum reflection of free fermionic fields
from a smooth $CP$-varying mass wall. This simple example is of
relevance for electroweak baryogenesis (see
refs.~\cite{FarSha93,FarSha94,CliKaiVis96} and section
\ref{sec:EWBG}), where the complex spatially varying mass function
arises from Yukawa couplings to the vacuum expectation value of the
Higgs field(s):
\beq
m(z) = y \phi(z)\,,
\label{massaprof}
\eeq
where $y$ is a Yukawa coupling and $\phi(z)$ is a complex scalar
field corresponding to the total effect of the VEVs of (possibly)
multiple Higgs fields~\cite{CliKaiVis96}. For the complex mass wall we have
used the following parametrization (see Fig.~\ref{fig:CPwall}):
\beq
|\phi(z)| = \frac{1}{2}\big(1-{\rm tanh}(z/\ell_w)\big)\,,
\qquad
\arg[\phi(z)] = \frac{1}{2}\Delta\theta \big(1+{\rm tanh}(z/\ell_w)\big) \,,
\label{phiprof}
\eeq
where $\ell_w$ is the width of the wall, and $\Delta \theta$ is the
total magnitude of the change of the phase of $\phi$. The problem is
described by the same integrated evolution equation as for the
fermionic Klein problem, Eq.~(\ref{rho_eq_z}), and the connection relations
Eq.~(\ref{rhocomp_STA}) between the components of the chiral matrix
$\rho_s$ and the on-shell functions $f_\alpha$ with the replacements
$V(z) \rightarrow 0$ and $m_{R,I} \rightarrow m_{R,I}(z)$.  
For the asymptotic boundary conditions we set the incoming
flux from deep in the symmetric phase at $z\rightarrow\infty$ to unity
and take no incoming flux from deep in the broken phase
at $z\rightarrow -\infty$, and correspondingly set the asymptotic
coherence to zero as well:
\begin{eqnarray}
f^s_- =  1, &\quad & z \rightarrow \infty
\nonumber \\
f^s_+ = f^s_1 = f^s_2 = 0, &\quad & z \rightarrow -\infty \,.
\label{bc}
\end{eqnarray}
Let us summarize the results of the numerical calculations presented
in ref.~\cite{paper1}. The calculations were performed using
Eq.~(\ref{rho_eq_z}) in Bloch representation: $\rho_s \equiv
\frac12(\langle g^s_0 \rangle + \langle \vec{g}^s \rangle \cdot \vec\sigma)$,
 together with the relations Eq.~(\ref{rhocomp_STA}) and the boundary
 conditions Eq.~(\ref{bc}). In figure \ref{fig:fprofs} we plot the results 
 for the case of $s=1$ and $q/|m_{-\infty}| = 0.088$, where $q \equiv (\omega^2 -
 |m_{-\infty}|^2)^{1/2}$ is the asymptotic momentum in the broken
 phase. In the left panel we show the values of the
 on-shell functions $f_\alpha$ while in the right panel we plot
 the components of the chiral matrix $\rho$. We can see that in the
 symmetric phase (to the right from the wall) the
system is a coherent superposition of left and right moving
 states with opposite $k_z$-momenta, and the $k_z$=0-shell functions are
 oscillating coherently. In the broken phase however, all but the
 $f_-$-function die off and the state soon becomes a pure transmitted
 left moving state. Note that the physical flux is conserved:
 $f_+(\infty) + f_-(-\infty) = f_-(\infty) \equiv 1$. For the chiral 
 components we see that the imaginary part of $\rho_{\rm LR}$ goes
 to zero when $z \rightarrow -\infty$, as a result of our choice that
 $m$ becomes asymptotically real in the broken phase. The diagonal
 components of $\rho$ become large in the broken phase, because they
 represent chiral densities that are enhanced in the
 regime of small local velocity $v_z = k_z/\omega$ due to flux conservation.

\begin{figure}
\centering
\includegraphics[width=0.65\textwidth]{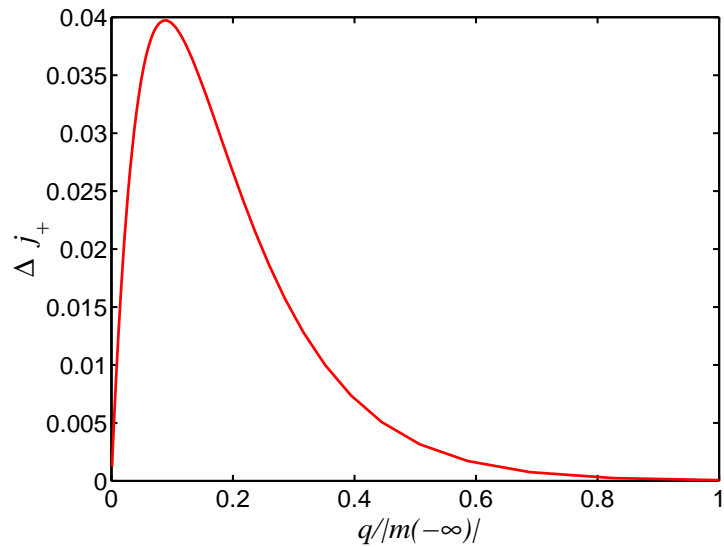}
    \caption{Shown is the current asymmetry of reflected
             states as a function of the asymptotic momentum to mass
             ratio in the broken phase. The wall parametrization is
             the same as in figure \ref{fig:fprofs}.}
    \label{fig:deltaj}
\end{figure}

In figure \ref{fig:deltaj} we show the particle-antiparticle
flux-asymmetry $\Delta j_+ \equiv (f_+ - \bar f_+)_{z=\infty}$ as a
function of $q/|m_{-\infty}|$. For antiparticles we simply need to
make a replacement $m\rightarrow m^*$ in all equations. The
characteristic peaked shape of the flux-asymmetry can be understood as follows: 
The reflection amplitudes for both particles and antiparticles tend to
unity in the limit of total reflection $q \rightarrow 0$, so the
asymmetry must tend to zero. For large $q$ on the other hand, the
(anti)particles start to behave classically compared to the width (and
height) of the wall, and hence the reflection amplitudes and the
asymmetry start to decrease exponentially.   

These results have been derived earlier using the Dirac equation approach (see
\eg ref.~\cite{CliKaiVis96}). Here we just wanted to demonstrate how the
results are obtained using our formalism. We emphasize that for noninteracting
fields our method is {\em
  exact}, as in the case of Klein problem. That is, all gradients of the
mass are vanishing asymptotically
at $z \rightarrow \pm \infty$, and consequently the mean field
approximation becomes exact in that limit. So the $f_\pm$-functions
indeed correspond to the physical fluxes asymptotically, and this is
the only place where this identification is needed in our
analysis. Between the boundary regions the $f$-functions are just a
parametrization of the chiral matrix $\rho$. However, the whole point of
our approach is the ability to include the effects of collisions
together with nonlocal coherence. With collisions the picture changes
as we need the on-shell identifications of the $f$-functions for the
reliable computation of the collision term\footnote{Also the
  corrections from mass gradients that arise from the gradient
  expansion of the collision term become important}. For this reason
our method becomes an approximation in the case with collisions, with
the better {\em quantitative} results expected the thicker the wall
is. But even in the thin wall limit we expect to get at least the correct {\em
  qualitative} behaviour, yet it would not be surprising if the
results would prove to be quite accurate also quantitatively, since
the region of (rapidly) varying mass is very
narrow in this case, and outside the wall region the mean field
approximation should work well. Moreover, the $f$-functions are
presumably smooth also in the wall region, as suggested by the plots
in Fig.~\ref{fig:fprofs}, thus the overall error in the dynamics
arising from the (approximate) computation of the collision term in
that region should be rather small.

%
%

\section{Preheating of decaying fermions}
\label{sec:ferm_preheating}

As a first example of our formalism including collisions, we consider
the production of fermionic particles at preheating, during which the fermion is
subjected to decays. 
A similar fermion production scenario in out-of-equilibrium conditions
has been recently studied using the 2PI approach in ref. \cite{BerPruRot09},
where a complete numerical next-to-leading order calculation in a
$1/N$-expansion of the 2PI effective action was performed for a
$SU(2)_L \times SU(2)_R \sim O(4)$ symmetric theory with Yukawa
interactions between fermions and bosons.
As discussed in section \ref{sec:preheating},
in preheating the (coupled) fermions obtain a rapidly oscillating
effective mass due to the couplings to the inflaton condensate, for
which we use here a simple cosine function (here $t$ corresponds to a
time coordinate in the conformal frame, and the expansion of the
universe is neglected during the preheating) \cite{PelSor00,GarProSch04}:
\beq
m(t) = m_0 + A \cos(2\omega_\varphi t) + i B\sin(2 \omega_\varphi t),
\label{osc_mass_ferm}
\eeq
where $m_0$, $A$, $B$ and $\omega_\varphi$ (the inflaton oscillation
frequency) are real constants. The decay of the fermion into daughter
particles is modelled here by a left-chiral non-diagonal Yukawa
interaction defined by the Lagrangian:
\beq 
{\cal L}_{\rm int} = - y\; \bar \psi_L \phi \, q_R + h.c. 
\label{interaction}
\eeq
where $\psi$ is the fermion field considered, $q$ is some
other fermion (quark) field and $\phi$ is a complex scalar field. For
simplicity, we assume that the fields $q$ and $\phi$ form a thermal
background \ie they are in thermal equilibrium throughout the
preheating. A more realistic but complex scenario could be modelled by
taking into account the full dynamics of the daughter particles with
appropriate equations of motion. 
By the assumption of thermal equilibrium for the daughter fields, the self
energies $\Sigma^{<,>}$ for the fermion $\psi$ are related by the
KMS relation $\Sigma^>(k) = e^{\beta k_0}\Sigma^<(k)$, and the
interaction width $\Gamma$ can be written in the form \cite{paper2}
\beq 
  \Gamma(k) = \frac{1}{2}(1+e^{\beta k_0} )i\Sigma^<(k) = 
         \left[\Gamma_0 \,\gamma^0 - \Gamma_3 \,
                      \big(\hat{k}\cdot \vec{\gamma}\big) \right] P_L
                    \,,
\label{self3}
\eeq
where $P_L=\frac12(1-\gamma^5)$ is the left chiral projector and
$\Gamma_{0,3}(k)$ are real functions. Using these expressions
and the mean field spectral function Eq.~(\ref{specA3})
we find that the zeroth order gradient contribution to the collision
term ${\cal C}_h^-$ in the dynamical equation
(\ref{AntiHermitian22_2}) is given by
\beq
{\cal C}_h^- = - i\left\{D , \, g^<_h - (g^<_h)_{\rm
    eq} \right\}\,,
\eeq
where
\beq
 D \equiv \frac{1}{2}(1+\rho_3) \,\Gamma_h 
 \qquad {\rm with} \qquad \Gamma_h \equiv \Gamma_0 - h\Gamma_3  
\label{D_def_eqn}
\eeq
and $(g^<_h)_{\rm eq}$ is the thermal equilibrium limit of $g^<_h$
with $f^h_\pm=f_\pm^{\rm eq} \equiv 1/(e^{\pm\beta \omega_{\vec
    k}}+1)$. Upon integration over $k_0$ we get then the matrix equation
(\ref{rho_coll_intHOMOG}):
\beq
\partial_t \rho_h = -i[H, \rho_h] - i\langle{\cal C}_h^-\rangle\,,
\label{rho_coll_intHOMOG2}
\eeq
where the collision integral is now given by
\beqa
i\langle{\cal C}_h^-\rangle &\equiv& \int \frac{{\rm d}k_0}{2\pi}
      \left\{D \,, \, g^<_h - (g^<_h)^{\rm eq} \right\}
\nonumber \\[3mm]
&\; = \;&
       \Gamma_{m0} \,
              \left( \begin{array}{cc}
                   (f^h_0 - f_0^{\rm eq}) - h\frac{k}{\omega} (f^h_3 -
                   f_3^{\rm eq}) &
                   \quad \frac{m}{2\omega}(f^h_3 -  f_3^{\rm eq}) \\[2mm] 
                   \frac{m^*}{2\omega}(f^h_3 -  f_3^{\rm eq}) & \quad 0
                      \end{array} \right)
     \nonumber \\[3mm]
          &&\hskip -0.6truecm -\,\,h\Gamma_{m3}
                  \left( \begin{array}{cc}
                   (f^h_3 - f_3^{\rm eq}) - h\frac{k}{\omega}(f^h_0 -
                   f_0^{\rm eq}) & \quad \frac{m}{2\omega}(f^h_0 -
                   f_0^{\rm eq}) \\[2mm] \frac{m^*}{2\omega}(f^h_0 -
                   f_0^{\rm eq}) & \quad 0
                         \end{array} \right) 
     \nonumber \\[3mm]
          &&\hskip -0.4truecm \, + \,\,\,\Gamma_{00}\, 
                  \left( \begin{array}{cc}
                   h\frac{m_R}{k}f^h_1 -
                   h\frac{m_I}{k}f^h_2 & \quad
                   \frac{1}{2}(f^h_1 - if^h_2) \\[2mm]
                   \frac{1}{2}(f^h_1 + if^h_2) & \quad 0
                         \end{array} \right) 
             \,,
\label{coll_int}
\eeqa

\hskip 0.5truecm

\noindent
where we use the shorthand notations $f^h_{0,3} \equiv f^h_+ \pm f^h_-$
and $k \equiv |\vec{k}|$. The $\Gamma_i$-functions appearing as
coefficients are now projections on the different singular shells:
\beqa
\Gamma_{m(0,3)}(|\vec{k}|,t) &\equiv& \Gamma_{0,3}(k_0 = \omega(t),|\vec{k}|) \qquad\quad
(\textrm{positive mass-shell}) \nonumber \\
\Gamma_{00}(|\vec{k}|) &\equiv& \Gamma_0(k_0 = 0, |\vec{k}|) \qquad\qquad \;\;\
(\textrm{$k_0 = 0$ -shell})\,.
\label{gamma_shell}
\eeqa
The explicit expressions for these $\Gamma$-functions are not
presented here, but they can be read from the relations
$\Gamma_{0,3}(k) = \frac{1}{2}(1+e^{\beta k_0})i\Sigma^<_{0,3}(k)$ and
the equations (6.23-6.28) in ref. \cite{paper2}.
Note that $\Gamma$ on the negative mass-shell is related to that on
the positive with $\Gamma_{0,3}(-k_0,|\vec{k}|) = \pm
\Gamma_{0,3}(k_0,|\vec{k}|)$, and furthermore that $\Gamma_3$ vanishes
on the $k_0 = 0$ -shell.

\begin{figure}
\centering
\includegraphics[angle=270,width=0.9\textwidth]{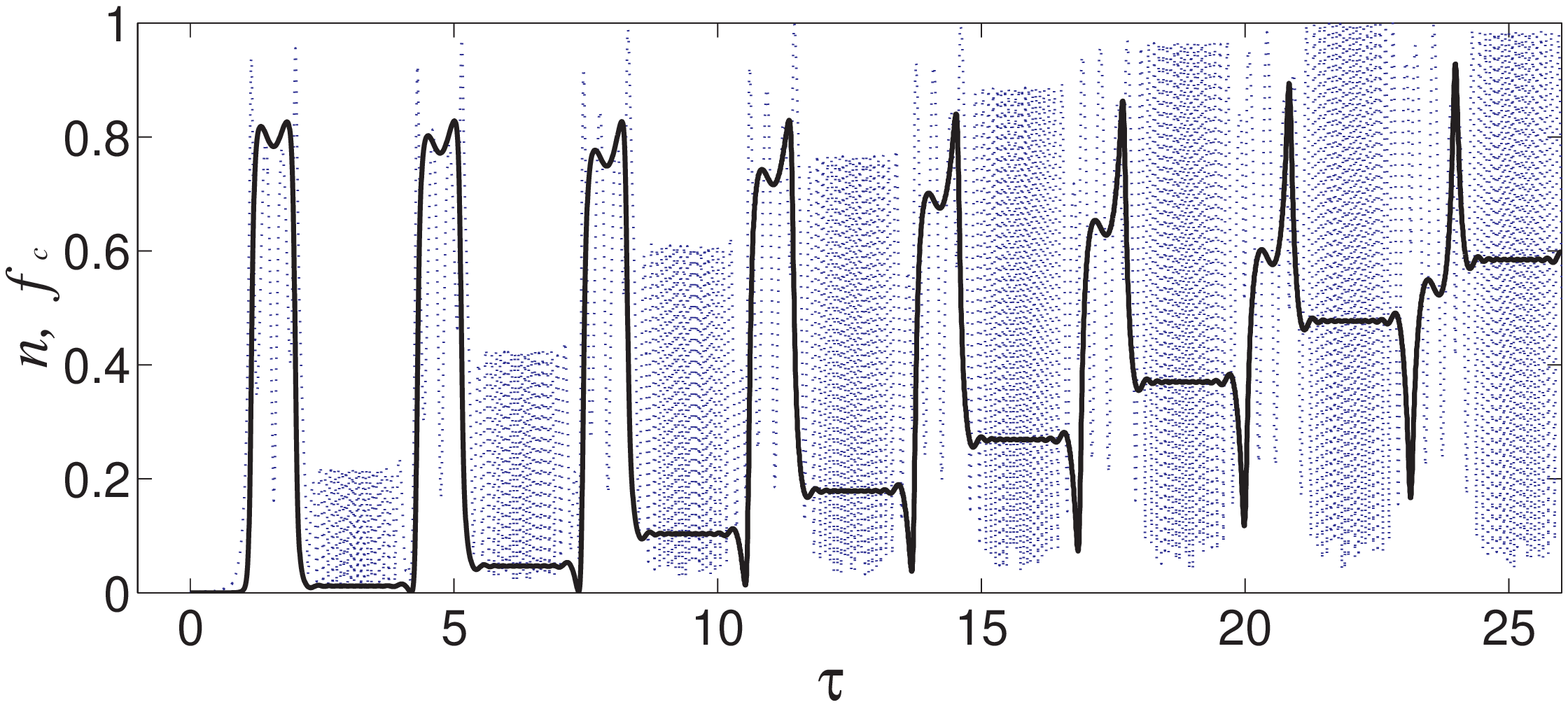}
\vskip-0.62truecm
\includegraphics[angle=270,width=0.9\textwidth]{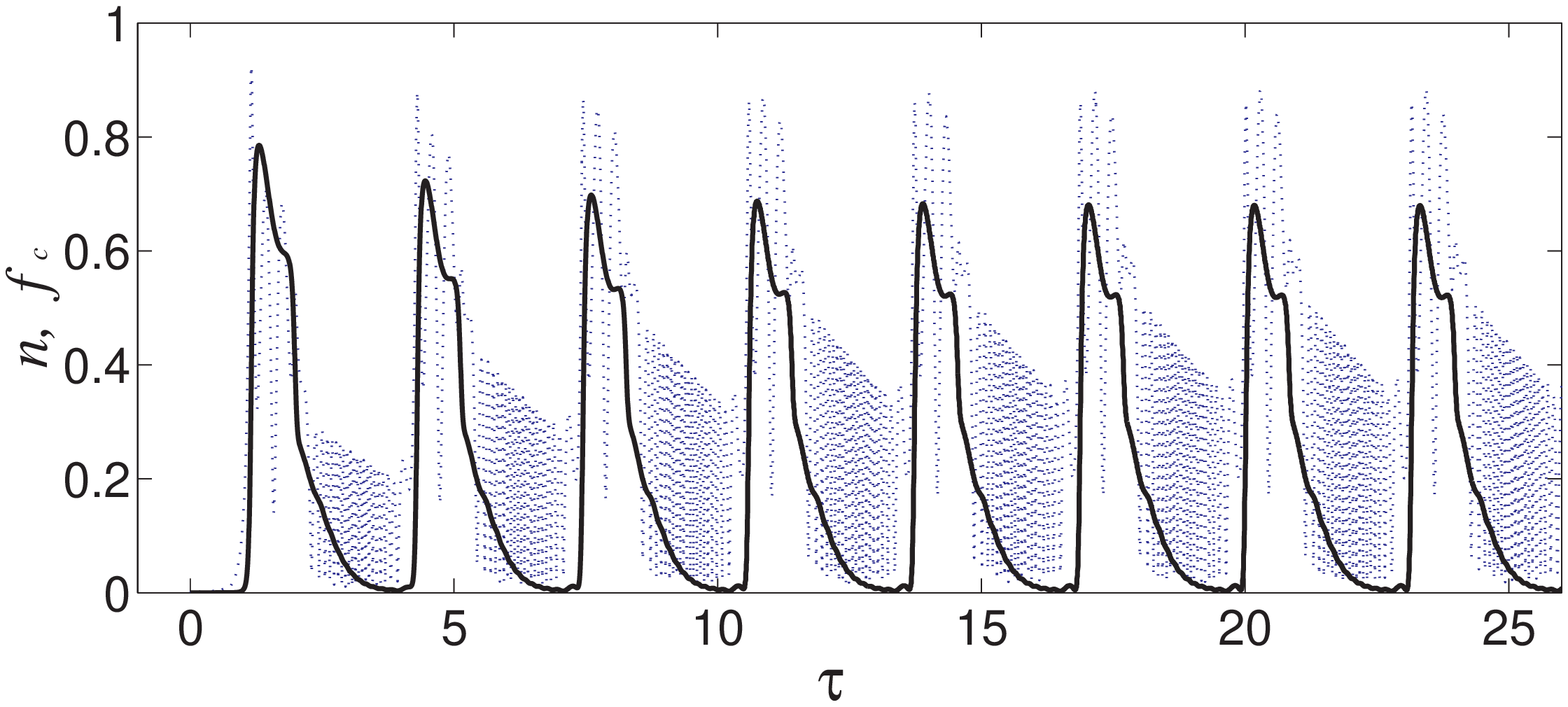}
\caption{Shown is the mean field number density $n_{{\vec k} h}$ of
  produced fermions, and the total amount of coherence $f^h_c\equiv
((f^h_1)^2 + (f^h_2)^2)^{1/2}$ (thin dotted blue line) for the negative
helicity state $h = -1$. Effects of inflaton oscillations are modelled
by a varying mass $m(t) = (10 + 15 \cos(2 \omega_{\varphi} t) + i
\sin(2 \omega_{\varphi} t))T$. The upper panel corresponds to the
collisionless case, while in the lower panel the collision
terms of Eq.~(\ref{coll_int}) were used, with the parameters $|\vec
k|=T$, $y=5, m_q = 0.02 T$, $m_\phi =0.1 T$, and $\omega_{\varphi}=
T$, where the temperature $T$ sets the scale. Initially, at $\tau  \equiv
\omega_{\varphi} t = 0$ the fermion system is taken to be an
uncorrelated vacuum state.} 
\label{fig:particle_ferm}
\end{figure}

Let us now summarize the results of the numerical calculations for
particle production through a parametric resonance. A more detailed
discussion is found in ref. \cite{paper2}. In the calculations we have used
the equations (\ref{rho_coll_intHOMOG2})-(\ref{coll_int}) with the
connection relations (\ref{rhocomp_HOM}), and the oscillatory mass
function Eq.~(\ref{osc_mass_ferm}). We have studied the time-evolution
of the fermionic particle number and coherence with varying
interaction strengths. In all cases, the initial condition for the
evolution has been an uncorrelated vacuum state with $n_{\vec{k}h} \equiv {\bar
  n}_{\vec{k}h} = f_{1,2} = 0$. The results of our calculations for
helicity $h=-1$ are presented in Fig.~\ref{fig:particle_ferm}. In the
upper panel we have considered the noninteracting case with
$\Gamma=0$. We see that the produced particle number (thick black
line) as well as the ``total amount'' of coherence $f_c \equiv
\sqrt{f_1^2 + f_2^2}$ (dotted blue line) increases steadily as a
function of time. This increase takes place around the resonance peaks
while between the peaks the particle number is essentially
constant and the coherence oscillates with a constant amplitude,
saturating to a maximum after a few resonance crossings. 
This picture of parametric resonance depends sensitively on the parameters of
the mass oscillation and the size of momentum $|\vec{k}|$ and helicity
$h$. For example for the opposite helicity $h=+1$ with otherwise the
same parameters we would get a completely different figure without a
clear resonance behavior.

\begin{figure}
\centering
\includegraphics[angle=270,width=0.9\textwidth]{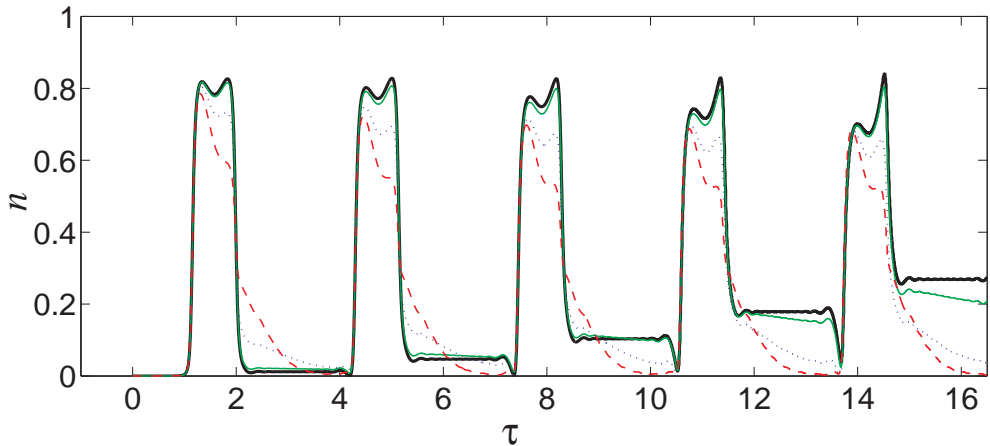}
\caption{Shown is the number density $n_{{\vec k} h}$ in the same setting
  as before with changing interaction strengths. The thick black line is
  the free field case with a coupling constant $y = 0$. The other lines
  are interacting cases, with $y = 1$ (green line),  $y = 3$ (blue dotted
  line) and  $y = 5$ (red dashed line).}
\label{fig:numberdensity_ferm}
\end{figure}

In the lower panel of Fig.~\ref{fig:particle_ferm} we have considered
the case with interactions, $\Gamma \neq 0$, with otherwise the same
parametrization as in the noninteracting case. The difference compared
with the noninteracting cases is quite dramatic. Now we see that the particle
number drops between the resonance peaks, which results from the
decays to the daughter particles, only to be regenerated again in the
next resonance crossing. Also the growth of the coherence is now
damped in comparison with the noninteracting case, and both the
particle number and the coherence evolution settle into a stationary
pattern after a few oscillation periods. In
Fig.~\ref{fig:numberdensity_ferm} we plot the particle number
evolution for varying interaction strengths, again with otherwise the
same parametrization as before. We see that the same pattern remains
with increase in particle number during the resonance peaks, while
the magnitude of damping (decays) between the peaks depends on the
strength of the interactions, as expected.

These results show that if the fermion has strong enough
interactions with other fields during the preheating, the effects
(especially decoherence) on the amount of produced particles can be
significant. Caution must be exercised in that our
approximation scheme with singular shell decompositions is not
guaranteed to produce correct quantitative results in this example,
because the rapidly oscillating mass function gives rise to big
gradient corrections. However, at least the qualitative picture is
expected to be correct.
%

%
%

\section{Coherent production of decaying scalar\\ particles}
\label{part_production_sca}

In this section we consider a similar coherent particle
production with decays as in the previous section
\ref{sec:ferm_preheating}, but now for scalar fields. 
A comprehensive study of resonant particle production using
the 2PI approach has been carried out in ref.~\cite{BerSer03a}, where a
complete numerical next-to-leading order calculation in a
$1/N$-expansion of the 2PI effective action was performed for an
$O(N)$-symmetric scalar theory. We are not trying to model the scalar
preheating scenario here, we simply want to compare the behaviour and
the results of decaying scalar fields with that of fermions. For this purpose we
take the mass term driving the scalar particle production to the
absolute value of the fermionic mass in Eq.~(\ref{osc_mass_ferm}):
\begin{equation}
m^2(t) \equiv \big|m_0 + A \cos(2\omega_\varphi t) +
i B\sin(2 \omega_\varphi t) \big|^2\,,
\label{secondmassterm}
\end{equation}
where  $m_0$, $A$, $B$ and $\omega_\varphi$ (oscillation frequency of the
driving field $\varphi$) are real constants. For the
decays we use the following Yukawa interaction:
\begin{equation} 
{\cal L}_{\rm int} = - y\, \bar\psi \psi \phi\,,
\label{interactionA}
\end{equation}
where $\phi$ is the real scalar field considered and $\psi$ is some
fermion field, which is assumed to be in thermal equilibrium
throughout the evolution. Hence the scalar field self
energies $\Pi^{<,>}$ are related by the KMS relation $\Pi^>(k) = e^{\beta
  k_0} \Pi^<(k)$ and the interaction width $\Gamma$ is given by
\begin{equation}
\Gamma(k) = \frac{1}{2}(1+e^{\beta k_0} ) i\Pi^<(k) \,.
\label{gammakms}
\end{equation}
Using these expressions and the mean field spectral function in
Eq.~(\ref{spectral_function_sca}) we find that up to first order 
gradients in the collision term $i{\cal C}_{\rm coll}^\phi$, the collision
integrals in the dynamical equations (\ref{rho_Eq_Coll1}) are given by    
\begin{eqnarray}
\left<{\cal C}^+\right> 
&=& \frac12 \partial_{k_0}\Gamma_0 \partial_t f_c
\nonumber\\
 \left<{\cal C}^-\right> 
&=& - \frac{1}{2\omega_{\vec k}}\Gamma_m\left[(f_+ + f_-) - (f^{\rm eq}_+ +
  f^{\rm eq}_-)\right]
\nonumber\\
\left<k_0{\cal C}^-\right>
&=& - \frac12\Gamma_m\left[(f_+ - f_-) - (f^{\rm eq}_+ - f^{\rm eq}_-)\right] \,,
\label{coll_terms_thermal}
\end{eqnarray}
where we have now defined $f^{\rm eq}_\pm \equiv 1/(e^{\pm\beta \omega_{\vec
    k}}-1)$, and we have neglected terms of order\footnote{These terms are
  neglected also indirectly, by recursive use of equations of motion
  (\ref{rho_Eq_Coll1})} ${\cal
  O}(\Gamma^2,\Gamma \partial_tm^2)$. The $\Gamma_i$-functions appearing in
Eq.~(\ref{coll_terms_thermal}) are again projections onto the mass-
and the coherence shells:
\begin{eqnarray}
\Gamma_m(|\vec{k}|,t) &\equiv& \Gamma(k_0=\omega_{\vec k}(t),|\vec{k}|)
\qquad\quad (\textrm{positive mass-shell})
\nonumber\\ 
\partial_{k_0}\Gamma_0(|\vec{k}|)
&\equiv& \partial_{k_0}\Gamma(k_0=0,|\vec{k}|)\qquad\quad \;\,
(\textrm{$k_0 = 0$ -shell})\,. 
\label{collfunctions}
\end{eqnarray}
As in the previous section, we present no explicit expressions
for these $\Gamma_i$-functions here, but they can be read from the
equations (\ref{gammakms}) and (A.5-A.11) in ref.~\cite{paper3}. Note
that $\Gamma$ on the negative mass-shell is again simply related to
that on the positive mass-shell: $\Gamma(-k_0,|\vec{k}|) =
-\Gamma(k_0,|\vec{k}|)$ and it vanishes on $k_0=0$-shell. This is the
reason that we need to include the first order gradients of the
collision term in the scalar case, as we see that
$\partial_{k_0}\Gamma_0$ is the lowest order nonvanishing $k_0=0$-shell
contribution, giving rise to decoherence effects. Using the relations
(\ref{rho-f_HOM}) we can now express the on-shell functions $f_{\pm,c}$ in terms of the moments
$\rho_{0,1,2}$ to find a closed set of equations for these moment functions:
\begin{eqnarray}
\frac14 \partial_t^2 \rho_0 + \omega_{\vec k}^2 \rho_0 - \rho_2 &=& -\frac12\partial_{k_0}\Gamma_0\,\partial_t\rho_0
\nonumber\\
\partial_t\rho_1 &=& -\frac{1}{\omega_{\vec k}}\Gamma_m \left(\rho_1 - \rho_{1,{\rm
      eq}}\right)
\nonumber\\
\partial_t\rho_2 - \frac12 \partial_t(m^2) \rho_0 &=& -\frac{1}{\omega_{\vec k}}\Gamma_m \left(\rho_2 - \rho_{2,{\rm
      eq}}\right)\,,
\label{rho_Eq_Coll2}
\end{eqnarray}
where $\rho_{i,{\rm eq}}$ are the the thermal equilibrium values for
the moments with $f_\pm=f^{\rm eq}_\pm$ and $f_c=0$.

\begin{figure}
\centering
\includegraphics[width=0.85\textwidth]{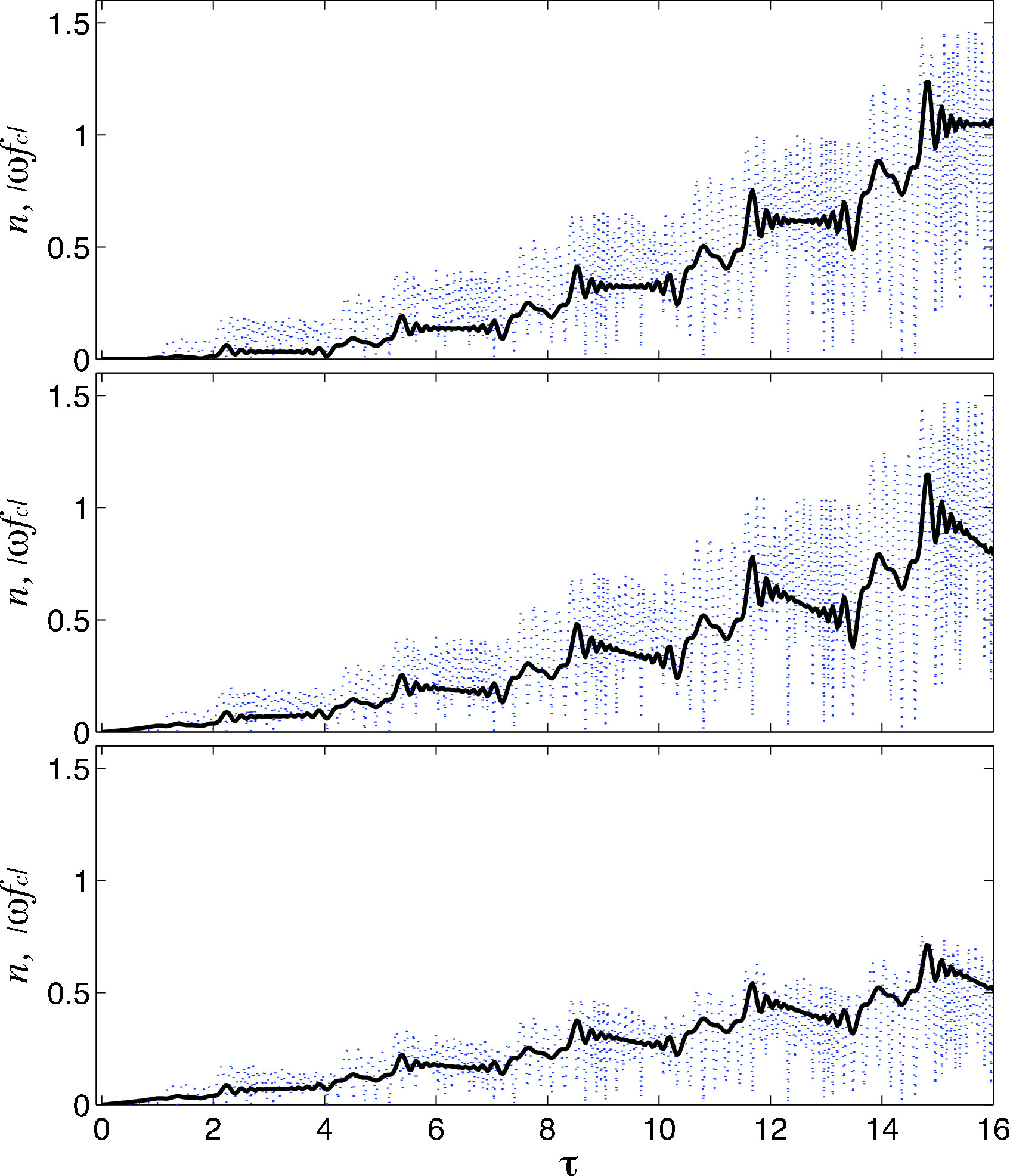}
\caption{Shown is the number density $n_{{\vec k}}$ (thick solid line)
  and the coherence function $f_c(|\vec k|)$ (dotted(blue) line) with
  changing interactions. The driving mass function is taken to be
  $m^2(t) = |(1 + 1.5 \cos(2 \omega_{\varphi} t) + i\, 0.1 \sin(2
  \omega_{\varphi} t))T|^2$. The upper panel corresponds to the case
  without collisions. In the central panel we have included the 
  collision terms on the mass-shells, but kept $\partial_{k_0}
  \Gamma_0 = 0$. In the lowest panel the full interaction terms were
  kept for all shells. For parameters we have used $|\vec k| = 0.6\,
  T$, $y = 1$, $m_\psi=0.1\, T$ and $\omega_\varphi = 0.1 T$, where
  the temperature $T$ sets the scale. Initially, at $\tau \equiv
  \omega_\varphi t = 0$ the system is in the adiabatic vacuum.}
\label{fig:numberdensity1}
\end{figure}

Let us now summarize the results of the numerical calculations for
particle production through a parametric resonance. A more
detailed discussion can be found in ref. \cite{paper2}. It
turns out that the equations of motion (\ref{rho_Eq_Coll2}) become
numerically unstable for rapidly oscillating driving mass terms in
Eq.~(\ref{secondmassterm}). For that reason we have written
Eq.~(\ref{rho_Eq_Coll2}) in a different form with nonlinear terms and
an integration constant, given by Eqs.~(8.3)-(8.5) in ref. \cite{paper3}, for
which the stability problems do not occur. Using these equations we
have performed similar calculations as for fermions in section
\ref{sec:ferm_preheating} \ie we have studied the time-evolution
of the particle number and coherence from an uncorrelated vacuum
state $n_{\vec{k}} \equiv {\bar n}_{\vec{k}} = f_c = 0$, with varying
interaction strengths. 

The results of the calculations are presented
in Fig.~\ref{fig:numberdensity1}. In the upper panel we have considered
the noninteracting case with $\Gamma=0$. The increase of 
the number density (thick solid line) is seen to be accompanied by a steady
growth of the amplitude of the coherence (thin dotted blue
line). Similar to fermions there are regions (resonance ``peaks'') 
where the increase takes place, while between the peaks the particle
number and the amplitude of the coherence are relatively constant. In
the middle panel we considered the case where only the mass-shell
collision terms were included but we set artificially
$\partial_{k_0}\Gamma_0\equiv 0$, while in the lowest panel all
collision terms were included properly. 
In these cases with interactions we see again that the
decrease of the particle number (decays to daughter particles) and of the
amplitude of the coherence (decoherence) takes place between the
resonance peaks. By comparing the two lowest panels we can conclude
that the dominant effect of interactions comes from the coherence
shell contribution, $\partial_{k_0}\Gamma_0$, which would be absent in
the standard quantum Boltzmann approach. 
\begin{figure}
\centering
\includegraphics[angle=270,width=0.85\textwidth]{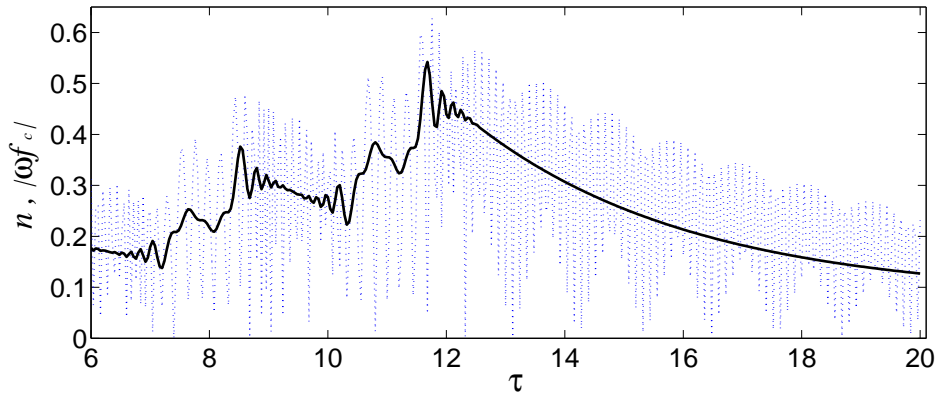}
\caption{The same configuration as in the lowest panel of
  Fig. \ref{fig:numberdensity1} but with the driving mass term smoothly
  set to a constant $m = 2.5 \,T $ after $\tau > 4 \pi$. }
\label{fig:decoherence}
\end{figure}

To see the effects of thermalization and decoherence more clearly, we
consider finally a case where the driving term was switched off smoothly at a
specific moment. This scenario is depicted in
Fig.~\ref{fig:decoherence}. We can see a smooth thermalization of the
particle number towards the equilibrium value, accompanied by
decoherence towards zero in the form of damped oscillations of the
coherence solution $f_c$. From the master equations
(\ref{rho_Eq_Coll2}) we can identify that the corresponding
thermalization and decoherence rates are given
by $\Gamma_m$ and $\partial_{k_0} \Gamma_0$, respectively.
This example of thermalization and decoherence of a highly coherent
initial state provides a more suited application for our eQPA scheme,
as the big gradients of the background field are no longer a
problem. It would be interesting to study a more realistic example of
scalar field thermalization using our methods, \eg with a model of quartic self
interaction.

\subsection{Resummation of the coherence contributions in\\ the collision
  term}
\label{sec:resummation}

In the examples of sections \ref{sec:ferm_preheating} and
\ref{part_production_sca} the gradient expansion of the collision term
was truncated to the lowest order that includes decoherence. For
fermions this was the zeroth order and for scalars the
first order. However, as discussed in section \ref{sec:validity}, also
the higher order coherence contributions in the collision term should
be taken into account, since their expansion is not (completely)
controlled by small parameters. In this section we
present an analysis that shows how the coherence contributions in the
collision term can be resummed for a scalar field with a thermal
interaction. The results of this section were not presented in
refs.~\cite{paper2,paper3}, and a new publication \cite{HKR4} on these
and the corresponding results for fermions is being prepared.

To be specific, we want to expand the collision term to leading order
in $\Gamma$ (and $\partial_t m^2$): ${\cal O}_1 \equiv {\cal
  O}(\Gamma, \partial_t m^2)$, neglecting the terms of order
${\cal O}_2 \equiv {\cal O}(\Gamma^2, \Gamma\,\partial_t m^2)$ and
higher. We begin by an observation that the on-shell functions obey
the zeroth order equations:
\beqa
\partial_t f_\pm &=& {\cal O}_1
\label{f_eq_leading1}
\\
\partial_t^2 f_c + 4 \omega_{\vec{k}}^2 f_c &=& {\cal
  O}_1\,,
\label{f_eq_leading2}
\eeqa
which follow directly from Eqs.~(\ref{rho_Eq_Coll1}) and the relations
(\ref{rho-f_HOM}). Furthermore, the first part of the (standard) collision term
(\ref{collintegral}) gives
\beq
e^{-i\Diamond}\{\Gamma\}\{i\Delta^<\} =
\exp\Big(\frac{i}{2}\partial_{k_0}^\Gamma \partial_t^\Delta\Big)
\Gamma\,i\Delta^< + {\cal O}_2\,, 
\eeq
since $\partial_t \Gamma$ is proportional to $\partial_t m^2$ (as well
as to $\Gamma$) for the interaction with thermal background. When the
$k_0$-integration is performed for a term of order $n \geq 1$ of
this expansion, we see that the only leading order contribution is the
projection to the $k_0$-shell:
\beq
 \int \frac{{\rm d}
   k_0}{2\pi}\;\partial_{k_0}^n\Gamma\,\partial_t^n i\Delta^<
 = \partial_{k_0}^n\Gamma|_{k_0=0}\,\partial_t^n
 f_c + {\cal O}_2\,,
\eeq
where we have used Eqs.~(\ref{f_eq_leading1}). Now, using
Eq.~(\ref{f_eq_leading2}) we find that to zeroth order all the
derivatives $\partial_t^nf_c$ are proportional to either $f_c$ or
$\partial_t f_c$: $\partial_t^{2n} f_c = (- 4 \omega_{\vec{k}}^2)^n
f_c + {\cal O}_1$ and $\partial_t^{2n+1} f_c = (- 4
\omega_{\vec{k}}^2)^n \partial_t f_c + {\cal O}_1$.
Using these relations we get then
\beqa
\int \frac{{\rm d}
   k_0}{2\pi}\;e^{-i\Diamond}\{\Gamma\}\{i\Delta^<\} &=&
\cosh(\omega_{\vec{k}}\partial_{k_0})
  \Gamma|_{k_0=0}\,f_c +
  i\frac{1}{2\omega_{\vec{k}}}\sinh(\omega_{\vec{k}}\partial_{k_0})
  \Gamma|_{k_0=0}\,\partial_t f_c  
\nonumber\\
&& + {\rm regular} + {\cal O}_2\,,  
\eeqa
where regular refers to standard first order mass-shell
contribution. The second part of the collision term
(\ref{collintegral}), proportional to the spectral function ${\cal
  A}$, gives only the regular contribution, since the spectral
function does not involve the coherence contributions in our
approach. Putting things together, we find the following generalizations of Eq.~(\ref{rho_Eq_Coll2}):
\begin{eqnarray}
\frac14 \partial_t^2 \rho_0 + \omega_{\vec k}^2 \rho_0 - \rho_2 &=&
-\frac{1}{2\omega_{\vec{k}}}\sinh(\omega_{\vec{k}}\partial_{k_0})
\Gamma|_{k_0=0}\, \partial_t\rho_0
\nonumber\\
\partial_t\rho_1 &=& -\frac{1}{\omega_{\vec k}}\Gamma_m \left(\rho_1 -
  \rho_{1,{\rm eq}}\right) - \cosh(\omega_{\vec{k}}\partial_{k_0})
  \Gamma|_{k_0=0}\Big(\rho_0 - \frac{1}{\omega_{\vec{k}}^2}\rho_2\Big)
\nonumber\\
\partial_t\rho_2 - \frac12 \partial_t(m^2) \rho_0 &=& -\frac{1}{\omega_{\vec k}}\Gamma_m \left(\rho_2 - \rho_{2,{\rm
      eq}}\right)\,,
\label{rho_Eq_Coll3}
\end{eqnarray}
where the terms of order ${\cal O}_2$ are neglected. We see that the infinite
$k_0$-derivative expansions of $\Gamma$ in Eq.~(\ref{rho_Eq_Coll3}) are
not controlled by the small parameters $\Gamma$ and $\partial_t
m^2$; instead they provide expansions in powers of $\beta
\omega_{\vec{k}}$ or $\omega_{\vec{k}}/|\vec{k}|$ etc. that are
typically of order unity. Moreover,
because of the generally nontrivial $k_0$-dependence of the (thermal)
$\Gamma$, a brute force calculation of the (infinitely many) derivatives seems
impossible. However, it turns out that these expansions can be
resummed by a neat trick. That is, by denoting $\tilde\Gamma(r_0)$ the
Fourier transform of $\Gamma$ with respect to $k_0$ and using the
identity $\sinh(ix)=i\sin(x)$, we find
\beqa
\sinh(\omega_{\vec{k}}\partial_{k_0}) \Gamma|_{k_0=0} &=&
\bigg[\int {\rm d} r_0 \,\sinh(i\omega_{\vec{k}} r_0)
\tilde\Gamma(r_0) e^{i k_0 r_0}\bigg]_{k_0=0}
\nonumber\\
&=& \int {\rm d} r_0 \,\frac12 \big( e^{i\omega_{\vec{k}} r_0} -
e^{-i\omega_{\vec{k}} r_0}\big) \tilde\Gamma(r_0)
\nonumber\\
&=& \frac12 \big[\Gamma(\omega_{\vec{k}}) - \Gamma(-\omega_{\vec{k}})\big]\,,
\eeqa
and similarly
\beq
\cosh(\omega_{\vec{k}}\partial_{k_0}) \Gamma|_{k_0=0} = \frac12 \big[\Gamma(\omega_{\vec{k}}) + \Gamma(-\omega_{\vec{k}})\big]\,.
\eeq
For the thermal interaction considered we furthermore have:
$\Gamma(-\omega_{\vec{k}}) = -\Gamma(\omega_{\vec{k}})$, so that we
get simply: $\sinh(\omega_{\vec{k}}\partial_{k_0}) \Gamma|_{k_0=0} =
\Gamma(\omega_{\vec{k}}) \equiv \Gamma_m$
and $\cosh(\omega_{\vec{k}}\partial_{k_0}) \Gamma|_{k_0=0} = 0$. When
these amazingly simple results are substituted in Eq.~(\ref{rho_Eq_Coll3})
we get the final equations to the leading order as
\begin{eqnarray}
\frac14 \partial_t^2 \rho_0 + \omega_{\vec k}^2 \rho_0 - \rho_2 &=&
-\frac{1}{2\omega_{\vec{k}}} \Gamma_m \,\partial_t\rho_0
\nonumber\\
\partial_t\rho_1 &=& -\frac{1}{\omega_{\vec k}}\Gamma_m \left(\rho_1 -
  \rho_{1,{\rm eq}}\right)
\nonumber\\
\partial_t\rho_2 - \frac12 \partial_t(m^2) \rho_0 &=&
-\frac{1}{\omega_{\vec k}}\Gamma_m \left(\rho_2 - \rho_{2,{\rm eq}}\right)\,,
\label{rho_Eq_Coll4}
\end{eqnarray}
where the terms of order ${\cal O}(\Gamma^2, \Gamma\,\partial_t m^2)$
and higher have been consistently neglected. We see that due to the
resummation of the coherence contributions, the resulting equations
have actually simplified, as $\Gamma_m$ is now the only interaction
term to be evaluated. However, what is more important, these equations
provide a controlled expansion of the collision term to the leading
order in $\Gamma$ and $\partial_t m^2$.
\begin{figure}[ht!]
\centering
\includegraphics[width=0.9\textwidth]{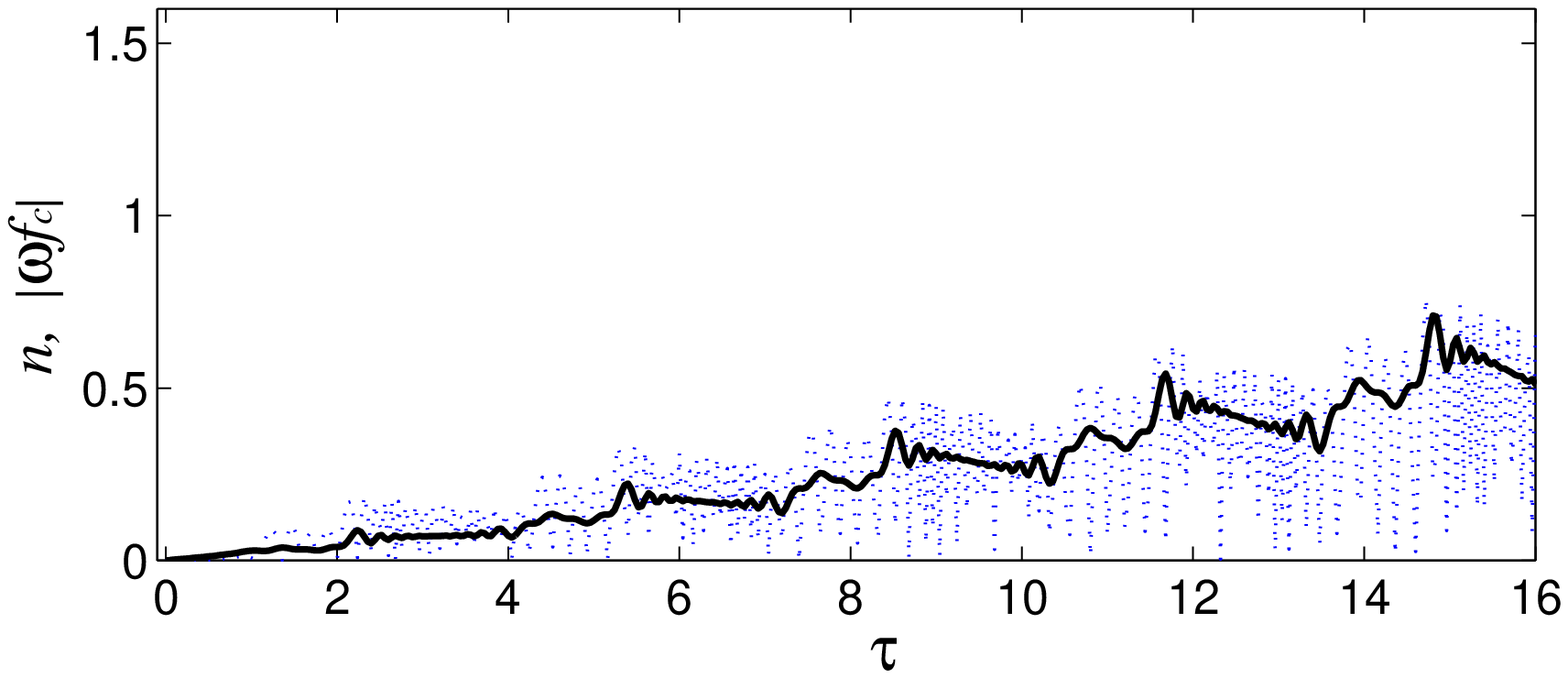} 
\vskip-0.73truecm
\includegraphics[width=0.9\textwidth]{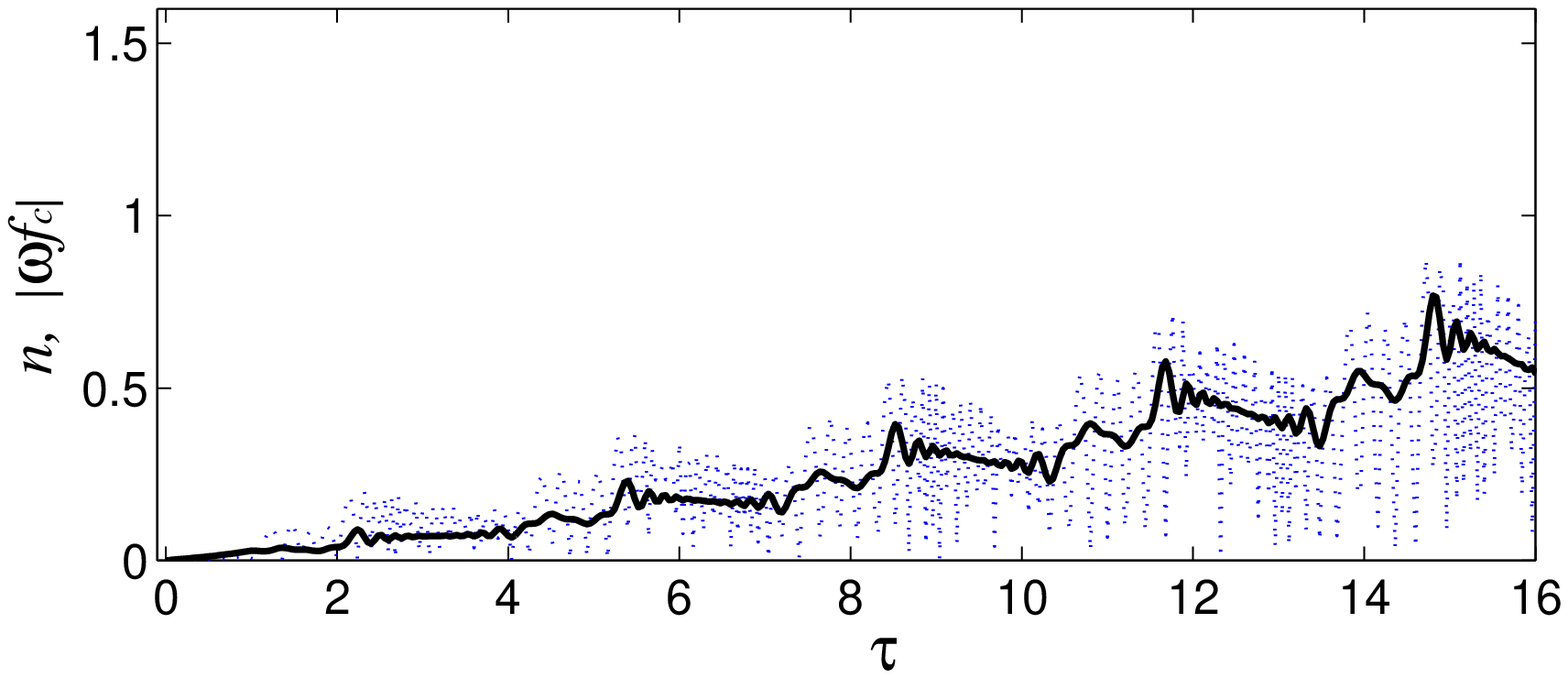}
\caption{The number density $n_{{\vec k}}$ (thick solid line)
  and the coherence function $f_c(|\vec k|)$ (dotted(blue) line) in
  the cases of non-resummed (upper panel) and resummed (lower panel)
  coherence shell collision terms. The configuration is the same as in
  the lowest panel of Fig.~\ref{fig:numberdensity1}.} 
\label{fig:numberdensity2}
\end{figure}

\begin{figure}
\centering
\includegraphics[angle=270,width=0.85\textwidth]{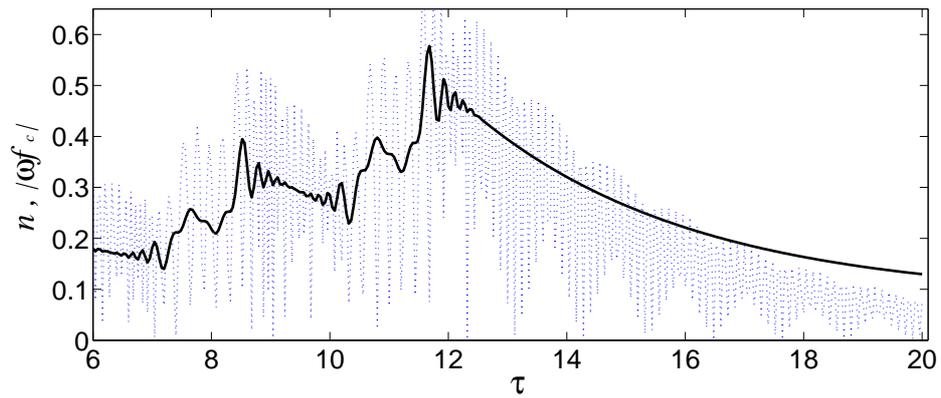}
\caption{The same configuration as in Fig.~\ref{fig:decoherence} but
  with a resummed coherence shell collision term.}
\label{fig:decoherence_resum}
\end{figure}

Let us now compare the numerical results obtained by using the
resummed equations (\ref{rho_Eq_Coll4}) to the original ones, presented
in Figs.~\ref{fig:numberdensity1}-\ref{fig:decoherence}. In figure
\ref{fig:numberdensity2} we show the results with the original
(non-resummed) collision term in the upper panel and the resummed
one in the lower panel. We have used the same configuration as in
Fig.~\ref{fig:numberdensity1}, so the upper panel here is just a copy
of the lowest panel in Fig.~\ref{fig:numberdensity1}. We see only a
slight difference in the results, with the net amount
of produced particles slightly increased in the resummed case. This is
explained in that the average value of the oscillating resummed
decoherence rate $\Gamma_m/\omega$ is approximately equal to
the constant non-resummed rate $\partial_{k_0} \Gamma_0$ in this
particular case. In the example of thermalization and decoherence,
however, a bigger difference can be seen. The case with the resummed
collision term and otherwise the same configuration as in
Fig.~\ref{fig:decoherence} is presented in
Fig.~\ref{fig:decoherence_resum}. We see that decoherence effects
are clearly stronger in the resummed case, because now $\Gamma_m/\omega$
freezes to a value that is significantly larger than $\partial_{k_0}
\Gamma_0$. The thermalization rate of the particle number, on the
other hand, is not changed at all, because the particle number and
the coherence solution decouple in the constant mass limit in the case
of a simple thermal interaction, and thus the (increased) decoherence
rate does not affect the thermalization of the particle number.

%% file: Conclusions.tex
%
%

\chapter{Conclusions and outlook}
\label{chap:discussion}

We have developed a novel approximation scheme (eQPA) that enables us to
include the effects of nonlocal quantum coherence in the standard
kinetic approach to nonequilibrium quantum dynamics. The key element
in our scheme is the finding of new singular shell
solutions in the phase space of 2-point Wightman function, located at
$k_0 = 0$ for spatially homogeneous problems and at $k_z = 0$ for a
static planar symmetric case. We have interpreted these new 
solutions as describing the nonlocal coherence between the ``opposite''
mass-shell excitations with either $k_0 = \pm\omega_{\vec k}$ or $k_z
= \pm k_m$, respectively. This nontrivial phase space
structure is then inserted in the dynamical Kadanoff-Baym equation for the
2-point correlator, leading to a closed set of transport equations for the
corresponding on-shell distribution functions $f$, that provides an
extension of the standard quantum Boltzmann equation to include
nonlocal coherence. 

We have shown how the eQPA scheme emerges from first principles in
chapters \ref{chap:basic_formalism}-\ref{chap:scheme}. We used the 2PI
effective action method to derive the self-consistent Schwinger-Dyson
equations for the 2-point correlation functions of the system. We then
performed a partial Fourier transformation to obtain the Kadanoff-Baym
transport equations for the Wightman function in the mixed
representation. When these equations were analyzed in detail in the
kinetic regime with weak interactions and a slowly varying background
field, the nontrivial singular shell picture with the new coherence
solutions was found.

The analysis for fermions was originally presented in
refs.~\cite{paper1,paper2}, and for scalar fields in
ref.~\cite{paper3}. In these papers we
have considered several applications to demonstrate the use of our
formalism. The (collisionless) Klein problem was considered in
refs.~\cite{paper1,paper3} for fermions and scalars, respectively. In
ref.~\cite{paper1} we also considered the (fermionic) quantum
reflection from a $CP$-violating mass wall, and a topic omitted in the
introductory part of this thesis, a generalization of our formalism to flavour
mixing. The coherent production of decaying fermionic and scalar
particles was considered in refs.~\cite{paper2,paper3},
respectively. Finally, the nonrelativistic limit of our formalism was
considered in ref.~\cite{paper3}.

In the collisionless examples of the Klein problem and the fermionic
quantum reflection from a smooth mass wall, presented in sections
\ref{sec:klein} and \ref{sec:mass_wall}, our formalism was seen to
reproduce the same (exact) results as the standard approach using
the Dirac and Klein-Gordon equations. Of course, the whole point of
our approach is the ability to include the effects of collisions
together with the nonlocal coherence. We studied the role of
collisions in the examples of coherent production of decaying
particles in sections \ref{sec:ferm_preheating} and
\ref{part_production_sca}. We saw a clear decrease in the amount of
produced particles due to smooth thermalization and decoherence
effects caused by the collisions. A caution must be given, in that the
parameters of the models were chosen such that the gradients of
the background field were not necessarily small, and thus the numerical
results of these particular examples are perhaps only qualitatively correct.
In section \ref{sec:resummation} we have shown how a resummation can
be carried out for the coherence contribution of the (infinite) gradient
expansion of the collision term. This procedure leads to a controlled
leading order result in powers of the interaction width $\Gamma$ and
gradients of the background field. These new results are not presented in
refs.~\cite{paper1,paper2,paper3} but will be published elsewhere
\cite{HKR4}, along with the corresponding results for fermions. 
 
One of the most interesting future applications for our eQPA scheme is a
realistic calculation of the baryon asymmetry in electroweak
baryogenesis, where the nonequilibrium dynamics of fermions in the
neighbourhood of an expanding $CP$-violating phase transition wall
needs to be considered. As discussed above, the relevant quantum
reflection effects from the wall have not yet been considered in a
consistent framework based on quantum field theory, including the
effects of decohering collisions. Our eQPA scheme provides a tool for
the study of such a problem with a smoothly varying background field
and well-defined asymptotic regions. Indeed, the $k_z=0$-shell
coherence solutions are just the missing piece for describing the
nonlocal quantum reflection effects. In the case of a thick wall with
small gradients, the mean field approximation works well
throughout, and in the opposite limit of a thin wall at least the
correct qualitative behaviour is expected. However, the
results in this case should be quite accurate also quantitatively, since the
region of (rapidly) varying mass is very narrow, and outside the wall
region the mean field approximation works well again.

Other interesting applications for our formalism include neutrino
flavour oscillations in inhomogeneous background, where the relevant
coherence effects occur between different flavour states travelling in
the same direction. More generally, we expect that a number of other
problems that can be studied in the standard kinetic
approach with the Boltzmann equations, could be studied using our eQPA scheme
including the coherence effects.